\documentclass[twocolumn]{aastex63}
\usepackage{listings}
\usepackage{mathtools}
\turnoffedit

\accepted{May 16, 2021}

\shorttitle{Young Stellar Populations}
\shortauthors{Kerr et al.}

\begin{document}

\title{Stars with Photometrically Young Gaia Luminosities Around the Solar System (SPYGLASS) I: Mapping Young Stellar Structures and their Star Formation Histories}

\correspondingauthor{Ronan Kerr}
\email{rmpkerr@utexas.edu}

\author[0000-0002-6549-9792]{Ronan Kerr}
\affiliation{Department of Astronomy, University of Texas at Austin\\
2515 Speedway, Stop C1400\\
Austin, Texas, USA 78712-1205\\}

\author{Aaron C. Rizzuto}
\altaffiliation{51 Pegasi b Fellow}
\affiliation{Department of Astronomy, University of Texas at Austin\\
2515 Speedway, Stop C1400\\
Austin, Texas, USA 78712-1205\\}

\author[0000-0001-9811-568X]{Adam L. Kraus}
\affiliation{Department of Astronomy, University of Texas at Austin\\
2515 Speedway, Stop C1400\\
Austin, Texas, USA 78712-1205\\}

\author{Stella S. R. Offner}
\affiliation{Department of Astronomy, University of Texas at Austin\\
2515 Speedway, Stop C1400\\
Austin, Texas, USA 78712-1205\\}

\begin{abstract}

Young stellar associations hold a star formation record that can persist for millions of years, revealing the progression of star formation long after the dispersal of the natal cloud. To identify nearby young stellar populations that trace this progression, we have designed a comprehensive framework for the identification of young stars, and use it to identify $\sim$3$\times 10^4$ candidate young stars within a distance of 333 pc using Gaia DR2. Applying the HDBSCAN clustering algorithm to this sample, we identify 27 top-level groups, nearly half of which have little to no presence in previous literature. Ten of these groups have visible substructure, including notable young associations such as Orion, Perseus, Taurus, and Sco-Cen. We provide a complete subclustering analysis on all groups with substructure, using age estimates to reveal each region's star formation history. The patterns we reveal include an apparent star formation origin for Sco-Cen along a semicircular arc, as well as clear evidence for sequential star formation moving away from that arc with a propagation speed of $\sim$4 km s$^{-1}$ ($\sim$4 pc Myr$^{-1}$). We also identify earlier bursts of star formation in Perseus and Taurus that predate current, kinematically identical active star-forming events, suggesting that the mechanisms that collect gas can spark multiple generations of star formation, punctuated by gas dispersal and cloud regrowth. The large spatial scales and long temporal scales on which we observe star formation offer a bridge between the processes within individual molecular clouds and the broad forces guiding star formation at galactic scales.

\end{abstract}

\keywords{young stars --- open clusters and associations --- star formation --- T Tauri stars --- Sco-Cen --- Taurus --- Perseus --- Orion}

\section{Introduction} \label{sec:intro}

Most young stars are not found in isolation, instead residing in co-moving star clusters or associations (e.g., \citealt{lada03,Krumholz19}). These stellar overdensities are direct remnants of the molecular cloud that collapsed to create them, and as such they preserve significant information on the structure and dynamics of those parent clouds \citep{Elmegreen77,Krause20}. While studies of sites of active star formation are popular due to the presence of both young stars and dense gas for which exquisite dynamical studies can be performed \citep[e.g.,][]{Palla99,Hatchel05, Tobin09, Kirk13, Kirk17,Kerr19}, studies of clusters and associations provide a much longer-term view of star formation that has unique value. Rather than providing a snapshot in time, a detailed study of a young association can trace back tens of millions of years of star formation, enabling a complete study of a star-forming event from its onset to present day, and by extension the processes that drove the evolution of the population (e.g.,\citealt{deZeeuw99,Pecaut16, Wright18}). Furthermore, as these associations persist long after gas dispersal, the study of stellar populations through associations may also enable the identification of more rapid forms of star formation that do not have known equivalents in active star-forming sites.

Some processes in star formation require a complete record of the star-forming event, not just a snapshot, to be properly studied in nature. One example is sequential star formation, a process in which previous generations of stars compress the cloud beside them, which can then collapse to form stars, producing a self-sustaining cycle of star formation that can slowly propagate across an entire molecular cloud \citep{Elmegreen77}. Most cases where a sequential process has been suggested include just two generations of star formation: one recently-formed generation powering an H II region, and one site of active star formation triggered in a shell that the previous generation compressed \citep[e.g.,][]{Lee05,Maaskant11, Nony21}. Given that these processes are capable of continuing without limit as long as unused gas remains, the current view of sequential star formation has yet to explore large scales in both time and space. 

One of the greatest strengths in using clusters and associations as a record of past star formation lies in the time over which detailed information can be extracted. Having such an extensive star formation record allows for the study of long-lived star-forming processes, while also revealing unexpected anomalies in time, such as periods of dormancy in the star formation record. As simulations become increasingly sophisticated and begin to include more physical processes \citep[e.g.,][]{Grudic20}, having a robust record of star formation, complete with currently unexplained features can provide critical comparisons capable of testing new theoretical frameworks. 

While more spatially compact open clusters have long ago been discovered and catalogued \citep[e.g.,][]{Trumpler21,KleinWassink27,Mermilliod95}, much sparser stellar associations are considerably more challenging to identify due to relatively sparse on-sky densities and large spatial extents, often barring identification from on-sky density. For the nearest populations, stars with strong indicators of youth such as protoplanetary disks have typically been used as signposts for the identification of associated populations nearby, such as TW Hydrae and $\beta$ Pictoris \citep[e.g.,][]{Kastner97,Zuckerman01}. Beyond about 50-100 pc, however, recognizing stars with youth indicators and finding potential companions becomes increasingly difficult, leaving the population of low-mass, more distant associations largely unexplored. Due to their unbound nature, small velocities relative to a typical field star, and geometric effects from their wide spatial distributions, the kinematics of associations are often difficult to disentangle from the field. 
The effective identification of young stellar populations therefore requires both the suppression of older background populations and the use of accurate kinematics and 3-d spatial coordinates to properly group the young stars together and distinguish a young association from the field. Improvements in measurements of parallax and proper motions can therefore significantly expand our knowledge of these associations, and such developments frequently result in the addition of new members and the revision of associations' known spatial extents \citep[e.g.,][]{deZeeuw99,Preibisch08,Rizzuto15,Zari18,Kraus17,Luhman18}. With its unprecedented breadth of spatial, kinematic, and photometric data, Gaia Data Release 2 provides a data set capable of dramatically expanding our knowledge of young stellar populations \citep{GaiaDR218}. Gaia photometry can be combined with parallaxes to identify stars based on their high locations on the HR diagram, and those stars can then be clustered according to their spatial coordinates and transverse motions.

There have been multiple recent searches for stellar populations in the solar neighborhood enabled by Gaia DR2, including \citet{Sim19}, which reports the discovery of 207 new open clusters, and \citet{Kounkel19}, which notes the identification of 1901 stellar overdensities in Gaia DR2. However, these surveys included stars of all ages in their clustering, so young populations may not always stand out above the field density of the older stars. The work of \citet{Zari18} does focus on these younger populations, separating out pre-main sequence stars quite effectively and revealing multiple associations in the form of stellar overdensities. While \citet{Zari18} revealed substantial young stellar populations, it did not include a clustering analysis, and as such the structures present within those populations have yet to be rigorously defined. Some recent investigations have focused on young stellar populations while also performing a complete clustering analysis, however these have all been restricted to individual associations, such as \citet{Zari19}, which studies Orion, and \citet{CantatGaudin19}, which focuses on Vela. No paper to date has performed a spatially unbiased, all-sky search for young stellar populations with both robust youth diagnostics and a broad clustering analysis.

In this paper, we present the deepest comprehensive study of young stellar populations in the solar neighborhood to date. In Section \ref{sec:data}, we present the Gaia DR2 data set used for this work. We outline our methods for the identification of young stars, age estimation, and clustering in Section \ref{sec:style}, while a detailed analysis into the populations we identify is performed in Section \ref{sec:results}. Possible implications of this work with respect to the study of star formation progression in the solar neighborhood are discussed in Section \ref{sec:discussion}, while we conclude in Section \ref{sec:conclusion}.

\section{Data} \label{sec:data}

\subsection{Gaia Astrometry and Photometry} \label{sec:dat-gaia}

In this paper, we search for nearby young stars among a large sample drawn from Gaia Data Release 2 \citep{GaiaDR218,GaiaDR2ASLindegren}. When querying sources from the Gaia Archive, several search restrictions are included to ensure a manageable sample size, and to guarantee that all stars have high-quality photometric and astrometric measurements. First, we impose a search radius of parallax $\pi > 3$ mas, limiting our sample to stars at distances comparable to Taurus and Sco-Cen, two of the nearest young associations. Both Taurus and Sco-Cen are thought to extend to a maximum distance of approximately 200 pc \citep{Preibisch08, Galli19}, so a minimum parallax of 3 mas (d$<$333 pc) allows for the exploration of adjoining structures that may exist up to 100 pc beyond their known extents, in addition to numerous other structures, both known and unknown. This radius also includes nearly all of Perseus OB2 and the near edge of Orion \citep{Zari19,Bally08}, allowing for partial coverage in these more distant young associations. 

We then impose several restrictions based on quality indicators, following \citet{Arenou18}. Firstly we restrict the sample using the Unit Weight Error ($u$), which can be interpreted as a goodness of fit measurement to the astrometric solution in Gaia DR2. This value is given by $u = \sqrt{\chi^2/\nu}$, where $\chi^2$ represents the $\chi^2$ value between the source and a single-star astrometric solution\footnote{astrometric\_chi2\_al in the gaia archive}, and $\nu$ is defined as the number of good CCD observations used in the astrometric solution\footnote{astrometric\_n\_good\_obs\_al in the Gaia archive} minus five \citep{GaiaDR2ASLindegren}. We subsequently impose the following restriction on the Unit Weight Error: 

\begin{equation}
\begin{multlined}
u < 1.2 * \max[1, \exp(-0.2(G - 19.5))]
\end{multlined}
\end{equation}

\noindent as suggested in \citet{Arenou18}. \edit1{A different condition on astrometric goodness of fit was proposed by \citep{RUWELindegren18}, instead using the Re-normalized Unit Weight Error (RUWE) and the condition RUWE$<$1.4. This cut is somewhat more restrictive than our cut on $u$, producing approximately 5\% fewer stars than the cut used above. However, since we find no evidence of spurious astrometric solutions under the looser condition, we opt for the less restrictive cut on $u$.}

In addition, we impose a restriction on the BP/RP flux excess factor\footnote{phot\_bp\_rp\_excess\_factor in the Gaia archive} $E$, which is defined as the sum of the fluxes in the Gaia BP and RP photometric bands divided by the flux in the G band. The configuration of the Gaia photometric bands implies that $E$ should be slightly larger than 1 for stars with good photometric measurements, however the BP and RP bands are vulnerable to external contamination since they are based on integrated flux in a small field around the star rather than the profile fitting used for the G band \citep[][]{Evans18}. This contamination can manifest in the value for $E$, and contaminated sources can therefore be removed using the following restriction: 

\begin{equation}
\begin{multlined}
1.0 + 0.015(G_{BP} - G_{RP})^2 < E < \\1.3 + 0.037(G_{BP} - G_{RP})^2
\end{multlined}
\end{equation}

\noindent  This nearly matches with the restriction proposed in \citet{Arenou18}, with a slight modification to the factor in the upper bound,  where \citet{Arenou18} uses 0.06 rather than 0.037. We found that for known members of Upper Sco and Taurus \citep{Luhman18,Rizzuto15,Preibisch02,Preibisch98}, the slightly less restrictive conditions would occasionally show very old photometric ages for these young stars. The more restrictive condition removed nearly all such sources, significantly reducing our rates of false negatives among known young objects. 

We also impose restrictions on the number of visibility periods used in the astrometric solution, which are defined as groups of observations separated by at least four days. This requires that the astrometric solution is based on a strong baseline of measurements\edit1{, a condition that was shown in \citet{Arenou18} to significantly reduce the abundance of apparently spurious astrometric results:}.

\begin{equation}
\text{visibility\_periods\_used}>8
\end{equation}

\noindent Finally, we restrict the sample based on the parallax inverse fractional error\footnote{parallax\_over\_error in the Gaia archive} ($\pi/\sigma_\pi$) to exclude stars with a poorly constrained distance\edit1{, following the condition from Gaia's paper on observational HR diagrams \citep{ObsHR18}:}
\begin{equation}
\pi/\sigma_\pi>10
\end{equation}
\edit1{Parallaxes are important for generating the absolute magnitudes used in age estimation, and this condition ensures that the absolute magnitude uncertainty remains below $\sim$0.2 magnitudes, a limit comparable to the uncertainties in the reddening estimates used later in this paper.}

While the combination of cuts employed here excludes nearly half of known young stars in Taurus and Upper Sco, these restrictions are important to ensure that all absolute photometric measurements can be considered trustworthy, which is critical to the reliable derivation of stellar properties, particularly age. Once stars that pass these rigorous quality cuts are used to identify and define the extents of young groups and associations in space-velocity coordinates, the restrictions can be relaxed in future targeted studies, which can significantly increase the completeness. See Section \ref{sec:recovery} for a more complete and quantitative analysis into the completeness of our results.

\section{Methods} \label{sec:style}

The solar neighborhood contains vast quantities of stars, including nearly 5 million within our 333 pc search limit that survive our Gaia quality cuts. However, assuming a constant star formation rate over the last $\sim$10 Gyr, only about 1 in 200 stars will have formed in the last 50 Myr. The Sco-Cen association is thought to dominate nearby young populations, and it contains a membership of only $\sim$10000 stars covering a massive 80 degree-long swath along the plane of the Milky Way \citep{Preibisch02,Rizzuto15}, so field populations will dominate young structures in most cases. To isolate young populations from the field, photometry can often be used, as many low-mass young stars have yet to leave the Hayashi track \citep{Hayashi61} or have not yet settled onto the main sequence. Similarly, O and B stars are readily identified as young, as they do not live long enough to become old. However, in order to reliably identify young stars by photometric means, it is critical that we understand what young stellar populations can and cannot be confidently separated from the contamination of older field stars. 

In this section, we describe our approach for identifying and characterizing young stars in the Solar neighborhood. In Section \ref{sec:sampop}, we describe the implementation of a Bayesian statistical classification approach that uses a model stellar population to identify credibly young stars in the solar neighborhood, and we assess the success of this method in Section \ref{sec:recovery}. The rest of this section describes the methods used to further characterize stars and the larger groups they might belong to. Section \ref{sec:methods-ages} describes our method to obtain age estimates for individual stars, while we identify groups and other young structures in the Solar neighborhood using the HDBSCAN clustering algorithm in Section \ref{sec:clustering}, lastly computing more precise bulk ages for these larger groups in Section \ref{sec:groupage}.

\subsection{Generating the Model Population} \label{sec:sampop}

With the excellent photometric and astrometric data provided by Gaia, we are able to precisely determine the absolute magnitudes and colors of stars in the sample. However, complicating factors such as metallicity, multiplicity, and reddening modify the location of a star in absolute magnitude space, meaning that a star that appears to be photometrically young may be better explained by some combination of other factors \citep[e.g.,][]{Sullivan21}. We therefore need to generate posterior distributions in age for each star and marginalize over the other factors that may influence the photometric youth of a star, including reddening and extinction, multiplicity, metallicity, and stellar mass. To do this, we generate a simulated population of \edit1{10 million stellar systems, a sample size that balances parameter space coverage with computational limitations.} Each \edit1{system is assigned a multiplicity, and each star within is assigned an age, mass, and metallicity, as well as an extinction based on external reddening maps \citep{Lallement19}}. We then generate posterior distributions in age, as well as mass, corresponding to the location of real stars from Gaia DR2 in the model population. By integrating the age posterior over all ages less than 50 Myr, we can estimate the probability $P_{Age<50 Myr}$ that a star is genuinely young, allowing us to isolate young stellar populations in the solar neighborhood.

\subsubsection{Age} \label{sec:methods-agedist}

For the purposes of developing an age distribution, we can approximate the star formation in the solar neighborhood to be constant over its lifetime, for which we take the age of 11.2 Gyr from \citet{Binney00}. This is a simplification of the true star formation history of the solar neighborhood, however, as multiple studies have identified bursts in star formation activity on scales of $\sim$ 1 Gyr \citep[e.g.][]{Rowell13,Isern19}. Current literature also indicates the possible presence of larger-scale trends in the star formation rate, with most asserting some form of a decrease in star formation since the galaxy's earliest star formation bursts \citep[e.g.][]{Schonrich09,Aumer09}. Most studies that assert a decrease in the star formation rate take into account dynamical heating, a process by which older stars are raised higher in the disk, therefore excluding many of them from consideration. This effect works in opposition to a decreasing star formation rate, implying that despite a possible gradual drop in the star formation rate, the stars that remain within our Solar neighborhood are closer to having an even distribution in ages. 

We also assume that the age distribution is spatially isotropic, which allows us to generate a single representative population that can be applied to all of our Gaia stars, regardless of location. However, most recent star formation is thought to have occurred within $\sim$100-200 pc of the galactic plane \citep{Urquhart14,Anderson19}, with stars higher above the plane mainly consisting mainly of older objects, potentially raised into their current location by dynamical heating. As such, young stars will represent a significantly higher fraction of the total stars close to the disk rather than at high galactic latitudes. The assumption of an isotropic age distribution may therefore inflate the rate of false positive young star identifications near the Galactic poles by virtue of the absence of real young stars there, but our rate of recovery among genuinely young stars should be roughly constant across the entire population. 

Not all literature agrees with the conclusion of a SFR that is either decreasing or steady on average \citep[e.g.][]{RochaPinto00}, however the deviations that these alternative star formation histories might impose would be of order unity and therefore have a limited impact on our posterior distributions. On the main sequence, the photometry of stars is relatively constant, so there is usually little photometric difference between otherwise identical stars at, for example, 1\,Gyr and 10\,Gyr. Our objective in generating the population of older stars should therefore be to ensure that their numbers are approximately in proportion with their abundance in the solar neighborhood, meaning that any star formation bursts can be essentially averaged over. Any larger-scale SFR trends may slightly affect our results by means of changing the relative fraction of young and old stars, however these effects can be accounted for by modifying the probability cutoff for selecting candidate young stars in Section \ref{sec:Pcalc}. 

A star formation rate that is to first order uniform over the last $\sim$11 Gyr implies that stars formed within the last 50 Myr should account for less than one percent of the total stars in the solar neighborhood. Subsequently, a representative sample of ages would overwhelmingly skew towards very old objects, leaving a sparse sample of young stars. Since stars evolve quickly on the pre-main sequence, a sparse sample of model stars at these young ages would be insufficient to effectively assess whether any observed star is genuinely young. We therefore draw model stellar ages from a Log$_{10}$-uniform distribution spanning 1\,Myr to 11.2\,Gyr. This does introduce an exponential bias towards young objects, however that can be corrected using a prior, as described in Section \ref{sec:Pcalc}.

\subsubsection{Primary Masses} \label{sec:mass}

\begin{figure}[h]
\centering
\includegraphics[width=7.5cm]{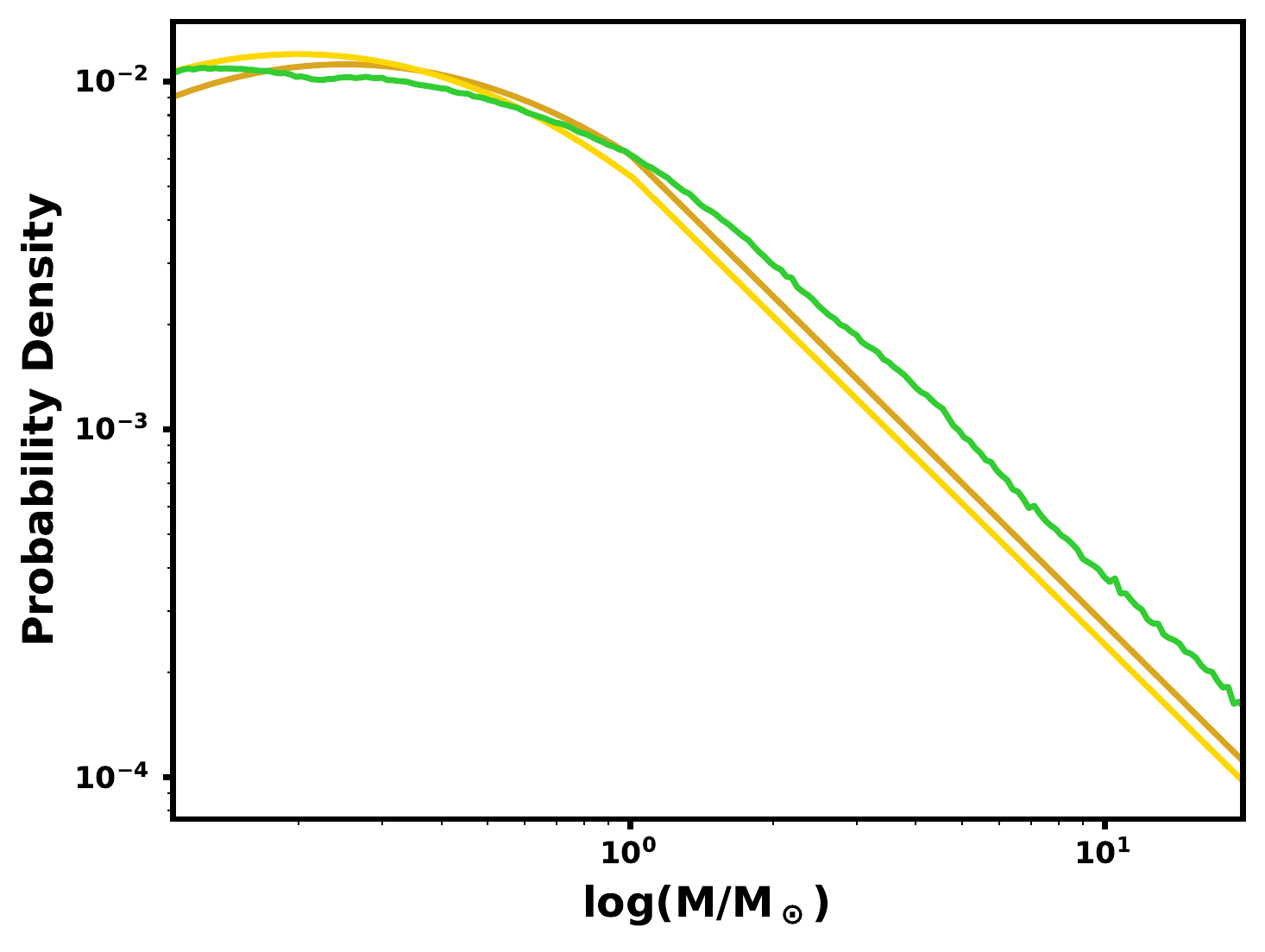}\hfill
\caption{The \citet{Chabrier05} individual object Initial Mass Function (bright yellow), which was used to set the primary masses in our model systems, plotted against the distribution of system masses in our model after binary and triple companions are added (green). The \citet{Chabrier05} system IMF is included for reference (dark yellow). Our model systems never deviate from the \citet[][]{Chabrier05} system IMF by more than 30\%, although our model does show a minor deficiency of low-mass binaries with components below the lower mass limit of our model grid \citep[][]{PARSECChen15}.}
\label{fig:IMF}
\end{figure}

The initial mass function, which displays how the abundance of stars varies with mass, is the basis for this project's mass generation. It follows a distribution that has been gradually revised since the first form presented by \citet{Salpeter55}, with the most recent revisions from \citet{Chabrier03} and \citet{Chabrier05} using a log-normal distribution at low mass (M $<$ 1 M$_{\odot}$), and the \citet{Salpeter55} power law distribution for higher masses. The IMF has been consistently shown to be invariant based on environment, so spatial variations to our mass generation are not of concern \citep{Chabrier05,Offner14}. Slightly different log-normal solutions have been presented for individual stars and complete stellar systems, and since we require masses of primaries that may or may not be in multiple systems, we make use of the \citet{Chabrier05} individual star Initial Mass Function (IMF) to generate the masses of primaries. Both the individual and system \citet{Chabrier05} initial mass functions are plotted in Figure \ref{fig:IMF}, alongside the distribution of system masses we reach after adding stellar companions in Section \ref{sec:multiplicity}.

The minimum mass we consider is 0.11 M$_{\odot}$, which is roughly the smallest mass that is available in all PARSEC v1.2S isochrones \citep{PARSECChen15}). Young stars of this mass also reach Gaia's approximate limiting magnitude of G$\simeq$21 near our maximum distance of 333 pc \citep{Arenou18}, so objects in this mass range will be accessible throughout our entire search radius. Our maximum mass was set to 20 M$_{\odot}$, which excludes a small subset of stars that are both very rare and very bright. Main sequence stars of more than 20 solar masses have absolute magnitude G$<$-3 \citep{PARSECChen15}, which corresponds to a magnitude of roughly 3 at the distance of Sco-Cen. Gaia photometry and astrometry in this range is known to be poor due to the start of pixel saturation at G$\sim$6 \citep{Arenou18}, so the contribution Gaia can make towards furthering the understanding of these very bright stars is relatively minimal. 

Due to a combination of the higher average mass of primaries relative to the complete population considered in the individual IMF, as well as the mass limit of 0.11 M$_{\odot}$ imposed on all model stars, our system mass distribution does not quite replicate the \citet{Chabrier05} system IMF. For stellar systems, which are discussed in depth in Section \ref{sec:multiplicity}, a minimum component mass of 0.11 M$_{\odot}$ implies a minimum binary mass is 0.22 M$_{\odot}$, and as a result this mass marks a small local minimum in the mass function in Figure \ref{fig:IMF}. As such, the discrepancies that do exist are mostly brought on by limitations in the masses of objects our system can create. However, despite these minor deviations, the final simulated system mass function remains comparable to the system IMF also presented in \citet{Chabrier05}, never differing by more than 30\%. We expect the effects of a slightly discrepant model IMF to be relatively minor in our final results. The overabundance of higher mass stars relative to lower mass binaries may overweight the contamination from subgiants, however, particularly in the parameter space occupied by the pre-main sequence, G and F stars follow evolutionary tracks well-separated from the early M stars, which show the greatest underabundances in our model \citep{PARSECChen15}. This suggests that situations where the relative abundances of these different mass model stars has a significant impact on our results will be very rare. Any such effects will be further minimized by our generous 10\% probability threshhold for the consideration of a star as young, and the reliance on kinematics to cull likely erroneously-detected field stars, as outlined in Section \ref{sec:clustering}.

\subsubsection{Metallicity}

While stars are thought to be more metal-rich on average at later formation epochs, the age-metallicity relationship in the solar neighborhood appears to be relatively flat beyond an initial early enrichment phase \citep[e.g.,][]{Lin20,Haywood19}, inferring that a single non-age-dependent metallicity distribution will be sufficient to represent the solar neighborhood. Regardless of whether metallicity is truly constant with time, it may not influence our detection of young stars, since the nearly age-independent photometry of main-sequence stars means that the age will weigh minimally on the photometry of these stars, regardless of metallicity. 

A possibly more consequential bias that a single metallicity distribution may introduce relates specifically to young stars, as most that are known have solar metallicities. Using the full range of possible metallicities to represent these stars may therefore present a wider range of possible photometry compared to what actually exists, particularly for low metallicities. This would result in an underestimation of $P_{Age<50 Myr}$, as these low-metallicity stars have photometry more consistent with older populations \citep{PARSECChen15}. However, since we wish to identify structures beyond those already known in the solar neighborhood, we have little cause to tether our populations to an assumption of solar metallicity. Even if the assumption of near-solar metallicity is universally appropriate, placing stars too low on the pre-main sequence due to low metallicity is a conservative error, and will therefore not lead to spurious assessments of youth. The resulting lower probability estimations can subsequently be negated using more permissive probability cuts for the identification of young stars. We therefore conclude that a non-age dependent distribution for metallicity would provide a reasonable approximation to reality. Data Release 2 of the GALAH survey \citep{GALAHDR2Buder18, Hayden19} provides an empirical sample containing over 62000 stars within 500 pc of the sun which we use as a representative sample for the metallicities of stars in the solar neighborhood.  

\begin{figure}[t]
\centering
\includegraphics[width=8.3cm]{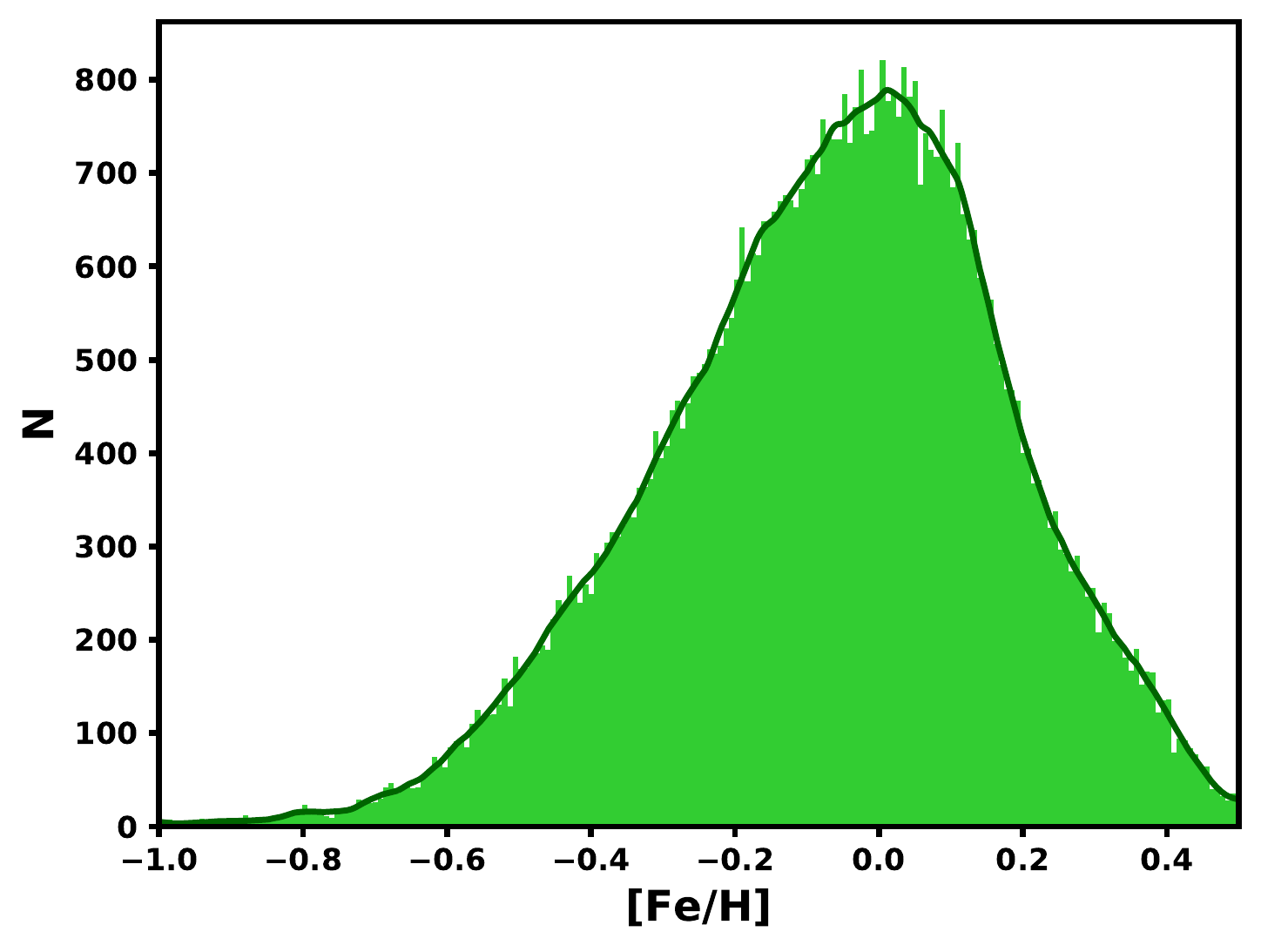}\hfill
\caption{The metallicity histogram for stars observed by the GALAH survey. The dark curve shows the smoothed distribution that we generated metallicities from (prior to normalization).}
\label{fig:FeH}
\end{figure}

Our metallicities were randomly generated from a slightly modified version of the GALAH metallicity distribution. We first restricted the metallicities to between [Fe/H] = -1.0  and 0.5. The lower limit was due to the negligible number of GALAH stars with metallicities in that range ($\sim$ 0.1\%), while the upper limit was chosen to comply with the metallicity range available in PARSEC v1.2S isochrones: -2.2$<$[Fe/H]$<$0.5  \citep{PARSECChen15}. There are more stars that exceed this high metallicity limit in the GALAH data compared to the low-metallicity limit, however the fraction of total stars occupied by these outliers remains under 0.5\% \citep{Hayden19}. Therefore, these extremely metal-rich stars are not sufficiently numerous for their exclusion to dramatically skew our results. We then binned the GALAH measurements with [Fe/H] between -1.0 and 0.5 into 200 evenly-spaced bins over that range, and smoothed the resulting distribution using a Savitsky-Golay filter with a 15-bin window length and third-order polynomial. This generated a smooth metallicity distribution consistent with the original histogram from which we can draw samples, which is shown in Figure \ref{fig:FeH}

\subsubsection{Multiplicity} \label{sec:multiplicity}

Binaries are quite common in the Milky Way, with a mass-dependent abundance fraction that increases in tandem with the mass of the primary \citep[e.g.,][]{Duchene13}. For stars less massive than the sun, between one fifth and one half of all stars are binaries, and depending on the distance many of these cannot be resolved by Gaia. While resolved binaries are not of concern for the purposes of this study, as Gaia is able to generate accurate photometry and astrometry for each resolved component, unresolved binaries create combined photometric detections that appear overluminous relative to comparable single stars. Dim companions have a relatively minor impact on the system photometry, however the presence of an unresolved equal-brightness binary can double the incident flux and subsequently lower the magnitude by up to 0.76 mag, a much more significant contribution. This brightness relative to the main sequence is comparable to what we see on the pre-main sequence, and therefore these older unresolved binaries can easily be mistaken for younger single stars. Therefore, to accurately assess whether a star is young, the abundance of multiple stars, distribution of companion masses, and companion separations must all be well-modelled to quantify the probability that stars in any region above the pre-main sequence are multiple systems rather than young stars. 

We therefore add multiples to our synthetic population, considering only single, double, and triple systems to avoid complications with the architecture of rare, higher-order systems. We begin with the statistics reported in \citet{Duchene13}, including the multiplicity frequencies (MF), which measure the occurrence frequency of multiple systems of any kind, and companion frequencies (CF), which considers the number of companions per primary star \citep[originally sourced from ][]{Raghavan10,Delfosse04, Dieterich12,Kouwenhoven05,Kouwenhoven07,Abt90, Sana12, Mason09, Chini12,Preibisch99}. 
While the MF describes the probability of a star being in some form of multiple system, the CF allows for conclusions to be drawn about the occurrence frequencies of both double and triple stars.

\begin{figure}[t]
\centering
\includegraphics[width=8cm]{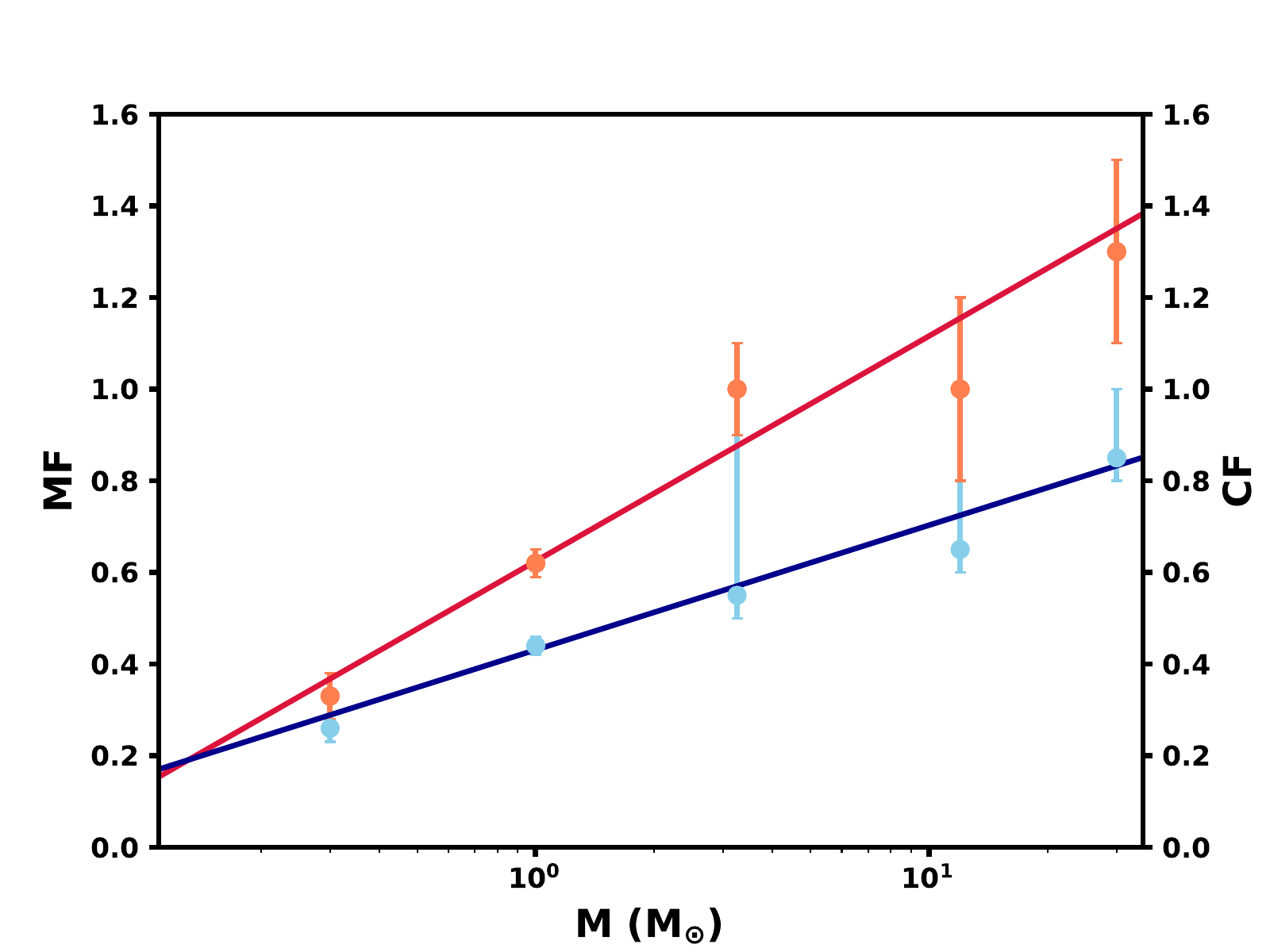}\hfill
\caption{The fits to the MF (blue) and CF (red) distributions as a function of primary mass. Note that the confidence intervals for the two rightmost points for the MF distribution are lower limits, and the dot provided is the chosen intermediate value. These points are heavily de-weighted compared to the well-constrained MF values at lower masses, and have minimal impact since our selected sample consists almost entirely of stars of spectral type G or later.}
\label{fig:CFMF}
\end{figure}

To generate a smooth distribution, we fit a logarithmic function as in equations \ref{eqn:logfit1} and \ref{eqn:logfit2} to the MF and CF values from \citet{Duchene13}, using reasonable intermediate values for the MF in cases where only upper limits are provided. These fits are shown in Figure \ref{fig:CFMF}, matching the linear trends in semi-log mass vs MF and CF that are followed by the \citet{Duchene13} values for MF and CF. The resulting solutions for CF and MF as a function of primary mass are as follows:
\begin{equation} \label{eqn:logfit1}
MF = 0.272*\text{log}_{10}M + 0.431
\end{equation}

\begin{equation} \label{eqn:logfit2}
CF = 0.491*\text{log}_{10}M + 0.625
\end{equation}

\begin{figure}[t]
\centering
\includegraphics[width=8.5cm]{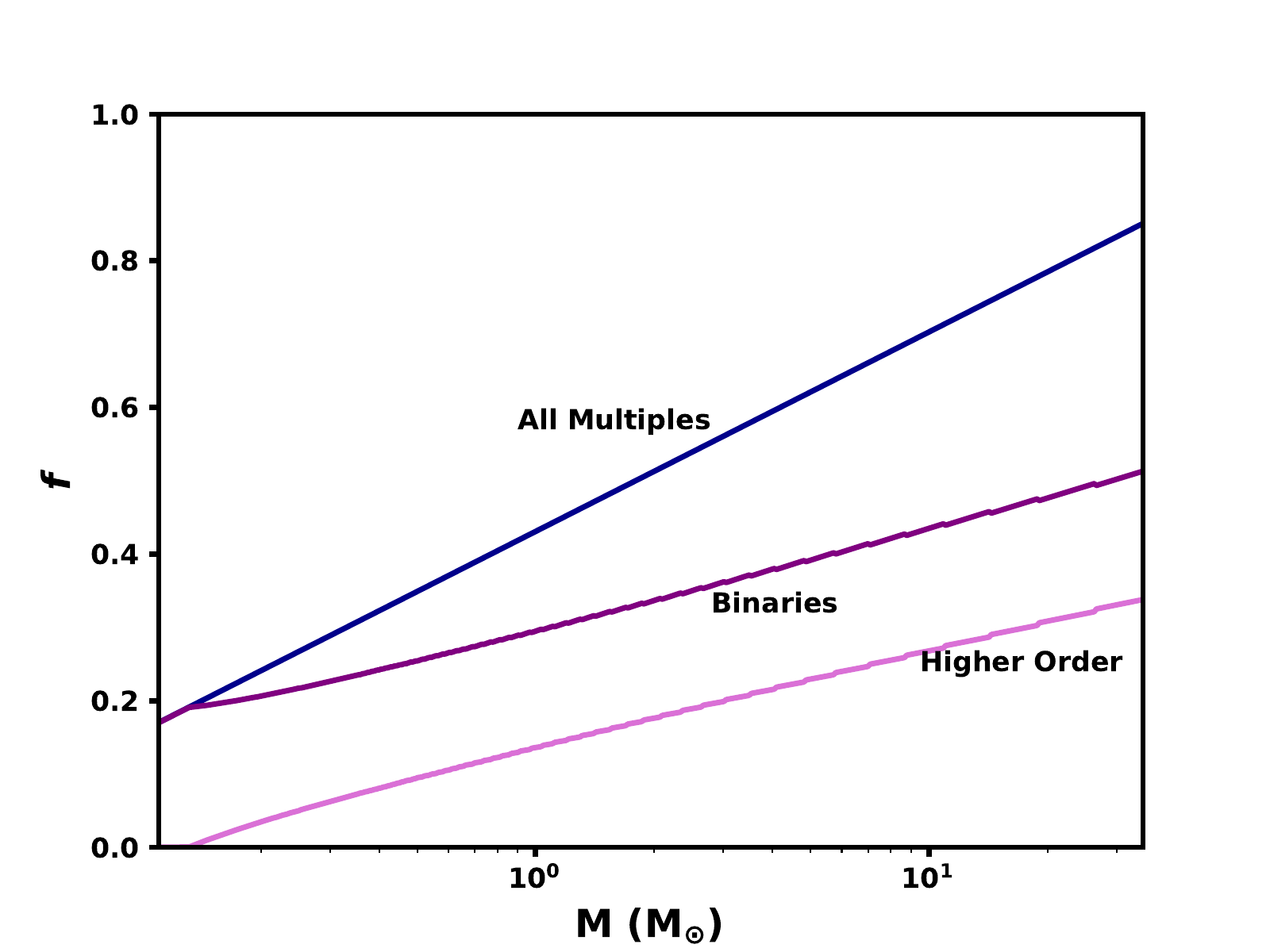}\hfill
\caption{The final distributions used for the binary, higher order (which are treated as triples), and total multiplicity fractions. The blue line is equivalent to the MF fit in Figure \ref{fig:CFMF}.}
\label{fig:binrate}
\end{figure}

Then, assuming that the probability of each multiplicity follows some geometric series $Ck^m$ \citep[as in][]{Duchene13}, where $m$ is the multiplicity of the stellar system, we fit for the unique solution for C and k that returns the appropriate values for CF and MF interpolated from our fit. The resulting abundance fraction distributions for multiples, binaries, and higher order systems are given in Figure \ref{fig:binrate}. The result has the correct edge behavior, as the double and especially triple rates drop dramatically as the primary mass approaches our minimum mass limit, and the multiple rate approaches but does not exceed one at the high-mass limit.

The probability distribution of the resulting companion masses is typically described as a function of the mass ratio between the primary and secondary components $m_s/m_p = q$, following a power law distribution of the form $P(q) = Cq^\gamma$ \citep{Kraus11,Rizzuto13}. We therefore generate secondary and tertiary components based on the mass ratio probability distributions presented in \citet{Kraus11} and \citet{Rizzuto13}, which cover $q$ values between 0.1 and 1. In order to exclude companions in the brown dwarf regime that are not massive enough to be included in the PARSEC isochrones, we multiply these probabilities by an additional factor, which limits the section of the mass ratio distribution from which companions can be drawn to only that region between the primary mass and the minimum mass accessible through the PARSEC isochrones. The corrective factor for a given primary mass is therefore the integral of the probability distribution over the window from the primary mass to the minimum PARSEC mass, normalized such that the sum over all $q$ is 1. Once this corrective factor to the multiplicity probabilities is applied, the primaries are separated into three populations, each with distributions for $q$ governed by a power law with a different exponent $\gamma$: $\gamma=0.4$ for M $<$ 0.7 M$_{\odot}$, $\gamma=0$ for 0.7-2.5 M$_{\odot}$, and $\gamma=-0.46$ for masses exceeding 2.5 M$_\odot$. The two lower mass ranges come from \citet{Kraus11}, and the upper mass bracket comes from \citet{Rizzuto13}. 

The generation of masses for all primaries and companions represents the final step necessary before computing simulated inherent photometric properties for these stars. The mass distribution of systems after the inclusion of companions is shown in Fig. \ref{fig:IMF}, plotted against the \citet{Chabrier05} IMFs.  

The separation of multiple system components must also be considered, as only spatially unresolved systems will display merged photometry. The stellar populations we wish to draw young stars from come from the Gaia mission, and since Gaia reports resolved stars independently, multiples that are resolved in Gaia are seen as a series of two or more singles. To generate approximate separations, we take the distribution of periods presented in \citet{Raghavan10}, randomly select an orbital period from that distribution, and compute an orbital semi-major axis from the mass of the system using Kepler's Third Law. A corrective factor of $\pi/4$ is applied to account for the projection of these orbits on the plane of the sky, which is modulated by the presence of randomly-oriented systems. Recent literature investigating the separations of M dwarf binaries has shown tighter lognormal distributions centered at slightly smaller periods for these stars \citep[e.g.,][]{Bergfors10}, however these results are not currently well-enough constrained for us to justify including a mass-specific period distribution. We may introduce revised separation distributions in a future iteration of this work, as the current implementation likely underestimates the abundance of unresolved binaries for M dwarfs. However, even if an updated form of the separation distribution were to imply that all M-dwarf binaries are unresolved, this would only represent a roughly 30\% increase to the unresolved rate for companions to a $\sim$0.2 M$_{\odot}$ star, so we do not expect significant effects from the use of the \citet{Raghavan10} period distribution.

Most Gaia stars with separations exceeding 1 arcsecond are resolved as independent sources \citep{rizzuto_zodiacal_2018, Ziegler18}, although this does not appear to be a hard boundary, as there is likely a population of systems at separations below 1 arcsecond that are neither unresolved nor reported independently, a factor which may reduce the mean separation at which this transition to independent identification occurs. However, even with that transition separation lowered to 0.2 arcseconds, the unresolved rate for an average 1 M$_{\odot}$ star only drops by 25\%, inferring that the use of 1 arcsecond as a hard limit for the transition between unresolved and resolved populations appears to be a reasonable approximation. We take a distance of 234 pc, which is the mean distance to the Gaia stars investigated, and break up test stars with separations exceeding 1 arcsecond at that distance. Stars with smaller separations are marked as unresolved systems, which are combined photometrically into a single star. For triple stars, separations are generated for each companion. If both are unresolved, then all three members are photometrically merged. If only one model separation is resolved, then the second component is counted as resolved from the primary, and the third is randomly assigned to either the primary or secondary. The result is that the system is now treated in our model as one unresolved double with a resolved single nearby. The choice to separate binaries at a single fixed distance, rather than make it adaptive to the distance of the target star, was an approximation made to reduce computational costs, however we do not expect the effects of this choice to be significant. At 10 pc, a companion to a 1 M$_{\odot}$ star would have an approximately 34\% chance of being unresolved, compared to about 72\% for a star at our field limit of 333 pc, and 68\% for our chosen distance at 234 pc. As such, the only major inaccuracies will be at very small distances, where we already expect our cluster detection to be less sensitive. Within the nearest 50-100 pc, geometric effects on the 2-d Gaia transverse velocities make it increasingly difficult to detect stars with common 3-d motions, making structures difficult to identify regardless of the recovery rate of their stars (see Section \ref{sec:clustering}).

\subsubsection{Generating Photometry of Simulated Stars and Systems} \label{sec:methods-genphot}

For each mass, metallicity, and age generated, we use isochrones to create corresponding Gaia G, G$_{RP}$, and G$_{BP}$ magnitudes. We make use of PARSEC version 1.2S\footnote{http://stev.oapd.inaf.it/cgi-bin/cmd} isochrones made freely available online \citep[see][]{PARSECTang14, PARSECChen15, PadovaBressan12}. 

Note that these isochrones do not include the white dwarf cooling sequence, and therefore these objects are treated as ``dead", with no photometric contribution. Since the photometry of these objects says little about the white dwarf's age relative to the formation of the system, this exclusion should have a minimal impact on the results. These objects also occupy a very different parameter space on the CMD, so they are easy to remove and nearly impossible to mistake for young main sequence stars. 

The isochrones cover a metallicity range evenly spaced between [Fe/H] = -1.0 to 0.5, and a less regular age selection, which was designed to optimize the sampling of parameter space between the ages from 6$<$log$_{10}$(age/yr)$<$10.049. The main concern when generating a grid to draw our synthetic population from is that with a sufficiently sparse sample in age, stars of a given mass would not have samples in between, for example, the main sequence and giant branch. This can result in situations where any interpolation on these photometric grids can end up deriving photometry somewhere between the main sequence and giant branch, which may not follow the evolutionary track stars follow between those sequences. To minimize artefacts caused by under-sampling, we require that for each slice of the grid in both mass and age, at least two points in age populate the horizontal branch, except for the most extreme massive stars. This requirement ensures that at least the lower red giant branch is well-covered by our grid, with the top of the RGB and Horizontal Branch being somewhat less well-covered. Further grid resolution beyond covering the lower RGB would only increase the computational cost without improving our calculations, as these more evolved stars do not overlap with the pre-main sequence stars in photometric parameter space.

The PARSEC v1.2S isochrones give a list of masses for each metallicity/age combination, and we regrid these onto a grid with 0.11$<$log(M/M$_\odot$)$<$20) and non-uniform point density. For masses, very sparse sampling is sufficient for stars less massive than the main sequence turnoff at the end of our age bracket, 11.2 Gyr. All stars below this point do not ever evolve off the main sequence, and therefore the only quick evolutionary process with sensitivity to mass that must be captured is the descent along the Hayashi track, which is significantly slower than evolution along the RGB and much less mass-sensitive \citep{Hayashi61, PARSECChen15}. Not many grid points are required to capture that evolution in the mass dimension, so only 100 linearly-spaced grid points are used for masses M $\la$ 0.8 M$_{\odot}$. We then use 4000 bins between 0.8 M$_{\odot}$ and 2.5 M$_{\odot}$, another 4000 between 2.5 M$_{\odot}$ and 10 M$_{\odot}$, and 600 more bins between 10 M$_{\odot}$ and 20 M$_{\odot}$, for a total of 8700 bins. These binning choices do result in the under-sampling of the more evolved stars in the population, however the pre-main sequence and stars sharing that parameter space are all well-covered.

\begin{figure}[t]
\centering
\includegraphics[width=8.2cm]{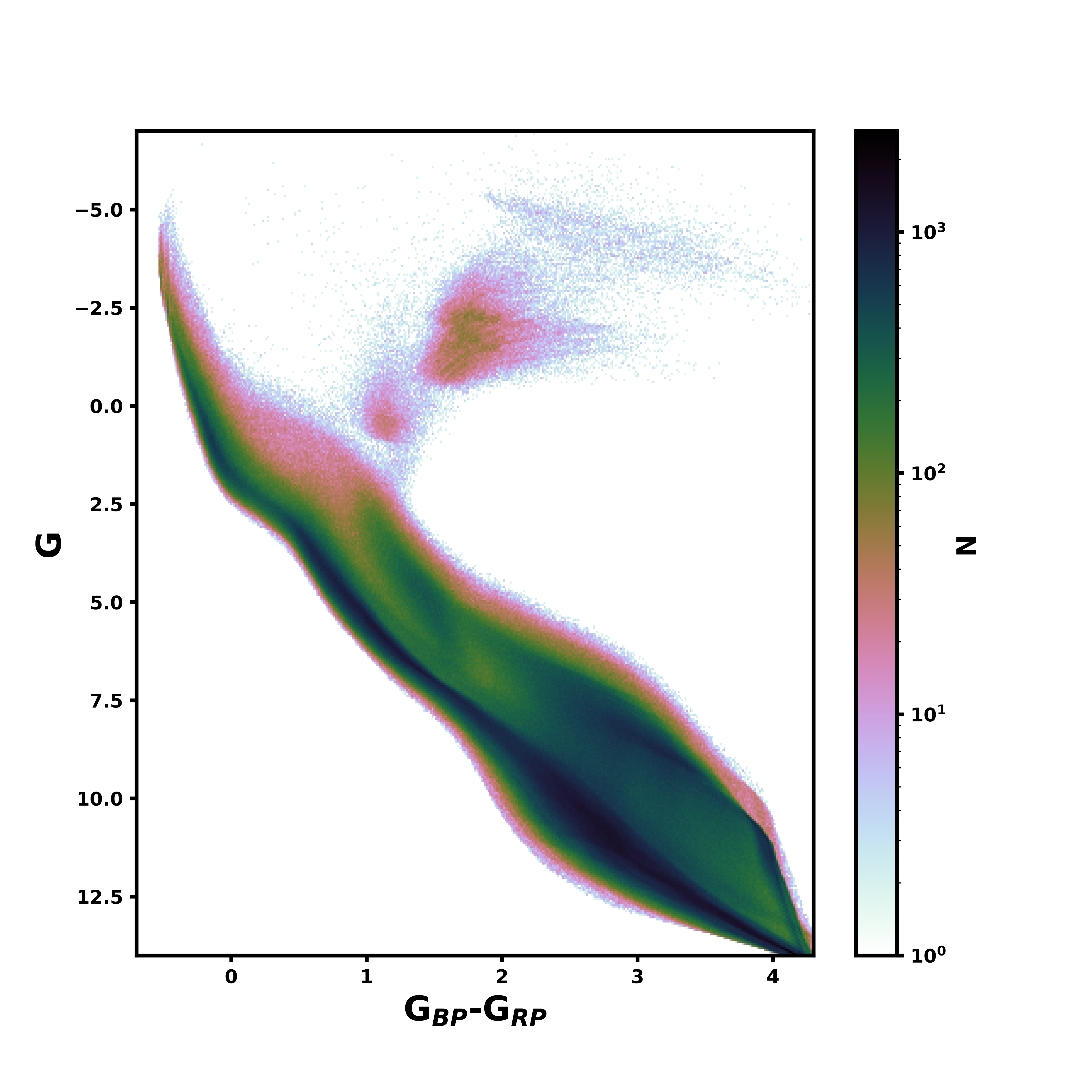}\hfill
\caption{A color-magnitude diagram for the stars generated by simulating a population with our randomly generated ages, masses, metallicities, and stellar multiplicities, then computing observables using the PARSEC isochrones.}
\label{fig:model_cmd}
\end{figure}

We assigned model stars random ages, metallicities, and masses according to the relevant distributions, and linearly interpolated  magnitudes for each star from our isochrone grid. When identified as part of an unresolved binary or triple system, we added the magnitudes of the component stars together, creating a merged photometric result for each unresolved system. This produced intrinsic synthetic G, BP, and RP magnitudes for any model star or system, resulting in the color-magnitude diagram presented in Figure \ref{fig:model_cmd}. While this completes the generation of our model population, these intrinsic magnitudes have yet to be modulated by the conditions in which a real Gaia star might exist, most notably extinction, which we address in the next subsection. 

\subsubsection{Extinction}

Significant interstellar extinction moves the main sequence of a stellar population on the HR-diagram to the right and downwards of where it would be without that extinction. Consequently, heavily extincted stars can appear younger than they would without the effects of extinction. Due to the significant regional variations in extinction, the use of a single, directionally independent distribution to modulate our simulated photometry would fail to differentiate between young stars and those that are merely heavily extincted. Instead, we make use of three dimensional, all-sky extinction maps, and generate a unique reddening distribution for each Gaia star based on that map. \edit1{Extinction values from these distributions are generated and independently applied to the model population for each Gaia star. }

We use the STILISM maps of reddening in the solar neighborhood from \citet{Lallement19}, which use Gaia DR2 and 2MASS photometry and distances to compute reddening for a large volume centered on the solar system. We interpolate reddening parameters for each Gaia star based on its on-sky location using a Monte-Carlo framework with distances drawn from the Gaia parallax and parallax uncertainty. This framework captures how Gaia distance uncertainties impact reddening uncertainties. 

An additional uncertainty modulation is added to each interpolated reddening result to reflect the inherent uncertainty in the maps themselves. \citet{Lallement19} reddenings often have very asymmetric uncertainties, so for each generated reddening and corresponding uncertainties, we select a random value from a normal distribution, multiplied by the uncertainty on that side of the reddening value recorded in the STILISM maps. These uncertainty results are added to each randomized reddening, allowing us to generate probability histograms for the reddening of all stars in the Gaia set. The reddenings of the simulated stars are therefore drawn from these histograms, separately for each Gaia star. The relationships presented in \citet{Wang19} were used to relate E(B-V) to extinction in the Gaia BP, RP, G filters. While these conversions are expected to differ somewhat by the effective temperature of the target, these effects will be negligible compared to the intrinsic sources of uncertainty in the \citet{Lallement19} reddening maps, especially given that the \citet{Wang19} reddening conversions are based on relatively cool red clump stars which are photometrically similar to many of the pre-main sequence stars we focus on in this paper. The addition of reddening completes a population of simulated stars with a reddening distribution reflecting the reddening expected of each Gaia star based on its location. 

\subsubsection{Generating Star Statistics}\label{sec:Pcalc}

The completed population of sample stars allows us to generate statistics for our population of Gaia DR2 stars. For each real star, we use Bayes' theorem to generate the probability that each sample star is consistent with the Gaia photometry according to the following formulation, as implemented in \citet{Huber16}:

\begin{equation}
p(y|g) \propto p(y)p(g|y) = \prod_{i}exp\left(-\frac{(g_i-x_i)^2}{2\sigma_{g, i}^2}\right)
\end{equation}

where y is a sample star with inherent properties $y$ = \{age, [Fe/H], mass, E(BP-RP), multiplicity\} and with observables $x$ = \{G, BP, RP\}. \edit1{Age, [Fe/H], mass, and multiplicity are all unknowns, while the extinction E(BP-RP) is a prior imposed by the reddening solutions from \citet{Lallement19}.} We then compare the observables of our sample stars to our Gaia star $g$ with the same observables, converting from apparent Gaia magnitude to absolute magnitude using the Gaia parallax measurements. 

These distances were generated by inverting the parallaxes, and while this is an imperfect estimate for high-relative uncertainty parallaxes, the distance estimates from \citet{BailerJones18} follow the inverted parallax very closely for the stars with $\pi/\sigma_\pi>10$ that we investigate here. The subscript $i$ multiplies over each observable in x, and $\sigma_{g,i}$ is the uncertainty in the Gaia measurement of the observable $i$, including both photometric and parallax uncertainties. 

\begin{figure}[t]
\centering
\includegraphics[width=8.0cm]{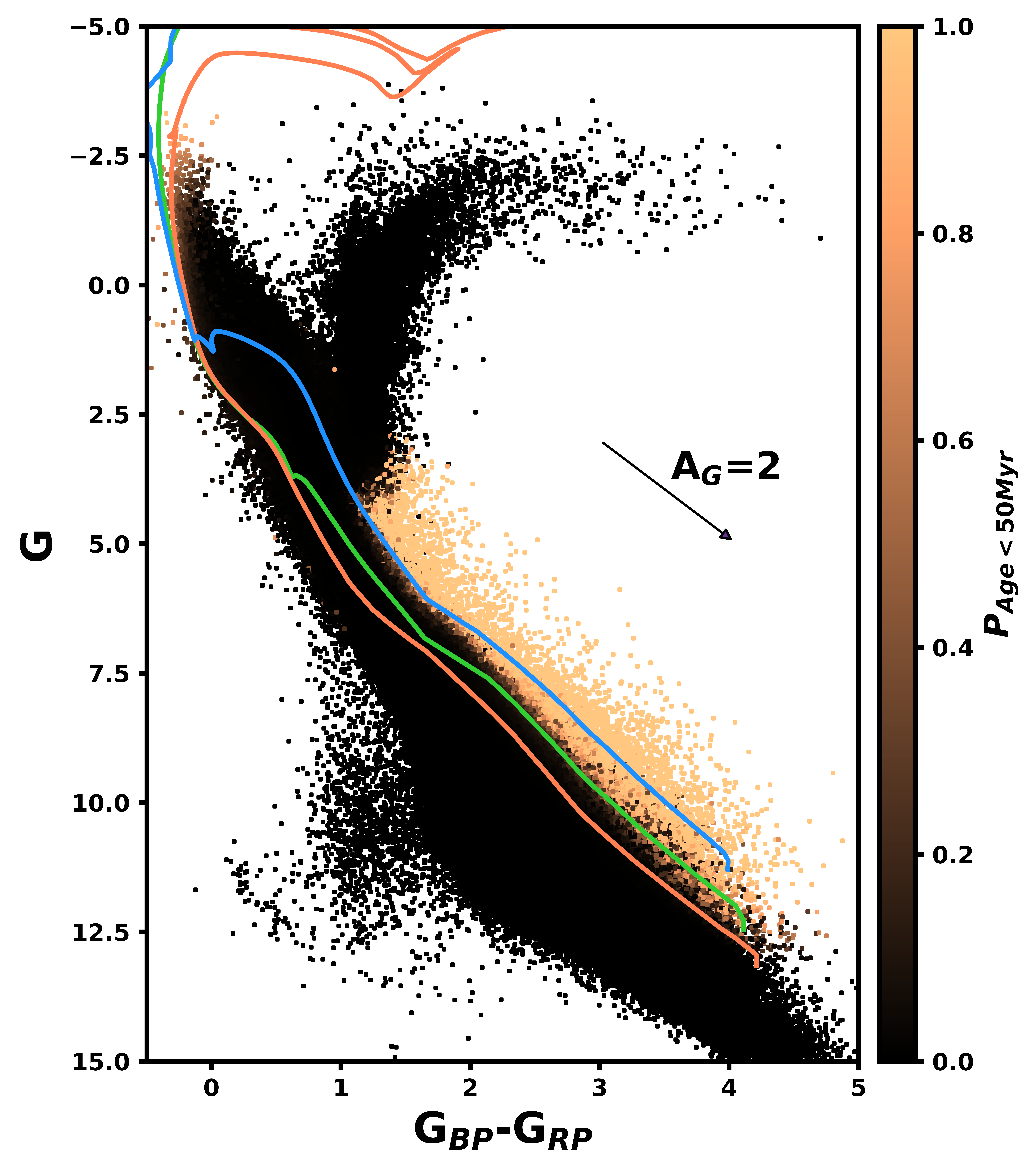}\hfill
\caption{A color-magnitude diagram containing all Gaia stars with more than 50 simulated stars within 1-sigma of their locations in the CMD. All magnitudes are absolute and dereddened. The colors represent the probability that the star is young, defined by this paper as younger than 50 Myr. The reddening vector for A$_G$=2 is labelled, showing the direction along which older stars may move to interfere with the pre-main sequence if not dereddened. \edit1{Isochrones are also included for ages of 5, 20, and 50 Myr. The orange region with high $P_{Age<50 Myr}$ values intersects with the 5 Myr isochrone from M dwarfs up to near the subgiant branch, while the 50 Myr isochrone only intersects for extreme low-mass M dwarfs, indicating a higher completeness for younger populations (see Sec. \ref{sec:recovery}).}}
\label{fig:Pages_CMD}
\end{figure}

In most cases, the prior $p(y) = 1$, owing to the consistency of our metallicity, mass, reddening, and binarity rates with true values. Age, however, does require a non-trivial prior, as we significantly over-sample young stars. We therefore add a prior equal to the age of the star, which nullifies the exponential bias cause by the uniform sampling in log-space, producing a flat probability distribution in age for the sample stars. To produce any probability distribution, we simply sum the $p(y|g)$ values for each model star into a histogram, and normalize over all sample stars. The probability of any given condition on a property, such as $P_{Age<50 Myr}$ can therefore be found by summing over the relevant bins. The resulting $P_{Age<50 Myr}$ values for each star in the Gaia sample are presented through the color-magnitude diagram in Figure \ref{fig:Pages_CMD}. Most stars have $P_{Age<50 Myr}$ similar to their neighbors, with some exceptions in cases with larger uncertainties, or cases where either the BP or RP filter is discrepant, which may not translate well to this 2-d plot from the 3-d input magnitude data set. 

\begin{figure}[t]
\centering
\includegraphics[width=8.8cm]{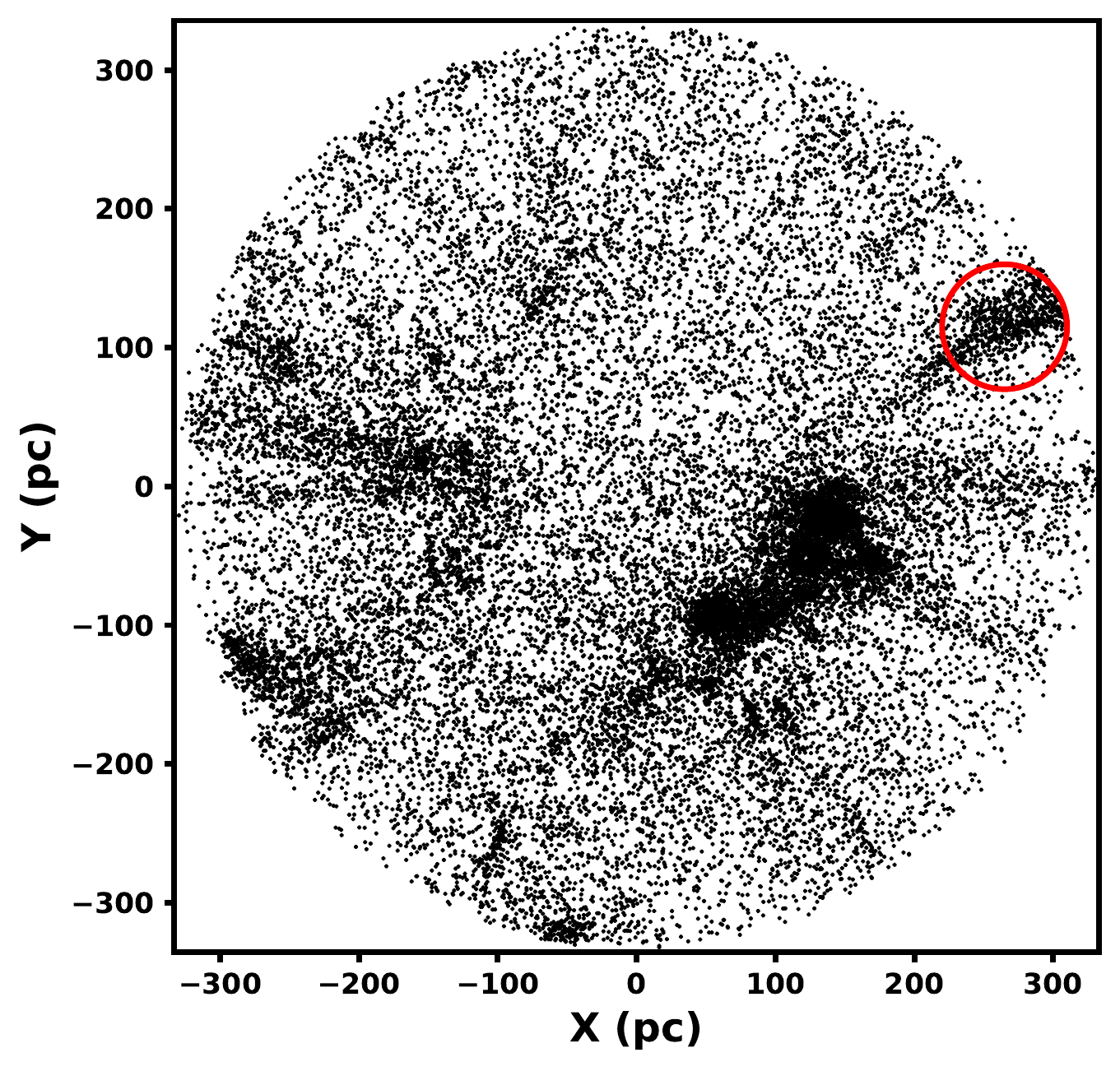}\hfill
\caption{All credible young stars identified by our search to have $P_{Age<50 Myr}>0.1$, plotted in X/Y galactic coordinates. The circled region is an example of a reddening anomaly, where the underestimated reddening causes stars to be marked as young. That particular region is in the direction of heavy reddening in Aquila and Serpens. This group is not identified in the clustering stage (see Fig. \ref{fig:allsky_clustering}), as the stars do not have common motions, and are likely dominated by reddened field stars.}
\label{fig:allsky_redcir}
\end{figure}

All Gaia stars with less than 50 simulated stars within 1-sigma of them in color-magnitude space were culled from the sample, as we identify them as inconsistent with the stars in our simulated population. As we do not consider white dwarfs and white dwarf-main sequence binaries in our model, most of these stars are excluded, as is evident from the near-absence of a white dwarf cooling sequence in Figure \ref{fig:Pages_CMD}. \edit1{For all Gaia sources well-sampled by our model population, we generated probability distributions for age, mass, metallicity, and multiplicity.} To extract a population of young stars from our sample, all objects with $P_{Age<50 Myr}<0.1$ were removed, leaving a population of 30518 credible young star candidates out of the full sample of nearly 5 million Gaia sources. These candidate young stars are presented in XY galactic spatial coordinates in Figure \ref{fig:allsky_redcir}, and compiled into a master candidate list in Table \ref{tab:master_ys}. As a result of the reddening vector direction shown in Figure \ref{fig:Pages_CMD}, the underestimation of reddening, especially near the subgiant branch, does have the ability to make older stars appear young. Minor anomalous reddening-related clumps do appear. While we do detect minor spatially clustered but kinematically scattered clumps of stars that appear consistent with anomalies from local reddening underestimates, many more young structures are visually identifiable, such as Sco-Cen, Orion, Perseus, and Taurus. 

\subsection{Recovery}\label{sec:recovery}

The method we employ in this paper searches the solar neighborhood for stars with a significant probability (P$>$0.1) of an age less than 50 Myr, and we should therefore have demonstrable capabilities to identify stars in that age range. Here we investigate the efficiency with which those nearby young stars are identified, and what kinds of stars are detectable for groups with different ages. There are two main ways in which stars might be missed in this survey: failure of Gaia photometric or astrometric quality cuts and misidentification by our pipeline as old. In this section we quantify the losses due to each of these factors, and provide insight into why some of these losses take place. 

The Gaia quality cuts, which remove stars based on factors like goodness of fit to the Gaia photometric and astrometric models, parallax error, and the number of Gaia visibility periods used (see Section \ref{sec:dat-gaia} for a full explanation of these) are necessary to ensure that our Gaia stars have reliable photometry and astrometry. However, it is nonetheless useful to investigate the properties of the stars lost to reveal any potential biases the cuts might impose. By comparing the samples of stars that pass and fail these quality cuts, we can reveal what properties are most likely to lead to the rejection of a star based on quality. We use a large sample of Taurus members drawn from \citep{Krolikowski21} to make this comparison, which has a wealth of information on $\sim$500 stars in the region, including spectral type, extinction, and binarity. All of these factors may influence the quality of any Gaia astrometric or photometric solutions. We find that our Gaia quality culling appears to correlate very little with stellar spectral type over the late G to M range, however those cuts do correlate strongly with binarity and extinction. Unresolved binaries are almost twice as likely to be removed by Gaia quality restrictions, with 59\% of singles passing these cuts, compared to only 31\% of multiples. Binaries are known to both introduce additional astrometric noise and skew the photometric solution \citep[e.g.,][]{Arenou18}, so their more frequent failure of quality cuts is expected. For heavily reddened populations with $A_V>3$, only 20\% of sources pass the quality cuts, compared to 67\% for $A_V<0.5$. This heavy loss rate in highly reddened locations is consistent with our results later in this paper, which include visibly incomplete samples of young stars in active star-forming sites such as Perseus and Lupus.

\begin{figure}[t]
\centering
\includegraphics[width=8.2cm]{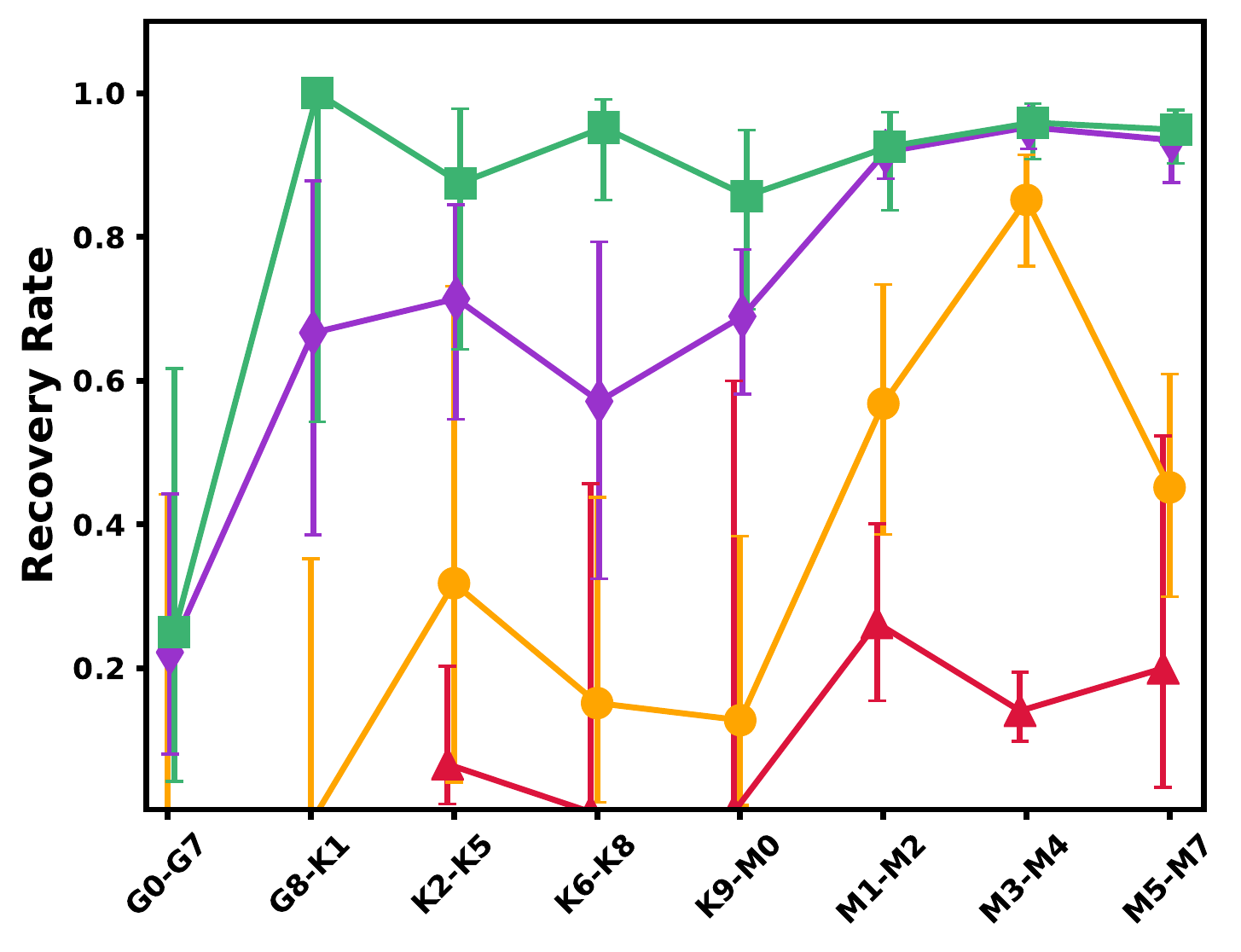}\hfill
\caption{Recovery rates after the imposition of Gaia quality cuts in Taurus (green squares), Upper Sco (purple diamonds), $\beta$ Pic (orange circles), and Tucana-Horologium (red triangles), binned by spectral type. In all bins, the order of the groups in recovery rate matches that of their relative ages, with the essentially newborn Taurus association having the most complete recovery. The error bars display the binomial 68\% confidence interval.}
\label{fig:recovery}
\end{figure}

Next we investigate recovery rates from our pipeline for the identification of young stars, which include only objects that pass the Gaia quality cuts. To calculate recovery rates, we gather lists of known members for four well-known associations: Upper Sco (in Sco-Cen), Taurus, the $\beta$ Pictoris Moving Group, and the Tucana-Horologium Moving Group. These groups cover nearly the full range of detectable cluster ages, spanning from Taurus, which is nearly newborn \citep[$<$ 5 Myr;][]{Kraus09}, to Tucana-Horologium, which has an age of approximately 45 Myr \citep{Kraus14,Bell15}. We can therefore use these stars as a strong representative sample of the populations we expect to identify in this paper.

As the youngest population we investigate in this section, stars in the Taurus Association should be the easiest to identify reliably as young through purely photometric methods. To verify this, we drew known members from the \citet{Esplin19} Taurus catalog, which compiled verified members from literature and introduced several new members. These members are focused on the young central subgroups in Taurus (i.e. Greater Taurus groups 8-11 in Section \ref{sec:sc_tau}), making them a more homogeneous sample for comparison relative to the more diverse sample from \citet{Krolikowski21}, which includes both the objects in \citet{Esplin19} and those in more peripheral regions, which may be older. \citet{Esplin19} identified members as candidates using proper motions and photometry, and then confirmed them using spectroscopic observations including Li I absorption. We found that 189 of 351 (54\%) stars in Taurus have Gaia detections that survive the quality cuts described in Section \ref{sec:dat-gaia}. While these restrictions did result in the loss of nearly half of the members, it is important for the reliability of our results that objects with poor Gaia data are excluded. Of the stars that survived the quality cuts, 172 of 189 were identified as young, or 91\%. We compile the recovery rates as a function of spectral class for Taurus and the other three test associations in Fig. \ref{fig:recovery}, where we show that the recovery rate in Taurus is essentially complete for spectral classes M and K, with a sharp sensitivity decline for G-type stars. This lowered sensitivity is caused by the photometric overlap between pre-main sequence stars earlier than about G8 and the subgiant branch on the HR diagram, which significantly reduces the probability that a star there is young. 

The Upper Sco Association is slightly older than Taurus (age $\sim$5-11 Myr; \citealt{Rizzuto15,Pecaut12,Preibisch99}), and we therefore expect less sensitivity to early type stars as members begin to merge onto the main sequence. The Upper Sco sample comes from \citet{Rizzuto15}, \citet{Preibisch02}, and \citet{Preibisch98}, making use of spectroscopic observations of Li and H$\alpha$ combined with complete kinematic solutions to confirm the membership of the stars \citep{Rizzuto15}. A total of 318 members out of the total list of 477, or 67\%, had a high-quality Gaia counterpart that passed our quality cuts. Of those high-quality Gaia detections of known Upper Sco members, 274 were identified as young by our method, or 86\%. As presented in Figure \ref{fig:recovery}, our completeness is best for the lowest-mass stars, which provide essentially complete results for M dwarfs up to about M1. For stars earlier than M1, the sensitivity drops gradually, with the same significant sensitivity drop we observed in Taurus appearing beyond spectral type G8. 

\edit1{To check the nature of contamination in our sample, we also investigated a subset of the \citet{Rizzuto15} catalog, which were identified as kinematically and photometrically consistent with Upper Sco but were rejected as young members due to the absence of a Lithium feature. Since our method is purely photometric, we identified most of these false positives as young candidates (71\%), however we also found that these stars have disproportionately high RUWEs, with 53\% having RUWE$>$1.1, compared to 43\% for our larger sample of candidate young stars. This is consistent with most of these being spectroscopic binary field interlopers, which appear to be the dominant source of contamination in our sample.}

The older $\beta$ Pictoris Moving Group (age $\sim$ 23 Myr; \citealt{Mamajek14}), hosts a more elusive population, as these stars sit on a pre-main sequence that is less well removed from the photometry of background stars. To check the recovery rate in $\beta$ Pic, we used the catalog from \citet{Shkolnik18}, which identified members using both 3-d motions and the presence of Li and H$\alpha$. 94 of 173 stars in the association, or 54\%, survived our Gaia quality cuts. The rate at which we recovered these young stars was considerably lower, however, with 44 out of 94 members being recovered, or 47\%. Rather than there being a sensitivity limit around G8, the brightest stars identified in $\beta$ Pic have spectral types in the early K, caused primarily by the more luminous stars having already settled close to the main sequence. Later K stars showed low recovery rates, suggesting that they are usually only marginally identified (see Fig \ref{fig:recovery}). Significantly better recovery rates were observed for mid-M stars, peaking at approximately 80\% for M3-M4, suggesting that our method remains effective at consistently identifying the less massive stars that remain well above the main sequence even at older ages. 

The Tucana-Horologium association is near the 50 Myr target age limit with an age of $\sim$45 Myr \citep{Kraus14}, and therefore we should expect significantly lower sensitivity to this population. The sample population of Tuc-Hor members that we used comes from \citet{Kraus14}, which, like for our $\beta$ Pic population, used a combination of 3-d kinematic information and stellar youth indicators to verify membership. 80\% of Tuc-Hor members (or 115 out of 143) survive the Gaia quality cuts. As expected, our recovery of Tuc-Hor members showed a significant decay in the recovery rate of stars overall, with no noteworthy sensitivity for stars earlier than a spectral class of about M1 (see Fig. \ref{fig:recovery}). The recovery rate for the later M-dwarf bins in Tuc-Hor is around 20\%, which, while significantly weaker than our recovery for younger stars, does still identify 17 Tuc-Hor members out of the complete sample of 115 across all spectral types, or 15\%, enough to make the population potentially identifiable.

The recovery rates among stars with quality Gaia detections in each of the four regions discussed here are compiled in Figure \ref{fig:recovery}, showing clearly how our recovery rates vary with spectral type, and with the age of the stars being observed. There we show how the effectiveness of our method at a given spectral class negatively correlates with the age of the region, with the nearly newborn Taurus Association being essentially complete from G8 to late M excluding the quality cuts, and the 45 Myr old Tuc-Hor Moving Group having extremely limited sensitivity essentially restricted to M dwarfs. For ages in between, sensitivity gradually falls off, with G and then K stars being lost first, and M dwarf sensitivity beginning to drop past $\sim$ 20 Myr as for our $\beta$ Pictoris sample. For Tuc-Hor, an association near our target age limit, we maintain a recovery rate of approximately 20\% among M stars, while only one earlier star is identified. Despite the restrictions to our sensitivity for these older groups, our ability to detect non-negligible numbers of members in even the relatively old Tucana-Horologium Moving Group suggests that large enough groups will remain identifiable up to the upper edge of our target age range at 50 Myr.

Despite the successful stellar recovery among these older populations, the composition of the stars recovered should be treated with caution, as we expect stars with features that artificially inflate a star's photometric youth, such as  unresolved binaries, to be detected much more easily than an average star in the samples included in this section. With an overall recovery rate around 15\%, as for Tuc-Hor, unresolved binaries may begin to dominate the recovered sample. Particularly in these older populations, demographic studies can be greatly expanded by using the locus of an association in space-velocity coordinates as defined by the limited sample we identify as young to locate and reintroduce probable members that are not recognized as photometrically young by our pipeline. The reintroduction of photometrically older probable members can not only expand the completeness of the original sample, but also help to suppress biases produced by the potential overrepresentation of binaries, such as in the derivation of group ages, an application we explore in Section \ref{sec:groupage}. 

\subsection{Age Estimation} \label{sec:methods-ages}

\begin{figure}[t]
\centering
\includegraphics[width=8.0cm]{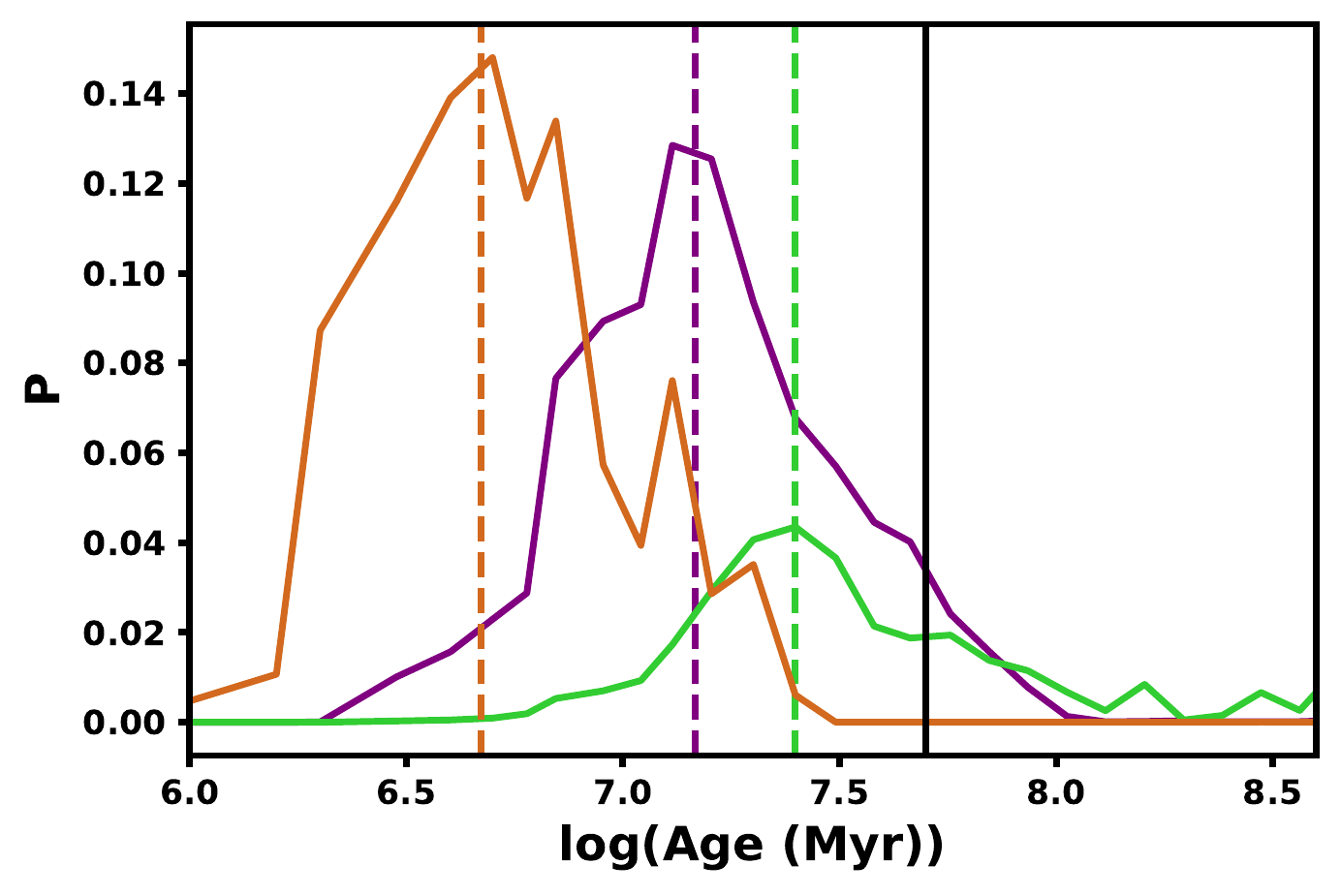}\hfill
\caption{\edit1{Log(age) probability distributions for three young star candidates. The peaks of the gaussian fits, which are used in age estimation, are marked with vertical dashed lines, and they closely reflect the mode of the distribution. The reduction in the peak height for the older sources is caused by the presence of a progressively larger high-age tail to older solutions which continues far to the right of the distribution shown. A larger tail reduces $P_{Age<50 Myr}$, which is defined as the sum of all bins younger than 50 Myr (marked by the solid black vertical line). }}
\label{fig:pdist}
\end{figure}

The primary objective of generating our model population and corresponding posterior distributions was the identification of young stars in the presence of diverse stellar populations in the solar neighborhood. However, the age distributions that we generate can also be used to derive approximate ages for most of the young stars in our sample. For candidate young stars, the age probability distribution from our pipeline \edit1{typically} consists of a peak at a young age, and a significant tail \edit1{to older ages populated by main sequence model stars with bright unresolved companions. A gaussian least-squares fit to that peak would provide a suitable first-order age estimate, emulating the mode of the distribution while avoiding influence from the high-age tail that may dominate estimates from the mean or median (see Fig. \ref{fig:pdist}). While these gaussian fits are effective in representing the age distribution peaks, the wide range of properties considered in these distributions, especially multiplicity, that make these peaks unlikely to reflect a star's true isochronal age. Much more accurate ages can be derived by limiting the range of metallicities and multiplicities that we consider, and using a population of model stars with known properties to create a map connecting gaussian least-squares fits to the age and mass probability distributions with a corresponding true age from the model.}

%However, these age fits alone would likely be insufficient because of the wide range of properties included in our model that influence brightness, which may introduce skewing effects for fits to a single star's probability distribution. Much more accurate ages can be derived by limiting the range of properties that we consider, most notably multiplicity and metallicity, and using model stars with known properties to create a map to connect existing fits to the probability distributions to age and mass with a corresponding true age from the model.

While binaries are very common, potentially representing a majority of young systems \citep[e.g.][]{Kraus11,Raghavan10}, their companions introduce a very wide range of photometric excesses, depending on the mass of the companion. As such, photometry is a much less consistent age diagnostic for multiple systems compared to stars without a companion. We therefore exclude binaries and higher-order systems from our individual age estimates, subsequently assuming that all Gaia sources have no companion. For single stars, the lack of consideration for binaries will ensure that our results are not influenced by the wide range of photometric solutions possible from the presence of an unresolved companion, making the age estimate of high quality. If a star does have a companion, it will contribute to a background of stars with younger age fits that are not consistent with one another, making them possible to clip out when looking at populations. 

We also fix the assumed metallicities of our candidate young stars to the solar value. This assumption has been shown to be appropriate for the vast majority of nearby young clusters and associations \citep[e.g.,][]{Almeida09, Mamajek13}, which have a uniformity in their metallicities consistent with the thorough mixing of the ISM on 100pc-level scales \citep[e.g.][]{deAvillez02}. Table 3 in \citet{Almeida09} presents metallicity values for most nearby young associations, with all solutions falling between [Fe/H]=-0.13 and [Fe/H]=+0.04, mostly within one standard deviation of solar. The $\alpha$ Persei cluster is one of the most metal-rich known groups accessible through our search, and it has a metallicity of only [Fe/H]=+0.18 \citep{Pohnl10}. Due to the mixing of the local ISM, we would not expect any of our groups to depart significantly from the trend of consistently near-solar metallicities for young nearby associations and clusters. Unlike these young groups, older clusters have a more diverse set of metallicities reflective of those in the rest of the local stellar population \citep[][see Fig. \ref{fig:FeH}]{GALAHDR2Buder18, Hayden19}. The Hyades and Praesepe have notably super-solar metallicities at [Fe/H]=+0.146 and  [Fe/H]=+0.21, respectively \citep[e.g.,]{DOrazi20,Cummings17}, while numerous other clusters are much more metal poor, such as M35 and NGC 2506 at [Fe/H]=-0.21 and  [Fe/H]=-0.52, respectively \citep{Bouy15, Friel93}. This contrast in the metallicity variation between the youngest stellar populations and various older populations in the solar neighborhood strengthens the decision to restrict the metallicity to solar for our in-depth look at young stars, while assuming a much broader prior metallicity distribution for our wider population. 

To extract age estimates from our output age probability distributions, we feed model stars with known properties through our pipeline for posterior distributions, and generate Gaussian least squares fits to the output age distributions. We generate these model stars at fixed Solar metallicity, chosen at random with uniformly distributed ages between 1 and 50 Myr, and masses from 0.1 to 1.3 M$_{\odot}$. This mass limit roughly corresponds to where stars at 1 Myr stop being clearly identifiable on the pre-main sequence as they cross the subgiant branch. The photometry of these test stars is allowed to vary according to the mean Gaia photometric uncertainty and mean reddening uncertainty in the group, providing the expected photometric variation for an individual star in that group. As for the Gaia sources, Gaussian distributions are fit to the resulting output age and mass distributions for each test star, and the input ages are binned according to the output fits for mass and age. For very young stars where the peak is in the youngest age bin, a Gaussian least squares fit will often view the probability distribution as the tail of an exponential with a peak below zero, and therefore we exclude the 1 Myr bin which often creates these anomalous fits, limiting the ages considered to between 2 and 50 Myr. We compute a median input age and standard deviation for the true ages of stars in each fit age and mass bin, which completes a map that links any given set of age and mass results from the main pipeline to a revised age result excluding any skewing effects related to metallicity and multiplicity.

The vast majority of corrected age solutions are younger than those derived from the peak of the Gaussian age distribution by less than $\sim$30-40\%, which is expected given that the addition of binaries adds a range of older possible solutions to the photometry of the target. Some of the older (age $>$30 Myr) and more massive stars near the limit of our method's sensitivity occasionally require larger correction to age of up to a factor of 2. A total of 25727 of the original 30518 credible young candidates generate quality age fits, or 84\%. Most of the stars without good fits have photometric ages below 2 Myr, a category that includes both genuinely young stars and objects lowered into that photometric age bracket by unresolved binarity or insufficiently corrected reddening. We also lose some objects near the limits for detectability where stars begin to merge into the main sequence.  Upon comparing the isochrones associated with these new revised ages to the photometry of these stars, \edit1{we find} that the age results are consistent with our \edit1{isochronal} model. Within Sco-Cen, many of the refined ages within subclusters host uncertainties comparable to internal variation of the age between members, which would be expected assuming a common formation time.

\edit1{These age estimates assume single stars with solar metallicity, so individual age estimates should be treated with caution when those details are not known}. The interference from binaries can however be reduced by considering groups of stars rather than individuals. Therefore, when we search for age gradients within a group, rather than considering the age fits of all stars individually, we consider the median of the 10 nearest neighbors in space-velocity coordinates. This is not a perfect solution for a group of stars, as \edit1{our young star candidates are typically photometrically younger on average compared to the population as a whole} (see Section \ref{sec:groupage}), however it is nonetheless a useful and efficient way to identify and visualize age patterns in a population. 

\subsection{Clustering with HDBSCAN} \label{sec:clustering}

The presence of structures in our young stellar population is clear through the overdensities visible in Figure \ref{fig:allsky_redcir}, however a consistent means of structure identification is necessary for a proper analysis. To this end, we employ HDBSCAN (Hierarchical Density-Based Spatial Clustering of Applications with Noise; \citealt{McInnes2017}), a hierarchical density-based clustering algorithm developed as an extension of the frequently-used DBSCAN clustering algorithm \citep{DBSCAN}. HDBSCAN has already been used quite successfully in literature for a variety of purposes, including in \citet{Kounkel19}, where it is used to identify tenuous stellar structures of all ages in their own Gaia DR2 sample. As such, there is strong precedent for its use in identifying groups and associations within large stellar populations. Unlike other clustering algorithms, which typically presuppose a density threshhold, shape, or some other expected cluster property, HDBSCAN uses only a minimum cluster size and a parameter with a smoothing effect on the density distribution. As such, it is an algorithm with next to no built-in assumptions, allowing for young structures to be identified without discrimination based on complicating factors such as unusual shapes or wildly varying scales. Furthermore, because the algorithm searches for structures at different scales, it also provides a unique opportunity to explore different levels of substructure within an association. Lastly, HDBSCAN enables clustering in an arbitrary number of dimensions, allowing for both XYZ spatial coordinates and l and b transverse velocities to be included in clustering simultaneously. In some locations with high localized reddening, the \citet{Lallement19} maps are insufficient to properly correct for reddening, resulting in anomalous overdensities without common velocities. Through the inclusion of velocities in our clustering, we are able to exclude such anomalies. 

HDBSCAN's search for overdensities makes use of a parameterized density metric, which is based on the distance to the $k$th nearest neighbor (referred to as the core distance, or $core_k($object$)$) \citep{McInnes2017}, where $k$ is the first of two key HDBSCAN input parameters\footnote{given by the HDBSCAN clustering parameter min\_samples}. Increasing $k$ effectively smooths the density distribution and causes smaller clumps of stars to be ignored, therefore serving as a proxy for how conservative the clustering will be. HDBSCAN generates a new mutual reachability distance metric from these core distances, which for objects $a$ and $b$ is defined as $d_{mreach-k}(a, b) = max\{core_k(a),core_k(b),d(a,b)\}$, where $d(a,b)$ is the distance from object $a$ to object $b$. Using this metric, HDBSCAN generates a minimal spanning tree, which is essentially a web linking each point in the data set through their nearest neighbor according to their mutual reachability distances. By removing links in the tree of sources with the lowest weights, corresponding to large mutual reachability distance to their neighbors, links connecting clusters in the web are broken, leaving behind a hierarchy of clusters that appear and fragment as the weight threshhold is raised\footnote{An excellent visualization of this process is given on HDBSCAN's website, at \url{https://hdbscan.readthedocs.io/en/latest/how_hdbscan_works.html}}.

The methods employed by HDBSCAN are comparable to those of DBSCAN, which defines the cores of clusters by the presence of more than some number of neighbors within a radius $\epsilon$, effectively taking only clusters that exist at some arbitrary chosen scale \citep{DBSCAN}. In HDBSCAN, rather than looking for clusters at a single $\epsilon$, clustering is simultaneously extracted at all scales, allowing the most persistent clusters to be identified, rather than those that happen to emerge at any requested scale \citep{McInnes2017}. Any clusters that come out of the tree with less than $N$ members, where $N$ is the second key HDBSCAN input parameter\footnote{given by the HDBCSAN clustering parameter min\_cluster\_size}, are thrown out and merged into their parent cluster: a larger-scale node on the clustering tree. The quality of clusters in the tree is set by their persistence, which relates to the range of scales over which the cluster is defined in the clustering tree. HDBSCAN also includes the value $\epsilon$ as an optional parameter\footnote{given by the HDBSCAN clustering parameter cluster\_selection\_epsilon}, setting a minimum threshold past which clusters can no longer be fragmented. This is useful in cases where large groups with significant substructure are present, as it allows for the entire region to be identified as a whole, without excessive fragmentation. HDBSCAN has two options for cluster selection - ``EOM" (excess of mass), which selects the best or most persistent clusters in the clustering tree, and ``leaf", which identifies only the nodes of the clustering tree, effectively selecting clusters from a maximally fragmented clustering solution \citep{McInnes2017}. EOM clustering is preferred in most cases, as the persistence parameter that defines an EOM cluster is used by HDBSCAN as a proxy for cluster quality, although leaf clustering can be useful when attempting to detect subclusters within a well-defined higher-level cluster. A more complete look at the methodology of HDBSCAN can be found in \citet{McInnes2017}, along with excellent visualizations of the tree clustering it implements.

An additional consideration is the incorporation of velocities in clustering. Since transverse velocity in km s$^{-1}$ and galactic XYZ coordinates in pc do not have matching scales, a corrective factor $c$ is required to \edit1{reflect the differing standard deviations of groups between distance and velocity coordinates. We therefore apply HDBSCAN to the following five-dimensional data set:}
\begin{equation}
    (\text{X}, \text{Y}, \text{Z}, c*v_{T,l},c*v_{T,b})
\end{equation}
The known young associations often have wildly different ratios between spatial extent and velocity dispersions, however ratios of 4 to 6 pc/km s$^{-1}$ are typical, approximately reflecting the ratios predicted for cloud scales of 25 to 100 pc in the Larson's Law relation \citep{Larson81}. While Larson's Law was designed to relate velocity dispersions to scales in clouds, associations are expected to inherit properties from their parent cloud, and maintain those properties for some time after formation, especially in loose associations where dynamical interactions are infrequent \citep{Larson79, Wright18,Ha21}. Size-velocity ratios of 4-6 pc/km s$^{-1}$ predict properties consistent with known groups such as Upper Sco and to a lesser extent the more dispersed regions of UCL and LCC \citep{Wright18}. This range of factors is also consistent with relative sizes of visually-identifiable Gaia clumps in spatial coordinates and in transverse velocity space. Any variation within this range can be interpreted as a choice to tweak the relative weight of velocity and spatial coordinates in the clustering, where a larger corrective factor weighs velocity more heavily by reducing the relative variability of the spatial component. 

For all-sky clustering, a size-velocity corrective factor of $c$ = 6 pc/km s$^{-1}$ was chosen, which weights kinematics slightly more heavily, reflecting the larger velocity dispersions expected in the larger groups such as Sco-Cen that we wish to identify at this level. A relatively large $\epsilon$ value of 25, in units of pc (or $c$ km s$^{-1}$ for velocity), supplements this choice, preventing fragmentation of groups below a scale comparable to a moderately-sized association like Upper Sco \citep{Wright18}. This choice ensures that groups like Sco-Cen can be treated like coherent singular groups, while also allowing stellar structures that link subtly separated groups to be detected, which would otherwise be incorporated into the background if those groups were recorded separately. While groups being linked by this condition does not guarantee that they share common formation origins, it does expose spatial and kinematic similarities that can be further explored in subclustering (see Section \ref{sec:dis-subgroups}). The input parameter $N$, representing minimum cluster size, was set to 10, as was $k$, which is set equal to $N$ by default. This choice avoids the inclusion of negligibly small clusters or high-order multiple systems, producing a set of largely visually convincing clusters. However, some groups with less internally consistent velocities or spatial distributions are also included, so we set a cluster persistence cut at a value at 0.015. This number was chosen to ensure that no groups with velocity distributions comparable to the field were included, most notably the reddening anomaly towards Aquila marked in Figure \ref{fig:allsky_redcir}, which is dense enough in space to be marked as a cluster without any persistence-based quality restrictions. Through this choice, our clustering results in this implementation are made to be intentionally conservative, with the objective of highlighting noteworthy structures in the Solar neighborhood.

\begin{figure*}[t]
\centering
\includegraphics[width=15.5cm]{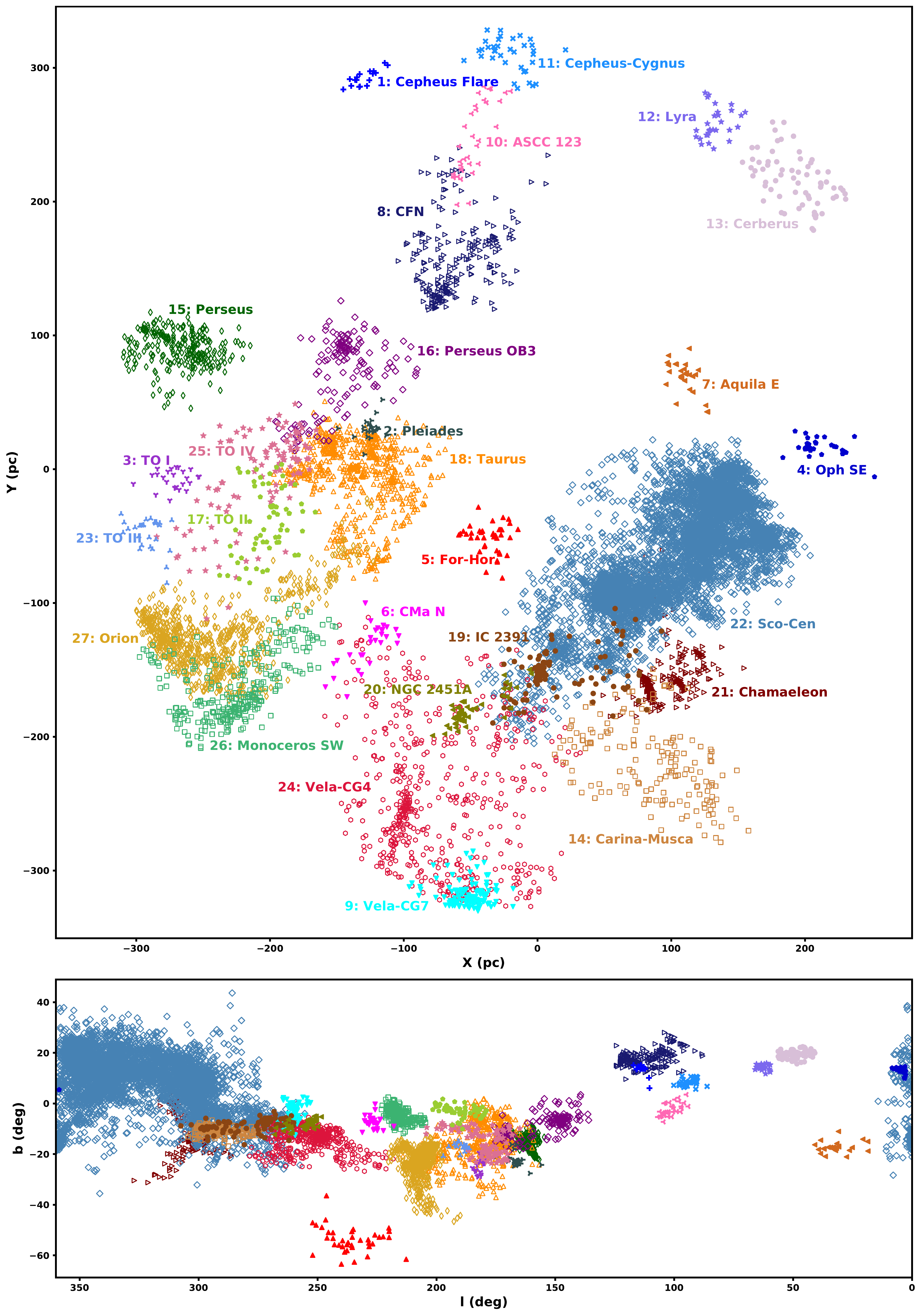}\hfill
\caption{The results of the all-sky clustering analysis from HDBSCAN in X/Y galactic spatial coordinates (upper panel) and l/b sky coordinates (lower panel). Each color corresponds to a separate top-level group identified by HDBSCAN, which are labelled according to the IDs and names in Table \ref{tab:allsky_clusters}. Groups with substructure are shown using \edit1{empty} icons, which are given complete subclustering analyses in Sections \ref{sec:sc_tau} (Taurus), \ref{sec:ori} (Orion), \ref{sec:perseus} (Perseus), \ref{sec:sc} (Sco-Cen and Chamaeleon), and \ref{sec:minorgroups} (all others). See Figure \ref{fig:allsky_redcir} for the X/Y distribution of stars prior to clustering.}
\label{fig:allsky_clustering}
\end{figure*}

We applied HDBSCAN using the above input parameters to members of our population of probable young stars\edit1{, with the additional requirement that} $\pi$/$\sigma_{\pi}>$25\edit1{. Clustering analyses in particular benefit from well-constrained distances, which prevent the erroneous merging of groups stretched along a similar line of sight, and this cut restricts the expected distance spreads down to about 13 pc, a scale comparable to those of many of the smaller groups in the solar neighborhood.} A total of 28340 stars entered into the final 5-d data set\edit1{, a 7\% reduction relative to the sample before the additional parallax restriction. }HDBSCAN identified 27 spatially and kinematically distinct young associations from this population, which are shown in XY galactic spatial coordinates in Figure \ref{fig:allsky_clustering}. These clustered sources account for 41\% of the total young population we identified, with the remaining sources likely being composed of a mixture of young stars ejected from larger associations, young stars in tenuous unrecognized associations, and false positives enabled by reddening anomalies and unresolved multiplicity. Sco-Cen was easily detected, representing by far the largest group with just under 7400 stars, nearly an order of magnitude larger than the next largest group, the near edge of Orion. Sco-Cen appears alongside a series of nearby but spatially distinct structures, such as the dense clusters in Chamaeleon. Perseus, Taurus, and the near edge of Orion are also identified in at least some form, with varying levels of internal division, along with some smaller known groups such as Perseus OB3 and the near edge of Vela. Many of the remaining groups appear to be previously unknown. 

\begin{figure}[t]
\includegraphics[width=8.2cm]{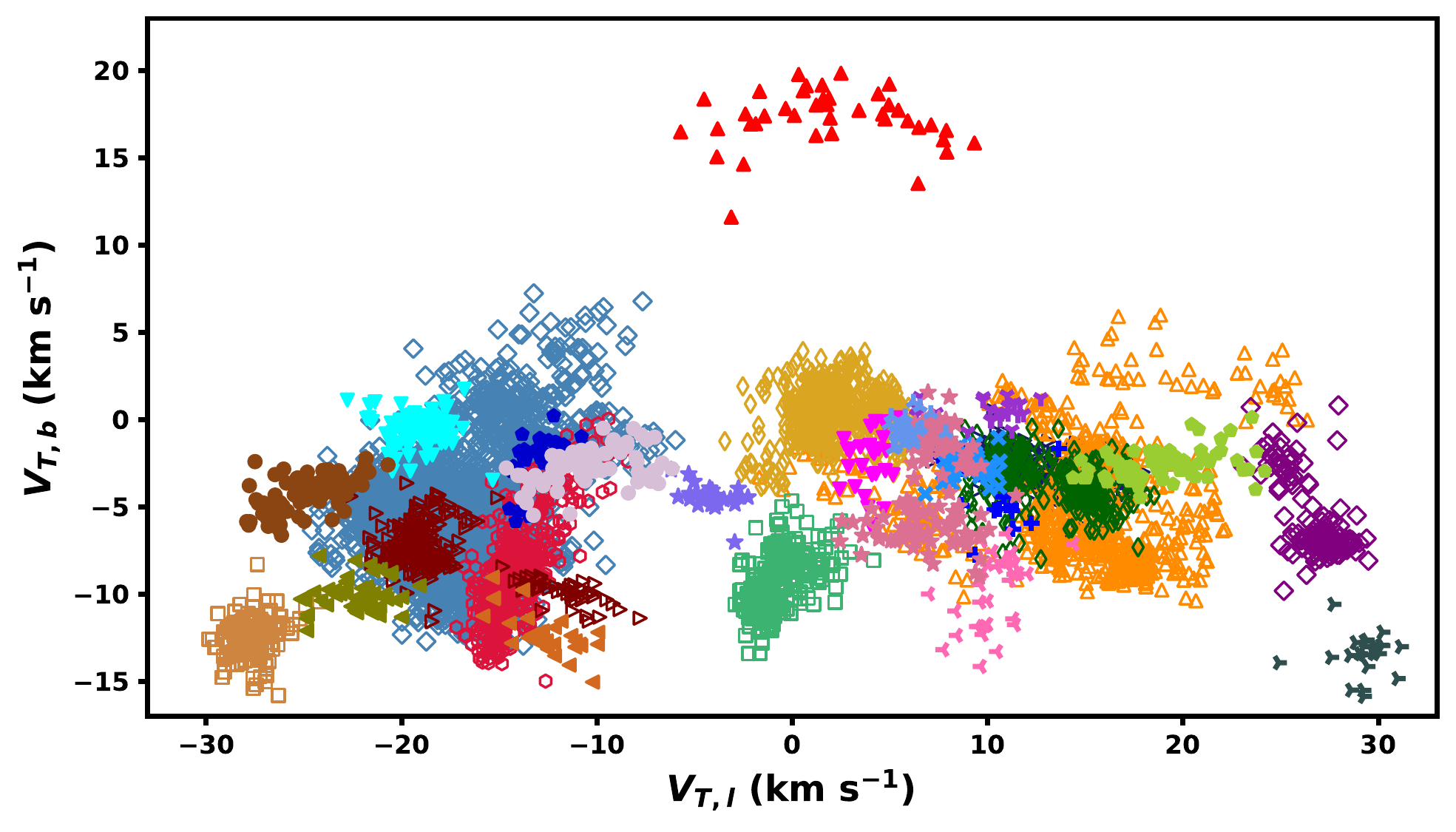}\hfill
\caption{The same as Figure \ref{fig:allsky_clustering}, but for transverse velocity.}
\label{fig:allsky_clustering_vel}
\end{figure}

To facilitate a more detailed study of Sco-Cen and other top-level groups we performed an additional round of clustering on groups that showed visible substructure, however the use of HDBSCAN there was modified slightly to enable the recognition of hierarchical structures in each region. Rather than identifying only the EOM clusters, as was done for top-level clustering, we also extract the smaller-scale leaf clusters, which might exist within a larger EOM cluster. The approach we employ fuses the two clustering methods, recognizing EOM clusters containing leaf clusters as intermediate-level groups, and the leaf clusters as subclusters within that larger EOM group. For this clustering, $N$ and $k$ were kept the same, while the space-velocity corrective factor $c$ was reduced to 5 to reflect the smaller scales, and $\epsilon$ was removed entirely, allowing for subclusters to be identified as separate regardless of the presence of nearby groups. The goal of this lower-level clustering is to identify all potential subgroups in a structure, regardless of scale or membership as part of a larger group. Ten of our 27 top-level subgroups have substructure, and within those ten groups this implementation identifies a total of 60 EOM clusters, with eight of those 60 further subdividing into a total of 27 additional smaller leaf clusters. We present the top level, EOM, and leaf clustering results for each candidate young star in Table \ref{tab:master_ys}, alongside $P_{Age<50 Myr}$ and the age solutions derived in Section \ref{sec:methods-ages}.

\begin{deluxetable*}{cccccccccccccc}
\tablecolumns{14}
\tablewidth{0pt}
\tabletypesize{\scriptsize}
\tablecaption{The full catalog of all Gaia objects we identify with $P(Age<50 Myr)>0.1$, with basic Gaia properties, $P(Age<50 Myr)$, and age as derived in Section \ref{sec:methods-ages} for each. The TLC, EOM, and LEAF columns mark the top level cluster, EOM cluster, and leaf cluster respectively that each star is a part of, if any.}
\label{tab:master_ys}
\tablehead{
\colhead{ID} &
\colhead{TLC\tablenotemark{a}} &
\colhead{EOM\tablenotemark{a}} &
\colhead{LEAF\tablenotemark{b}} &
\colhead{Gaia ID} &
\colhead{RA (deg)} &
\colhead{Dec (deg)} &
\colhead{$\pi$ (mas)} &
\colhead{g} &
\colhead{bp-rp} &
\colhead{$P(Age<50 Myr)$} &
\multicolumn{3}{c}{Age\tablenotemark{c} (Myr)} \\
\colhead{} &
\colhead{} &
\colhead{} &
\colhead{} &
\colhead{} &
\colhead{} &
\colhead{} &
\colhead{} &
\colhead{} &
\colhead{} &
\colhead{} &
\colhead{val} &
\colhead{+ err} &
\colhead{- err}
}
\startdata
0     &         -1 &   -1 &      &  3239529836638867840 &   73.8034 &   4.7175 &     4.170 &  17.80 &  3.50 &  0.34 &  28.3 &        8.4 &        5.7 \\
1     &         22 &   11 &      &  6099439604319602560 &  217.3498 & -44.7853 &     5.960 &  15.60 &  3.00 &  0.94 &  12.0 &        2.7 &        3.0 \\
2     &         -1 &   -1 &      &  4154795132763964032 &  278.9114 & -10.8489 &     3.020 &  14.40 &  2.40 &  1.00 &    &         &         \\
3     &         21 &    4 &      &  5774202930948040064 &  266.3687 & -82.1980 &     6.260 &  16.10 &  3.10 &  0.90 &  18.2 &        5.6 &        4.8 \\
4     &         -1 &   -1 &      &  3442403338420195840 &   82.6082 &  28.2391 &     3.460 &  11.90 &  1.40 &  0.23 &    &         &         \\
5     &         17 &   -1 &      &  3442418800302357248 &   83.0898 &  28.3245 &     5.220 &  16.10 &  3.00 &  0.89 &  15.6 &        3.3 &        5.3 \\
6     &         -1 &   -1 &      &  4154777227001098112 &  279.5766 & -10.6085 &     3.100 &  14.30 &  1.90 &  0.15 &    &         &         \\
7     &         -1 &   -1 &      &  1937498784187322624 &  351.8766 &  45.0985 &     3.830 &  16.70 &  2.90 &  0.24 &  22.8 &        9.6 &        4.3 \\
8     &         -1 &   -1 &      &  6195524512419762816 &  204.7577 & -21.6912 &    11.930 &  13.30 &  2.60 &  0.56 &    &         &         \\
9     &         -1 &   -1 &      &  1716855559591213824 &  192.1970 &  78.9078 &     7.670 &  15.10 &  2.90 &  0.46 &  19.9 &        4.8 &        4.8 \\
\enddata
\tablenotetext{a}{Objects with $\pi$/$\sigma_{\pi}<25$ were not subjected to clustering, and have a TLC and EOM markers of 0. Objects not assigned to a cluster are given the marker -1. }
\tablenotetext{b}{left blank if not related to a leaf cluster.}
\tablenotetext{c}{left blank if no solution reached.}
\tablenotetext{}{only a small subset of the 30518 total objects is shown here. The full table is available in the online version of this paper.}
\vspace*{0.1in}
\end{deluxetable*}

\subsection{Group Ages} \label{sec:groupage}

The ages we derive for entire groups of stars cannot be calculated in the same way as was done for individual stars, as the population we consider during clustering is, by design, skewed towards photometrically younger objects. In groups of fixed age, photometrically younger stars include objects with unresolved companions, or more generally stars on the high side of the expected photometric variation. This means that group ages calculated using our likely young stars alone will be skewed young, especially in groups older than $\sim$20 Myr \edit1{where only a limited section of the pre-main sequence is reliably detected as young (e.g., see Fig. \ref{fig:Pages_CMD})}. We can reduce this effect by reintroducing stars that our pipeline did not identify as young, and fitting an isochrone to \edit1{a cleaned subset of }the expanded population. This can be done by searching for objects that have spatial coordinates and kinematics consistent with the known members of that group, at least as close to the tenth-nearest HDBSCAN-identified member in space-velocity coordinates as the most peripheral HDBSCAN-identified member, mimicking the original HDBSCAN methods for finding members of clusters with $k$=10. This approach effectively uses the groups in the original clustering analysis as signposts from which more complete young stellar populations can be gathered and used for age estimation, removing influences from our pipeline's selection biases. The stars present in these extended populations are provided in Table \ref{tab:extpop}. 

For each candidate young cluster, we perform a single age fit based on the photometry of \edit1{candidate members}, including both the original candidate young stars and the extended population of comoving/cospatial objects mentioned above. Our methods employ a least squares fit, where we compare the photometry of these populations to a grid of solar metallicity PARSEC isochrones with ages between 0.25 and 80 Myr \citep{PARSECChen15}. Only stars within the range of G magnitudes occupied by the isochrone grid are included to enable reliable interpolation, a restriction that also helps to suppress some of the background contamination added upon reintroducing candidate members that are not photometrically young. We also \edit1{limit the range of BP-RP Gaia color to between 1.2 and 4} to restrict the sample to the pre-main sequence. Most of the groups we recover have pre-main sequences that fall comfortably within the range of photometry given by these 0.25-80 Myr isochrones, with the exception of the Pleiades, which we exclude from age estimation due to its considerably older known age \citep[e.g.,][]{Lodieu19}.

To reduce contamination from external sources, we restrict the sources included in our age fits according to the core distance (as defined in Section \ref{sec:clustering}). For the core distance restrictions, we define a weight parameter in which the group member with the smallest core distance in the association has a weight of 1, the member with the largest core distance has a weight of 0, and sources in between have weights spaced linearly according to the core distance between these extremes. For top-level clustering, a default condition of weight $>$ 0.3 was required for the inclusion of members in the age fit for the group, which in most cases ensures that a young pre-main sequence dominates background contamination. In more contaminated groups this was raised as high as 0.5. For subclusters, we often found that no weight cut was required due to tighter spatial and kinematic distributions that are less vulnerable to contamination, however we re-imposed the restrictions where necessary. Very few groups needed changes to these conditions, with some examples being the large and badly contaminated subgroups of CHA-1 and ORI-2, as well as Taurus-Orion IV and Cepheus Flare at the top level. 

\edit1{Since unresolved binaries appear photometrically younger compared to single stars, we attempt to remove those likely binary young star candidates from inclusion in our age fitting. We therefore} further restricted the population used for age fitting based on \edit1{the} Re-normalized Unit Weight Error (RUWE), \edit1{an astrometric quality measurement that has been shown to be a strong indicator of stellar multiplicity} \citep[e.g.,][]{Bryson20}. In most cases, we removed sources with RUWE$>$1.1, roughly corresponding to where the additional contribution to the RUWE curve from binaries becomes negligible in \citet{Bryson20}. This harsh RUWE cut works well to remove binary sequences, however in cases with consistently high RUWE, such groups with embedded sources (e.g. Perseus, see Sec. \ref{sec:perseus}), this condition occasionally removes nearly all stars. Therefore, in cases where more than two-thirds of stars are removed by RUWE$<$1.1, we relax the condition to RUWE$<$1.2. This value corresponds to approximately where Gaia populations are expected to become fully binary-dominated, and it is the recommended criterion for culling binaries in \citet{Bryson20}. This cut combined with the weight cut significantly reduced the number of new stars used in the fits, from a typically order-of-magnitude increase in populations upon the initial reintroduction of objects to a number more consistent with the expected number, which should be equal to the number of young candidates times the inverse of the expected recovery rate (see Section \ref{sec:recovery}).

\begin{figure}[t]
\centering
\includegraphics[width=4.1cm]{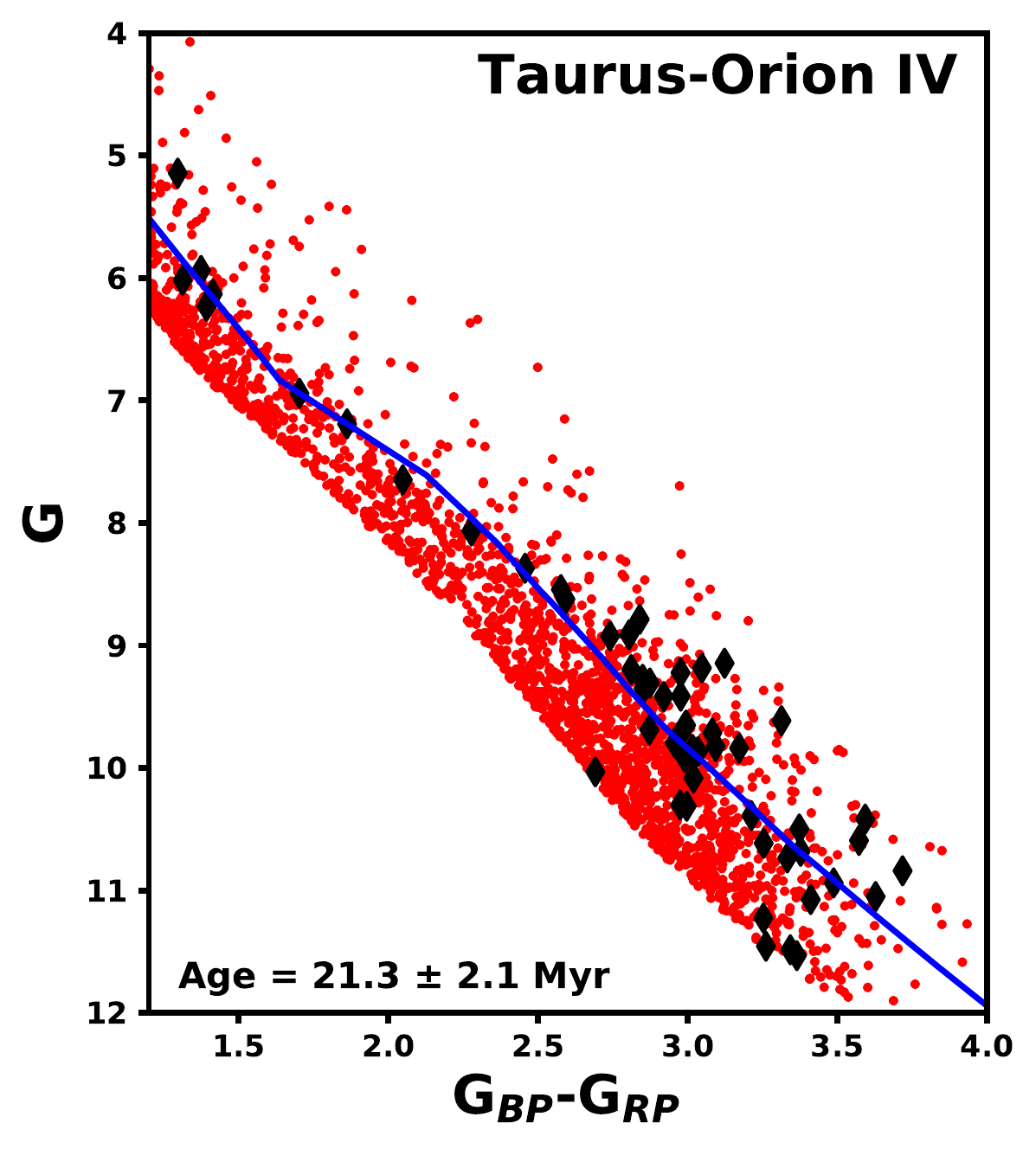}\hfill
\includegraphics[width=4.1cm]{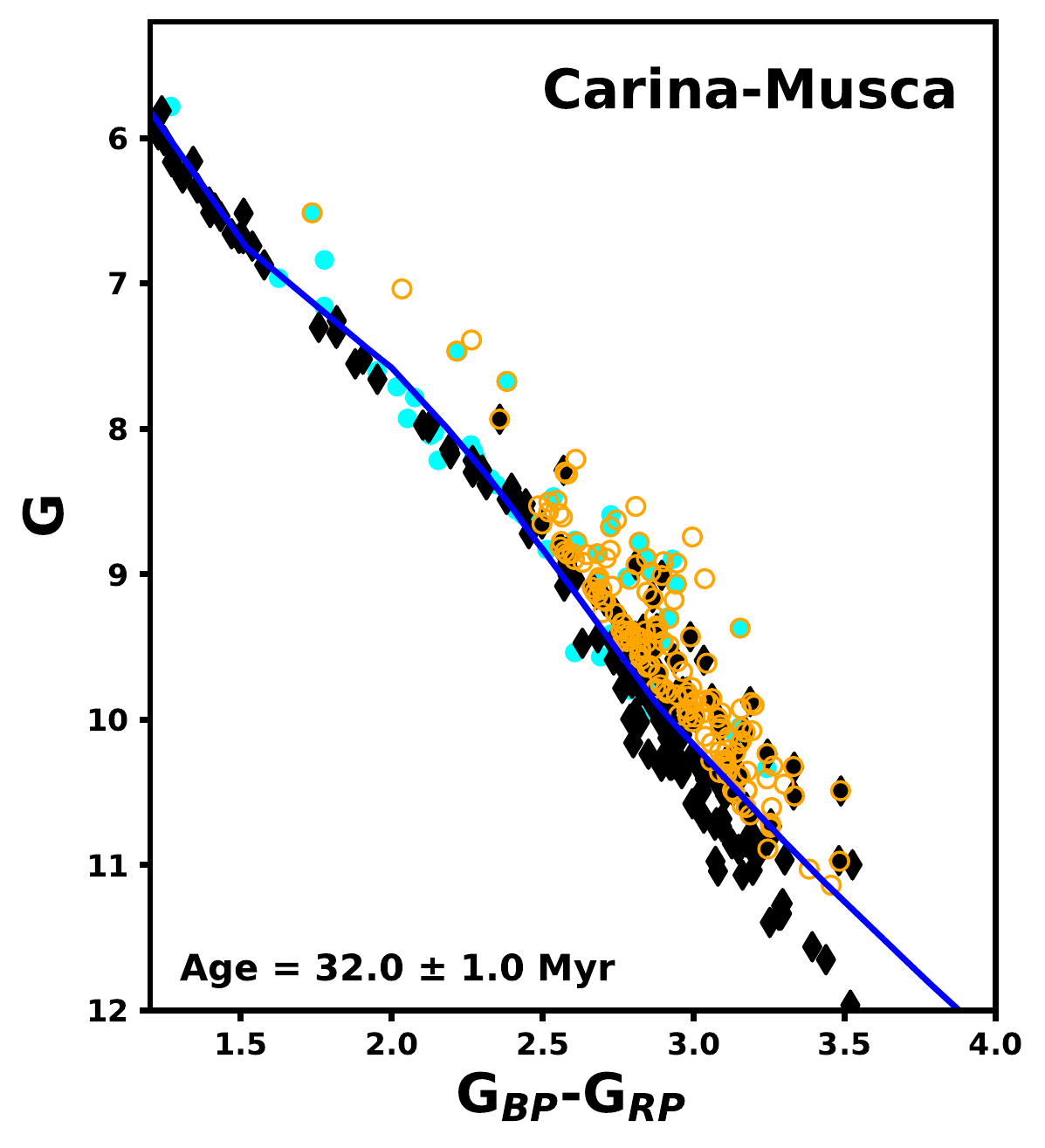}\hfill
\caption{Sample isochrone fits for two of our young associations. In the left panel, we display our handling of severe contamination in Taurus-Orion IV, where a weight cut raised to 0.5 resulted in a very clean pre-main sequence emerging from the noise. Objects that failed the weight cuts are marked with red circles, and the stars that pass are given by black diamonds. In the right panel, we present members of the relatively uncontaminated Carina-Musca association after applying weight cuts. Stars marked by large cyan circles fail the subsequent RUWE cut, while stars marked by black diamonds pass it. Stars that fail the RUWE cut are likely binaries, hence the existence of a major concentration of them along an apparent binary sequence. Only objects marked by \edit1{empty} orange circles are identified as young Carina-Musca members using our methods, highlighting the potential bias towards photometrically younger stars in that population that our age fitting method attempts to rectify. \edit1{Isochrone fits for all groups and regions are provided in Appendix \ref{appendix}.}}
\label{fig:CMDfit}
\end{figure}

Once these culling conditions were applied, visually convincing fits were produced for all groups in our sample, which neglect strong background contamination and visible binary sequences. Figure \ref{fig:CMDfit} shows a two examples of the results of this fitting. We generate fits for all HDBSCAN-identified groups without substructure, as well as a few cases that do show substructure but with a limited enough age spread for a global age to make sense. As we note throughout Section \ref{sec:results}, the vast majority of our age fits for known groups are consistent with literature values, however the uncertainties should be treated with caution, as we report standard errors that cannot capture the dominant systematic errors related to the choice of model and clipping, or any inherent age spread. Due to the 80 Myr isochrone limit for inclusion in fitting, some members in associations with ages $>$50 Myr where these isochrones become close together may be removed, so some additional caution should be applied to the age fits near the upper end of our age range. \edit1{Groups with a very small number of high-weight members should also be treated with additional caution, as their size makes them more vulnerable to the influence of a small number of field interlopers. These fits are all shown in Appendix \ref{appendix}. }

\begin{deluxetable*}{ccccccccccc}
\tablecolumns{11}
\tablewidth{0pt}
\tabletypesize{\scriptsize}
\tablecaption{The full catalog of Gaia stars with space-velocity positions consistent with our photometrically young populations. The TLC, EOM, and LEAF columns mark the Top Level cluster, EOM cluster, and LEAF cluster that each star is a part of, if any. We also include basic Gaia properties and the weight parameter described in Section \ref{sec:groupage}, which serves as a proxy for closeness of the group center.} \label{tab:extpop}
\tablehead{
\colhead{ID\tablenotemark{a}} &
\colhead{TLC} &
\colhead{EOM\tablenotemark{b}} &
\colhead{LEAF\tablenotemark{c}} &
\colhead{Gaia ID} &
\colhead{RA} &
\colhead{Dec} &
\colhead{parallax} &
\colhead{g} &
\colhead{bp-rp} &
\colhead{weight} \\
\colhead{} &
\colhead{} &
\colhead{} &
\colhead{} &
\colhead{} &
\multicolumn{2}{c}{(deg)} &
\colhead{(mas)} &
\multicolumn{2}{c}{(mag)} &
\colhead{}
}
\startdata
0     &          1 &   -1 &      &  2297218611804127104 &  309.8679 &  79.8920 &     3.176 &  18.93 &  2.87 &    0.02 \\
1     &          1 &   -1 &      &  2226243658763815552 &  336.1287 &  69.4314 &     3.035 &  17.39 &  3.61 &    0.23 \\
2     &          1 &   -1 &      &  2226228579136886144 &  336.4913 &  69.3191 &     3.296 &  13.51 &  1.23 &    0.08 \\
3     &          1 &   -1 &      &  2286800434190203776 &  334.4768 &  80.5854 &     3.191 &  17.79 &  2.69 &    0.09 \\
4     &          1 &   -1 &      &  2297464666187525120 &  324.2971 &  80.7429 &     3.391 &  11.98 &  0.83 &    0.01 \\
5     &          1 &   -1 &      &  2212189293119019264 &  343.6841 &  66.3300 &     3.134 &  17.04 &  2.58 &    0.02 \\
6     &          1 &   -1 &      &  2226302723153297664 &  336.9986 &  69.7687 &     3.066 &  18.36 &  3.76 &    0.27 \\
7     &          1 &   -1 &      &  2286846781182031744 &  339.9039 &  80.6398 &     3.300 &  15.34 &  1.98 &    0.08 \\
8     &          1 &   -1 &      &  2229484297488789248 &  338.5122 &  70.9616 &     3.439 &  17.85 &  2.65 &    0.05 \\
9     &          1 &   -1 &      &  2229446020739020032 &  336.3050 &  70.5487 &     3.399 &  17.84 &  2.56 &    0.05 \\
\enddata
\tablenotetext{a}{Note that in complex environments, extreme outliers may be shared between two clusters. These cases receive one entry for each possible solution.}
\tablenotetext{b}{Objects not assigned to an EOM cluster are given the marker -1. }
\tablenotetext{c}{Left blank if not related to a leaf cluster.}
\tablenotetext{}{only a small subset of the 94424 total objects is shown here. The full table is available in the online version of this paper.}
\vspace*{0.1in}
\end{deluxetable*}

\section{Results} \label{sec:results}

By applying HDBSCAN clustering to our sample of young stars, we have identified extensive young structures in the solar neighborhood. These include 27 top-level regions (see Fig. \ref{fig:allsky_clustering}), with properties provided in Table \ref{tab:allsky_clusters}. Many of these regions have significant internal structure, often with well-separated subclusters, or subclusters widely separated in age. Performing a subclustering analysis on each of these regions allows for insight into their star formation histories. We therefore manually inspect each region for visible substructure, and apply HDBSCAN subclustering on the ten regions where substructure is visually evident. While we do not assert an absence of subclustering in the remaining regions, most are too small or sparse for a subclustering analysis to be useful, and will therefore likely require a more targeted census before substructure can be ascertained. Details on this HDBSCAN implementation are described in Section \ref{sec:clustering}. Further detail on individual complex regions will be left to future publications, as will precise characterization of some of the previously unknown or little-known groups that we identify for the first time in this publication. These lesser-known groups are discussed in Section \ref{sec:minorgroups}, along with a discussion of their relations to past literature. Most are assigned a unique name in Table \ref{tab:allsky_clusters} to reflect their distinctiveness as significant nearby young associations.

\begin{deluxetable*}{cccccccccccccccc}
\tablecolumns{16}
\tablewidth{0pt}
\tabletypesize{\scriptsize}
\tablecaption{Properties of the nearby clusters identified using HDBSCAN. NC in the age column indicates that we find the group to be non-coeval, see the age column in the relevant subclustering table for the ages of the subgroups (e.g., Tables \ref{tab:tau-sc}, \ref{tab:ori-sc}, \ref{tab:sc_sc}, and \ref{tab:minor_sc}).} 
\label{tab:allsky_clusters}
\tablehead{
\colhead{TLC} &
\colhead{Name} &
\colhead{N} &
\colhead{RA} &
\colhead{Dec} &
\colhead{l} &
\colhead{b} &
\colhead{$D_{sky}$\tablenotemark{a}} &
\colhead{d} &
\colhead{$\mu_{RA}$} &
\colhead{$\mu_{Dec}$} &
\colhead{$V_{T, l}$} &
\colhead{$V_{T, b}$} &
\colhead{$\sigma_{V_T}$\tablenotemark{b}} &
\colhead{Age} \\
\colhead{} &
\colhead{} &
\colhead{} &
\multicolumn{2}{c}{(deg)} &
\multicolumn{2}{c}{(deg)} &
\colhead{(deg)} &
\colhead{(pc)} &
\multicolumn{2}{c}{(mas/yr)} &
\multicolumn{2}{c}{(km/s)} &
\colhead{(km/s)} &
\colhead{(Myr)}
}
\startdata
1  & Cepheus Flare       &   17 &  339.5 &  73.9 &  113.9 &  13.4 &    5.3$\times$2.2 &     329.3$\pm$3.2\tablenotemark{c}  &   6.8 &    0.9 &   9.9 &  -4.0 &   2.0$\times$1.4 &  11.2$\pm$3.3 \\
2  & Pleiades\tablenotemark{e}            &   23 &   57.4 &  24.9 &  166.4 & -22.4 &    7.6$\times$4.4 &     139.9$\pm$6.7  &  18.9 &  -44.7 &  29.2 & -13.4 &   1.3$\times$1.2 &  $>$80 \\
3  & Taurus-Orion I\tablenotemark{$\dagger$}      &   24 &   65.9 &  13.5 &  181.6 & -24.4 &    5.3$\times$2.5 &     299.2$\pm$10.7 &   4.5 &   -5.2 &   9.7 &   0.4 &   2.0$\times$0.6 &  10.6$\pm$2.2 \\
4  & Ophiuchus Southeast\tablenotemark{$\dagger$} &   31 &  257.5 & -18.3 &    4.5 &  12.7 &    4.7$\times$1.7 &     217.5$\pm$14.2 &  -5.4 &  -11.6 & -13.0 &  -2.2 &   1.6$\times$0.7 &  17.5$\pm$1.9 \\
5  & Fornax-Horologium\tablenotemark{$\dagger$}   &   40 &   54.2 & -34.5 &  235.1 & -53.7 &   20.1$\times$9.6 &     107.8$\pm$10.7 &  34.7 &   -3.9 &   1.9 &  17.3 &   3.9$\times$1.7 &  38.5$\pm$3.6 \\
6  & CMa North\tablenotemark{$\dagger$}           &   28 &  102.3 & -15.3 &  226.4 &  -7.4 &    5.5$\times$4.8 &     184.9$\pm$19.7 &  -0.2 &   -5.4 &   4.1 &  -2.3 &   1.7$\times$0.8 &  30.2$\pm$2.7 \\
7  & Aquila East\tablenotemark{$\dagger$}         &   30 &  297.8 &  -9.0 &   31.6 & -17.3 &   12.2$\times$3.8 &     136.6$\pm$6.5  &   8.5 &  -25.9 & -12.7 & -12.2 &   1.9$\times$0.8 &  20.2$\pm$1.5 \\
8  & Cepheus Far North\tablenotemark{$\dagger$}   &  219 &  329.6 &  74.7 &  111.7 &  17.8 &   16.9$\times$5.7 &     179.1$\pm$26.3 &  14.8 &    6.1 &  14.1 &  -2.9 &   2.4$\times$0.9 &  23.8$\pm$1.4\tablenotemark{d} \\
9  & Vela-CG7            &  111 &  124.9 & -43.7 &  260.8 &  -4.2 &    7.3$\times$4.6 &     323.7$\pm$9.0\tablenotemark{c}  &  -7.3 &   10.1 & -19.2 &  -0.5 &   1.2$\times$0.9 &  14.4$\pm$1.2 \\
10  & ASCC 123            &   34 &  334.0 &  53.8 &  101.3 &  -2.2 &    8.0$\times$3.5 &     251.1$\pm$24.4 &  12.0 &   -2.0 &  10.3 &  -9.8 &   2.2$\times$1.1 &  49.5$\pm$2.2 \\
11 & Cepheus-Cygnus\tablenotemark{$\dagger$}      &   36 &  311.7 &  56.6 &   93.8 &   8.4 &    5.9$\times$2.6 &     314.2$\pm$11.7\tablenotemark{c} &   5.1 &    3.6 &   8.9 &  -2.7 &   1.0$\times$0.7 &  32.6$\pm$1.2 \\
12 & Lyra\tablenotemark{$\dagger$}                &   29 &  282.7 &  32.8 &   62.7 &  14.4 &    3.5$\times$1.9 &     300.1$\pm$11.9 &   1.7 &   -3.9 &  -4.1 &  -4.4 &   1.0$\times$0.6 &  31.0$\pm$1.8 \\
13 & Cerberus\tablenotemark{$\dagger$}            &   66 &  271.4 &  22.1 &   48.3 &  19.5 &    9.3$\times$2.9 &     308.6$\pm$13.3\tablenotemark{c} &  -1.0 &   -7.3 & -10.4 &  -2.7 &   2.2$\times$0.9 &  30.0$\pm$0.8 \\
14 & Carina-Musca\tablenotemark{$\dagger$}        &  168 &  158.8 & -69.4 &  291.0 & -10.8 &   15.0$\times$3.3 &     241.8$\pm$33.4 & -25.8 &    3.1 & -27.7 & -12.5 &   1.1$\times$0.9 &  32.0$\pm$1.0\tablenotemark{d} \\
15 & Perseus             &  264 &   58.4 &  32.9 &  161.4 & -15.9 &    6.4$\times$4.1 &     294.8$\pm$22.4\tablenotemark{c} &   5.3 &   -8.1 &  13.0 &  -3.6 &   2.5$\times$1.2 &  NC  \\
16 & Perseus OB3         &  145 &   56.1 &  44.1 &  153.2 &  -7.7 &   21.2$\times$6.7 &     167.5$\pm$18.6 &  22.5 &  -26.6 &  26.7 &  -5.8 &   2.4$\times$0.9 &  NC  \\
17 & Taurus-Orion II\tablenotemark{$\dagger$}     &   64 &   89.5 &  18.3 &  190.6 &  -2.8 &   13.1$\times$4.1 &     207.4$\pm$17.2 &   7.6 &  -18.3 &  19.1 &  -2.6 &   2.5$\times$0.8 &  41.5$\pm$1.7 \\
18 & Greater Taurus      &  674 &   73.3 &  19.1 &  181.6 & -14.9 &   23.9$\times$13.5 &     142.8$\pm$24.9 &   7.6 &  -22.1 &  14.5 &  -4.7 &  5.0$\times$3.4 &   NC \\
19 & IC 2391             &  101 &  139.1 & -58.0 &  276.7 &  -8.3 &   23.2$\times$4.2 &     157.7$\pm$15.8 & -26.7 &   18.9 & -24.8 &  -4.3 &   1.8$\times$0.7 &  45.0$\pm$1.7 \\
20 & NGC 2451A           &   51 &  116.7 & -40.9 &  255.1 &  -8.0 &   10.0$\times$2.4 &     189.0$\pm$14.0 & -22.2 &   16.4 & -22.3 & -10.1 &   1.4$\times$0.9 &  48.5$\pm$1.3 \\
21 & Chamaeleon          &  259 &  188.0 & -74.3 &  301.8 & -13.0 &   14.9$\times$11.6 &     190.1$\pm$14.5 & -18.2 &   -5.0 & -17.7 &  -7.7 &   3.3$\times$1.2 &   NC \\
22 & Sco-Cen             & 7394 &  223.0 & -40.5 &  329.1 &   9.4 &   50.4$\times$20.0 &     137.8$\pm$24.3 & -18.5 &  -19.3 & -18.0 &  -6.0 &   2.6$\times$2.0 &   NC \\
23 & Taurus-Orion III\tablenotemark{$\dagger$}    &   27 &   76.4 &  11.8 &  189.4 & -17.2 &    5.0$\times$2.5 &     309.6$\pm$10.1 &   1.9 &   -3.5 &   5.9 &  -0.5 &   0.7$\times$0.6 &  15.2$\pm$2.8 \\
24 & Vela-CG4            &  551 &  108.7 & -41.3 &  252.8 & -13.9 &   21.5$\times$8.8 &     265.7$\pm$45.0\tablenotemark{c} & -11.0 &    7.5 & -14.1 &  -8.7 &   2.9$\times$0.9 &  33.7$\pm$0.9\tablenotemark{d} \\
25 & Taurus-Orion IV\tablenotemark{$\dagger$}     &  108 &   73.1 &  19.4 &  181.0 & -15.0 &   18.1$\times$8.7 &     218.8$\pm$27.4 &   1.4 &   -7.6 &   7.1 &  -3.9 &   2.8$\times$1.6 &  21.3$\pm$2.1 \\
26 & Monoceros Southwest\tablenotemark{$\dagger$} &  292 &  100.3 &  -4.6 &  215.9 &  -4.4 &    8.5$\times$4.1 &     279.0$\pm$35.2\tablenotemark{c} &  -6.6 &   -2.7 &  -0.7 &  -9.3 &   1.7$\times$0.9 &  25.5$\pm$0.6\tablenotemark{d} \\
27 & Greater Orion       &  757 &   79.7 &  -5.5 &  207.1 & -23.0 &   12.5$\times$9.1 &     299.1$\pm$34.0\tablenotemark{c} &   1.0 &   -1.3 &   2.1 &   0.4 &   1.8$\times$1.3  &  NC \\
\enddata
\tablenotetext{a}{On-sky spatial extent in galactic l/b, in the form of the RMS in major axis $\times$ minor axis, when fit with a bivariate Gaussian.}
\tablenotetext{b}{The radial extent of the velocity distribution, in the form of the RMS in semi-major axis $\times$ semi-minor axis, when fit with a bivariate Gaussian.}
\tablenotetext{c}{has a member within 1 pc of the search horizon, and therefore members are likely present beyond that limit.}
\tablenotetext{d}{Has substructure, but approximately coeval.}
\tablenotetext{e}{Older cluster - Binary sequence dominant in detection.}
\tablenotetext{\dagger}{Name is new - limited to no coverage in the literature or left unnamed in the literature that does exist.}
\vspace*{0.1in}
\end{deluxetable*}

\subsection{Greater Taurus} \label{sec:sc_tau}

\begin{deluxetable*}{cccccccccccccccc}
\tablecolumns{16}
\tablewidth{0pt}
\tabletypesize{\scriptsize}
\tablecaption{Clustering within the Greater Taurus. GT is the Greater Taurus ID, a number from EOM clustering used for identification. Leaf clusters within an EOM cluster are given a subgroup ID, which is given in the LEAF column. NC in the age column marks subclustered groups with non-coeval populations.} \label{tab:tau-sc}
\tablehead{
\colhead{GT} &
\colhead{LEAF} &
\colhead{Name} &
\colhead{N} &
\colhead{RA} &
\colhead{Dec} &
\colhead{l} &
\colhead{b} &
\colhead{$D_{sky}$\tablenotemark{a}} &
\colhead{d} &
\colhead{$\mu_{RA}$} &
\colhead{$\mu_{Dec}$} &
\colhead{$V_{T, l}$} &
\colhead{$V_{T, b}$} &
\colhead{$\sigma_{V_T}$\tablenotemark{b}} &
\colhead{Age} \\
\colhead{} &
\colhead{} &
\colhead{} &
\colhead{} &
\multicolumn{2}{c}{(deg)} &
\multicolumn{2}{c}{(deg)} &
\colhead{(deg)} &
\colhead{(pc)} &
\multicolumn{2}{c}{(mas/yr)} &
\multicolumn{2}{c}{(km/s)} &
\colhead{(km/s)} &
\colhead{(Myr)}
}
\startdata
1 &      & $\mu$ Tau  &  17 &  59.0 &   9.6 &  180.0 & -32.0 &   8.5$\times$2.8 &     153.5$\pm$6.1 &  23.7 &  -23.3 &  24.1 &  1.9 &   1.3$\times$1.0 &  52.2$\pm$3.4 \\
2 &      &            &  25 &  80.3 &  -6.2 &  208.1 & -22.9 &   6.0$\times$3.5 &     144.2$\pm$6.2 &  -1.1 &   -3.0 &   1.5 & -1.6 &   0.8$\times$0.6 &  31.0$\pm$2.5 \\
3 &      &            &  46 &  77.6 &  -2.0 &  202.6 & -23.2 &  11.1$\times$5.2 &     165.5$\pm$8.2 &  10.0 &  -14.2 &  13.6 &  1.5 &   2.4$\times$1.3 &  NC \\
3 &    A &            &  13 &  78.1 &   2.0 &  199.1 & -20.9 &   5.2$\times$1.6 &     166.9$\pm$3.6 &   8.6 &  -14.1 &  13.1 &  0.3 &   0.7$\times$0.6 &  19.3$\pm$3.9 \\
3 &    B &            &  12 &  80.8 &  -8.4 &  210.5 & -23.4 &   2.0$\times$1.2 &     157.2$\pm$2.3 &   8.3 &  -12.4 &  11.0 &  1.5 &   0.4$\times$0.4 &  13.8$\pm$2.2 \\
4 &      &            &  51 &  68.0 &  19.7 &  177.6 & -18.8 &  11.2$\times$6.6 &     121.6$\pm$7.8 &   0.7 &  -15.1 &   7.0 & -5.3 &   1.6$\times$1.0 &  NC  \\
4 &    A &            &  16 &  70.6 &  20.4 &  178.7 & -16.6 &   7.0$\times$2.9 &     119.8$\pm$4.4 &  -1.1 &  -16.1 &   6.7 & -6.2 &   0.7$\times$0.5 &  37.8$\pm$3.2 \\
4 &    B &            &  11 &  61.1 &  20.1 &  172.7 & -23.6 &   3.3$\times$2.8 &     120.5$\pm$3.3 &   4.8 &  -14.0 &   7.7 & -3.5 &   0.5$\times$0.3 &  29.8$\pm$8.3 \\
5 &      & Theia 93   &  55 &  80.7 &  24.2 &  181.3 &  -6.8 &   4.3$\times$3.1 &     177.3$\pm$7.4 &   1.8 &  -18.1 &  13.5 & -7.2 &   0.9$\times$0.6 &  8.5$\pm$1.3 \\
6 &      & 118 Tau    &  33 &  84.2 &  23.0 &  184.0 &  -4.8 &   5.8$\times$4.5 &     108.8$\pm$3.6 &   6.1 &  -37.6 &  18.1 & -7.7 &   0.8$\times$0.6 &  16.8$\pm$2.0 \\
7 &      & 32 Ori     &  11 &  81.5 &   8.3 &  195.3 & -14.7 &   3.6$\times$3.3 &     102.0$\pm$5.1 &   8.8 &  -33.2 &  15.9 & -4.6 &   0.7$\times$0.6 &  27.2$\pm$3.8 \\
8 &      & L1527\tablenotemark{c}      &  87 &  66.6 &  26.4 &  171.5 & -15.5 &   5.5$\times$1.8 &     130.7$\pm$4.5 &   7.9 &  -23.8 &  14.2 & -6.2 &   1.7$\times$0.7 &   4.5$\pm$0.9\tablenotemark{d} \\
8 &    A & L1529      &  30 &  68.3 &  24.5 &  174.0 & -15.6 &   3.1$\times$1.5 &     129.4$\pm$2.7 &   7.3 &  -21.5 &  12.9 & -5.1 &   0.7$\times$0.6 &   3.4$\pm$0.5 \\
8 &    B & B209/L1495 &  38 &  65.0 &  28.0 &  169.3 & -15.6 &   2.6$\times$0.7 &     131.2$\pm$3.3 &   8.6 &  -25.6 &  15.2 & -7.0 &   0.7$\times$0.4 &   3.8$\pm$0.8 \\
9 &      & L1544\tablenotemark{c}      &  77 &  77.1 &  29.0 &  175.4 &  -6.7 &   7.1$\times$3.7 &     159.2$\pm$5.6 &   4.4 &  -25.3 &  17.4 & -8.6 &   1.0$\times$0.6 &  NC \\
9 &    A & L1517      &  40 &  75.5 &  30.1 &  173.7 &  -7.1 &   4.1$\times$1.8 &     158.0$\pm$3.3 &   4.3 &  -25.1 &  16.9 & -8.8 &   0.7$\times$0.4 &   5.6$\pm$0.8 \\
9 &    B &            &  10 &  81.4 &  25.8 &  180.2 &  -5.4 &   1.6$\times$0.7 &     162.1$\pm$3.6 &   3.6 &  -25.5 &  17.8 & -8.5 &   0.4$\times$0.2 &  11.5$\pm$1.3 \\
10 &     & L1551      &  30 &  69.0 &  17.5 &  180.2 & -19.7 &   3.1$\times$1.8 &     145.6$\pm$3.8 &  12.3 &  -19.1 &  15.6 & -1.5 &   0.7$\times$0.4 &   7.0$\pm$2.0 \\
11 &     & B213/L1536 &  34 &  67.4 &  24.9 &  173.2 & -16.0 &   5.3$\times$1.1 &     159.2$\pm$4.2 &  10.7 &  -17.4 &  15.2 & -2.5 &   0.8$\times$0.7 &   3.3$\pm$0.9 \\
\enddata
\tablenotetext{a}{On-sky spatial extent in galactic l/b, in the form of the RMS in major axis $\times$ minor axis, when fit with a bivariate Gaussian.}
\tablenotetext{b}{The radial extent of the velocity distribution, in the form of the RMS in semi-major axis $\times$ semi-minor axis, when fit with a bivariate Gaussian.}
\tablenotetext{c}{This named structure is not included in any HDBSCAN-defined leaf subclusters. Names associated with subgroups are also contained in the higher-level EOM group.}
\tablenotetext{d}{Has substructure, but approximately coeval.}
\end{deluxetable*}

\begin{figure}
\centering
\includegraphics[width=7.5cm]{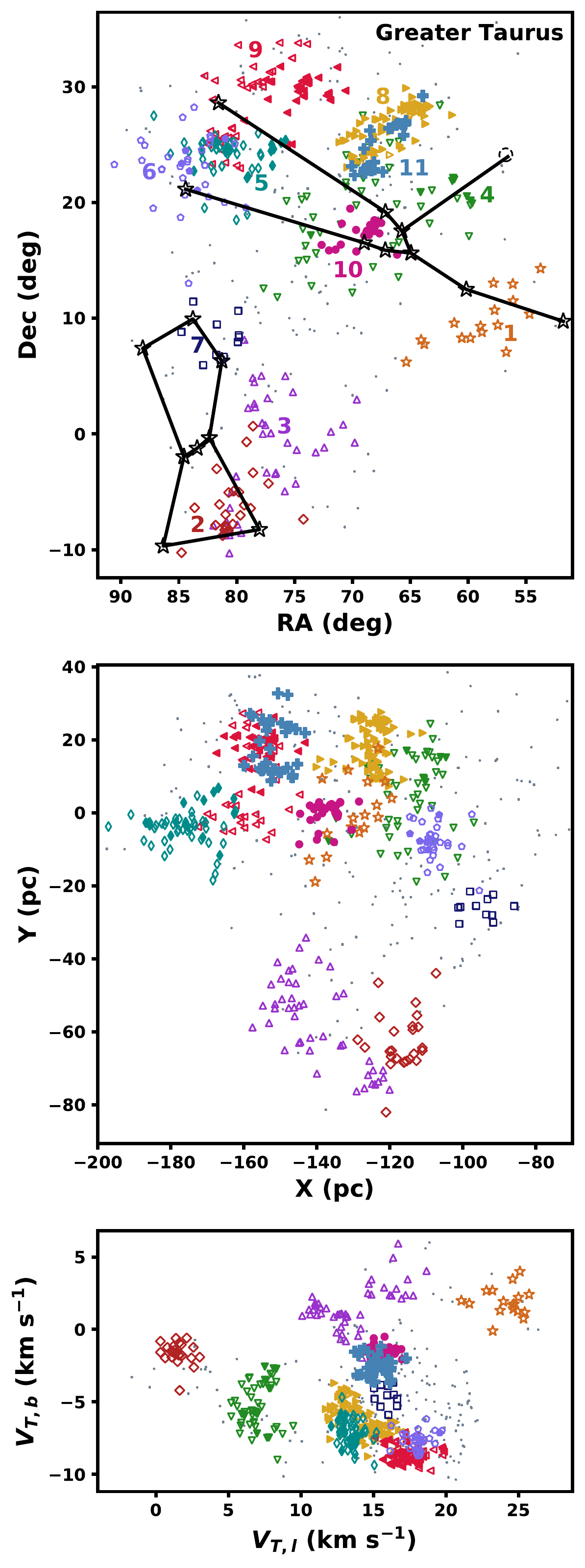}\hfill
\caption{Members of the EOM groups that HDBSCAN identifies in Greater Taurus in RA/Dec, galactic X/Y, and transverse velocity coordinates. \edit1{Filled} icons indicate known Taurus members and \edit1{empty} icons indicate objects not previously recognized as part of Taurus. Small grey dots mark unclustered Taurus members. Each group is labelled according to its Greater Taurus ID from Table \ref{tab:tau-sc} in the top panel. Our identification of Greater Taurus includes significant populations beyond the known membership. The main stars of the Taurus and Orion constellations are included for reference in the top panel, with the dotted circle attached to Taurus representing the Pleiades.}
\label{fig:TMC}
\end{figure}

\begin{figure}[h]
\includegraphics[width=8.5cm]{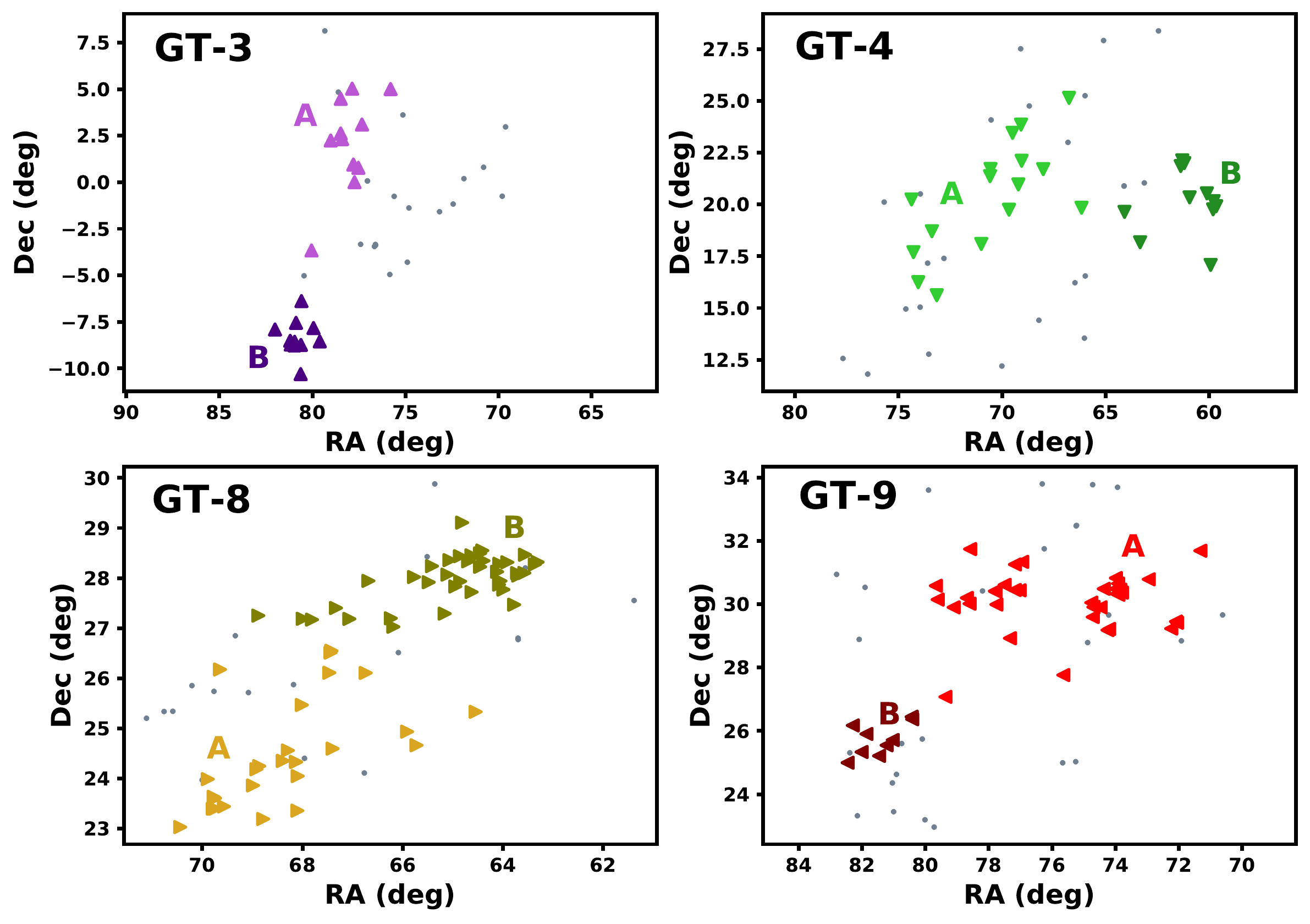}\hfill
\caption{Subclustering in the four Greater Taurus EOM clusters where it is present, displayed in RA/Dec sky coordinates. In all cases, unclustered members of the parent group are marked as small grey dots, and the subgroups are given as squares and diamonds. The ID of the parent EOM cluster is indicated on each panel.}
\label{fig:Tau_SC}
\end{figure}

The general direction towards the Taurus constellation is known to contain multiple nearby stellar populations. The largest of these is associated with sites of active star formation in the Taurus Molecular Cloud (TMC), which form an extensive network of young stars often referred to as the Taurus-Auriga Association or simply ``Taurus", as we refer to it here \citep[e.g., see][]{Kenyon08}. This complex represents one of the nearest and therefore most accessible sites of active star formation to the Sun ($\sim$140 pc, \citealp{Galli18}), and as such it has been subject to many recent studies to expand the known stellar populations, learn more about the structure and history of the region, and study star formation more broadly \edit2{\citep[][]{Luhman18,Galli19,Liu21,Krolikowski21}}. Through some of that work, kinematic substructure has recently been identified in the region, dividing the stellar populations into four different subgroups, each tied to a different set of clouds within the TMC \citep{Luhman18}. The lists of known and proposed Taurus members continue to expand, although the existence of older members beyond the well-established young ($<$10 Myr) populations remains controversial \citep{Kraus17,Luhman18}. Aside from the Taurus populations associated with the TMC, there are also many smaller and less-studied groups in the area, including the 118 Tauri, 32 Orionis, and $\mu$ Tauri Associations \citep{Mamajek16,Mamajek07,Gagne20}. Since all of these groups have only recently been fully characterized, any relations these associations might have to one another and the other major stellar populations in Taurus remain unsettled.

Using HDBSCAN, we identify an extensive contiguous network of young stars in the direction of Taurus, forming a top-level group that we refer to as ``Greater Taurus". Significant substructure is found within the region, breaking it into a total of 11 EOM groups which are displayed in Figure \ref{fig:TMC}. The velocity dispersions of Taurus subgroups are typically of order 1 km s$^{-1}$ or smaller, with a larger dispersion for the group as a whole, at about 4 km s$^{-1}$. However, most groups closely associated with the TMC are in a somewhat tighter distribution, with even the most extreme exemplars differing in transverse velocity by less than 10 km $^{-1}$. Outlying groups typically differ from the core TMC groups in transverse velocities by 5-15 km s$^{-1}$, however their distributions remain connected to each other and to the core Taurus groups by lower density distributions of stars in the intervening parameter space, hence HDBSCAN merges them into a single large-scale Greater Taurus group (see Fig. \ref{fig:TMC}). Since the area Taurus covers is quite large, projection effects account for a significant fraction of the velocity changes. Assuming a constant velocity equal to the Taurus UVW vector from \citet{Luhman09} would generate a 3 km s$^{-1}$ velocity dispersion from projection alone. There are, however, subgroups with notably different velocities compared to the projected Taurus UVW vector, so internal velocity spreads do exist. Known members in Taurus are marked using \edit1{filled} icons in Figure \ref{fig:TMC} as a way to compare our sample to the currently known extent of the region \citep{Kraus17,Luhman18,Esplin19, Galli19,Krolikowski21}. The subgroups we identify as part of Greater Taurus extend far beyond these known Taurus members. Higher-order subclusters are also found at the leaf level, with 4 of 11 EOM groups dividing further into 2 leaf groups each. Close-up views of the groups containing leaf subclusters are provided in Figure \ref{fig:Tau_SC}. 

Four of the groups that HDBSCAN identifies appear to be associated with the ongoing and recent star formation in the TMC \citep[e.g.,][]{Rebull11, Luhman18}. These groups contain the majority of currently known Taurus Association members (see Figure \ref{fig:Tau_SC}), and they all correspond to known populations of Young Stellar Objects (YSOs) in and around active dense clouds in the TMC. GT-8 contains YSOs associated with L1495 B209, L1527, and L1529; GT-9 contains L1517 and L1544; GT-10 contains L1551; and GT-11 contains B213 and L1536 \citep{Luhman18,Onishi02}. These four groups are similar to those proposed in \citet{Luhman18}, with the main differences being our separation between GT-10 and GT-11, which \citet{Luhman18} groups together into their ``blue" cluster, and our exclusion of their ``green" cluster as a distinct subgroup due to its very small membership. All four of our core TMC groups consist almost entirely of known Taurus member stars, with the notable exception of a southeastern extension to GT-9, which forms its own leaf cluster, GT-9B (recognized as Theia 68 in \citealt{Kounkel19}). The other GT-9 subgroup, GT-9A, consists of the known YSOs in the region around L1517. GT-8 also fragments into leaf clusters, with GT-8A consisting mainly of YSOs associated with L1529, and GT-8B consisting of B209 and L1495. All of these core TMC groups (aside from GT-9B) have derived ages between 3 and 7 Myr, consistent with our knowledge that active star formation is happening in these regions.

Distributed populations around Taurus have long been proposed, particularly in the direction of GT-1 \citep[e.g.,][]{Neuhaeuser95, Wichmann96,Magazzu97}, however these early papers have reached mixed results on potential association with the TMC \citep[e.g.,][]{Briceno97}. We also identify distributed populations in our work, forming an extensive network of peripheral groups that we associate with Greater Taurus, many of which are significantly older than the core TMC groups. Some of these groups, including GT-5, GT-6, and both subgroups of GT-4 contain proposed Taurus members, but as indicated in Figure \ref{fig:TMC}, their coverage in surveys of Taurus to date is far from complete \citep{Krolikowski21}. \edit2{The extended Taurus population and clustering given in \citet{Liu21} has however recently broadened coverage of populations around these groups, such as in GT-4, which loosely corresponds to \citet{Liu21} groups 12-14.} The youngest of these groups is GT-5 \citep[which has considerable shared membership with Theia 93 from][]{Kounkel19}, a subgroup with TMC-consistent kinematics that appears to represent a slight continuation of Taurus to farther distances. Others have ties to the TMC that are considerably less clear. GT-6 has proposed Taurus members in it, however it was previously identified independently as the 118 Tauri Moving Group by \citet{Mamajek16}, which has membership falling entirely within GT-6 and kinematically consistent objects. This group represents one of three we identify that have previously been identified independently from the core populations in Taurus, with the other two being GT-1 and GT-7. In GT-7, over half of our membership overlaps with that of the 32 Orionis Association \citep{Mamajek07,Bell17}, a somewhat older group where our derived age of 27.2 $\pm$ 3.8 Myr is in agreement with the 22$\pm$4 Myr isochronal age solution from \citet{Bell15}. The remainder of our membership matches spatially and kinematically with known 32 Orionis members, and while we only identify 11 members of this group, its established membership from \citet{Bell17} contains only 47 objects, making the sample we gather smaller, but consistent with recovery rates for groups in this age range (see Section \ref{sec:recovery}).

Our member stars for GT-1 match with those of the much older $\mu$ Tauri Association in both spatial coordinates and kinematics, where our derived age of $\sim$52.2$\pm$3.4 Myr is marginally within uncertainties of the 62$\pm$7 Myr solution derived by \citet{Gagne20}. The \citet{Gagne20} candidate member list contains all of our GT-1 members. With only 17 GT-1 members compared to $\sim$500 $\mu$ Tau candidates from \citet{Gagne20}, our detection of this group is limited, although this low recovery rate is likely primarily due to the very old age of the region (see Section \ref{sec:recovery}). A few GT-3 members in its far western extension are also included in the \citet{Gagne20} catalog, an overlap indicative of a possible link between GT-1 ($\mu$ Tauri) and GT-3. The cores of these two groups are separated by $\sim$60 pc in space, $\sim$10 km s$^{-1}$ in transverse velocity, and $>$30 Myr in age, and therefore their affiliation with one another is far from certain. However, the separation of these subgroups in velocity is consistent with projection effects, and a low-density, near-contiguous linear structure in the plane of the sky does appear to connect them, lending plausibility to the connection between these subgroups.

Aside from this slight overlap with $\mu$ Tau, the remainder of the stars in GT-3 and GT-2 do not match with any current Taurus catalogs related to the TMC, as expected given that these groups stretch up to 25 degrees south of the core TMC groups to the crowded fields in front of the Orion Association. GT-2 (along with GT-4) is grouped under \citet{Kounkel19} group Theia 44, but these groups are otherwise absent in literature. GT-3 divides into two subgroups: GT-3A to the north, and GT-3B to the south, with ages of $\sim$19 and 14 Myr, respectively. GT-3B is projected directly behind the somewhat older ($\sim$21 Myr) GT-2 group in the plane of the sky, however despite this close physical proximity between GT-3B and GT-2, they are separated by more than 10 km s$^{-1}$ in transverse velocity space, making close association in formation origin unlikely. GT-3, together with the GT-4 and GT-1, represent visible outliers from the relatively compact transverse velocity distribution occupied by the core TMC groups, which is less than 10 km s$^{-1}$ across. While GT-2 also qualifies as an outlier to the TMC in transverse velocities, it is actually not inconsistent with a common origin there, as its velocities are consistent with the projection of the Taurus UVW velocity vector at its location \citep{Luhman09}. GT-4 is unique in that it has an outlying kinematic signature despite it physically overlapping with GT-8 and GT-10. 

These distributed extensions to the Greater Taurus association demonstrate that an older population is present in the region. The youngest group we find outside of the well-characterized TMC regions is GT-5 at $\sim$8.5 Myr old, while all others have ages in excess of 10 Myr. These older regions range from the subgroups of GT-3, at $\sim$14-19 Myr, to $\mu$ Tau (GT-1), at $\sim$52 Myr old. While their inclusion in Greater Taurus is not necessarily certain for some of the peripheral subgroups (see Section \ref{sec:dis-subgroups}), the case of GT-9 is quite notable as an older, distributed population that is directly connected to the known elements of the Taurus association. This group consists of two leaf subgroups, GT-9A and GT-9B, which are contiguous with one another in spatial coordinates and have overlapping velocity distributions.  Subgroup GT-9A has a near-newborn age solution and contains known Taurus YSOs around L1517 \citep[e.g.,][]{Luhman18}, while GT-9B contains a dispersed population with an older age solution at $\sim$11.5 Myr old, forming a clear link between an older (age $>$10 Myr), distributed group and the current star formation events in Taurus. GT-6, GT-4A, and GT-4B also all contain members with known signs of youth and kinematics previously linked to Taurus \citep{Kraus17,Krolikowski21}, and with ages of $\sim$17, 38, and 30 Myr, respectively, these groups further enforce Taurus's connection to an older population. 

Our work supports recent studies that have also demonstrated the presence of this older distributed population, including \citet{Krolikowski21}, \citet{Kraus17}\edit2{, \citet{Zhang18}, and \citet{Liu21}}. Not all current literature promotes this conclusion, however, such as \citet{Luhman18}, which concludes that the proposed older members with age $>$10 Myr are likely not true Taurus members due to differing velocity distributions. Some of our proposed members have kinematics that make them potentially subject to this argument, likely requiring a full kinematic analysis and traceback as discussed in Section \ref{sec:dis-subgroups} to achieve a definitive conclusion on their association with the current star formation in the TMC. Other old subgroups, however, have clear kinematic overlap with known Taurus members near the TMC, and those with more discrepant velocities are linked by previously unknown low-density stellar distributions to the known Taurus groups, strengthening those connections. The presence of 9 groups and subgroups in the 10-40 Myr age range, as well as a much older group in $\mu$ Tauri with an age of $\sim$52 Myr, demonstrates that not only is there an older stellar population surrounding the currently known, essentially newborn TMC-assiocated groups, but that Greater Taurus may represent the oldest association in the solar neighborhood that still hosts active star formation.

\subsection{Greater Orion} \label{sec:ori}

\begin{deluxetable*}{cccccccccccccccc}
\tablecolumns{16}
\tablewidth{0pt}
\tabletypesize{\scriptsize}
\tablecaption{Clustering within Greater Orion. ORI is the Greater Orion ID, a number from EOM clustering used for identification. Leaf clusters within an EOM cluster are given a subcluster ID, which is given in the LEAF column.} \label{tab:ori-sc}
\tablehead{
\colhead{ORI} &
\colhead{LEAF} &
\colhead{Name} &
\colhead{N} &
\colhead{RA} &
\colhead{Dec} &
\colhead{l} &
\colhead{b} &
\colhead{$D_{sky}$\tablenotemark{a}} &
\colhead{d} &
\colhead{$\mu_{RA}$} &
\colhead{$\mu_{Dec}$} &
\colhead{$V_{T, l}$} &
\colhead{$V_{T, b}$} &
\colhead{$\sigma_{V_T}$\tablenotemark{b}} &
\colhead{Age} \\
\colhead{} &
\colhead{} &
\colhead{} &
\colhead{} &
\multicolumn{2}{c}{(deg)} &
\multicolumn{2}{c}{(deg)} &
\colhead{(deg)} &
\colhead{(pc)} &
\multicolumn{2}{c}{(mas/yr)} &
\multicolumn{2}{c}{(km/s)} &
\colhead{(km/s)} &
\colhead{(Myr)}
}
\startdata
1 &      &  &458 &  80.6 &  -3.6 &  205.8 & -21.4 &   8.3$\times$3.6 &     316.8$\pm$15.7\tablenotemark{c} &   1.1 &   -0.9 &  1.9 &  0.8 &   1.3$\times$1.0 &  13.3$\pm$0.3\tablenotemark{d} \\
1 &    A &  & 18 &  78.7 &  -7.1 &  208.2 & -24.7 &   2.1$\times$0.8 &     288.6$\pm$7.4  &   1.7 &   -3.0 &  4.7 &  0.2 &   0.6$\times$0.3 &  14.0$\pm$1.2 \\
1 &    B &  & 10 &  80.9 &  -1.5 &  203.9 & -20.1 &   1.1$\times$1.0 &     327.7$\pm$1.7\tablenotemark{c}  &   1.1 &   -0.5 &  1.5 &  1.2 &   0.5$\times$0.3 &  12.3$\pm$1.6 \\
1 &    C & 25 Ori  & 22 &  81.5 &   1.6 &  201.3 & -18.2 &   2.5$\times$0.4 &     328.4$\pm$2.4\tablenotemark{c}  &   1.5 &   -0.2 &  1.4 &  1.9 &   0.6$\times$0.3 &  13.2$\pm$0.8 \\
2 &      & Eridanus North\tablenotemark{$\dagger$} & 82 &  67.8 & -10.2 &  205.9 & -35.7 &   8.8$\times$5.1 &     227.8$\pm$18.2 &   2.3 &   -3.8 &  4.8 &  0.4 &   0.9$\times$0.6 &  20.0$\pm$1.8 \\
3 &      & Orion Southeast\tablenotemark{$\dagger$} & 81 &  87.5 &  -8.4 &  213.6 & -17.4 &   4.0$\times$2.6 &     294.1$\pm$8.5  &  -0.0 &   -0.6 &  0.7 & -0.4 &   1.1$\times$0.6 &  19.0$\pm$0.9 \\
\enddata
\tablenotetext{a}{On-sky spatial extent in galactic l/b, in the form of the RMS in major axis $\times$ minor axis, when fit with a bivariate Gaussian.}
\tablenotetext{b}{The radial extent of the velocity distribution, in the form of the RMS in semi-major axis $\times$ semi-minor axis, when fit with a bivariate Gaussian.}
\tablenotetext{c}{has a member within 1 pc of the search horizon, and therefore members are likely present beyond that limit.}
\tablenotetext{d}{Substructure present, but mostly coeval.}
\tablenotetext{\dagger}{name is newly assigned by this paper.}
\vspace*{0.1in}
\end{deluxetable*}

\begin{figure}
\centering
\includegraphics[width=8cm]{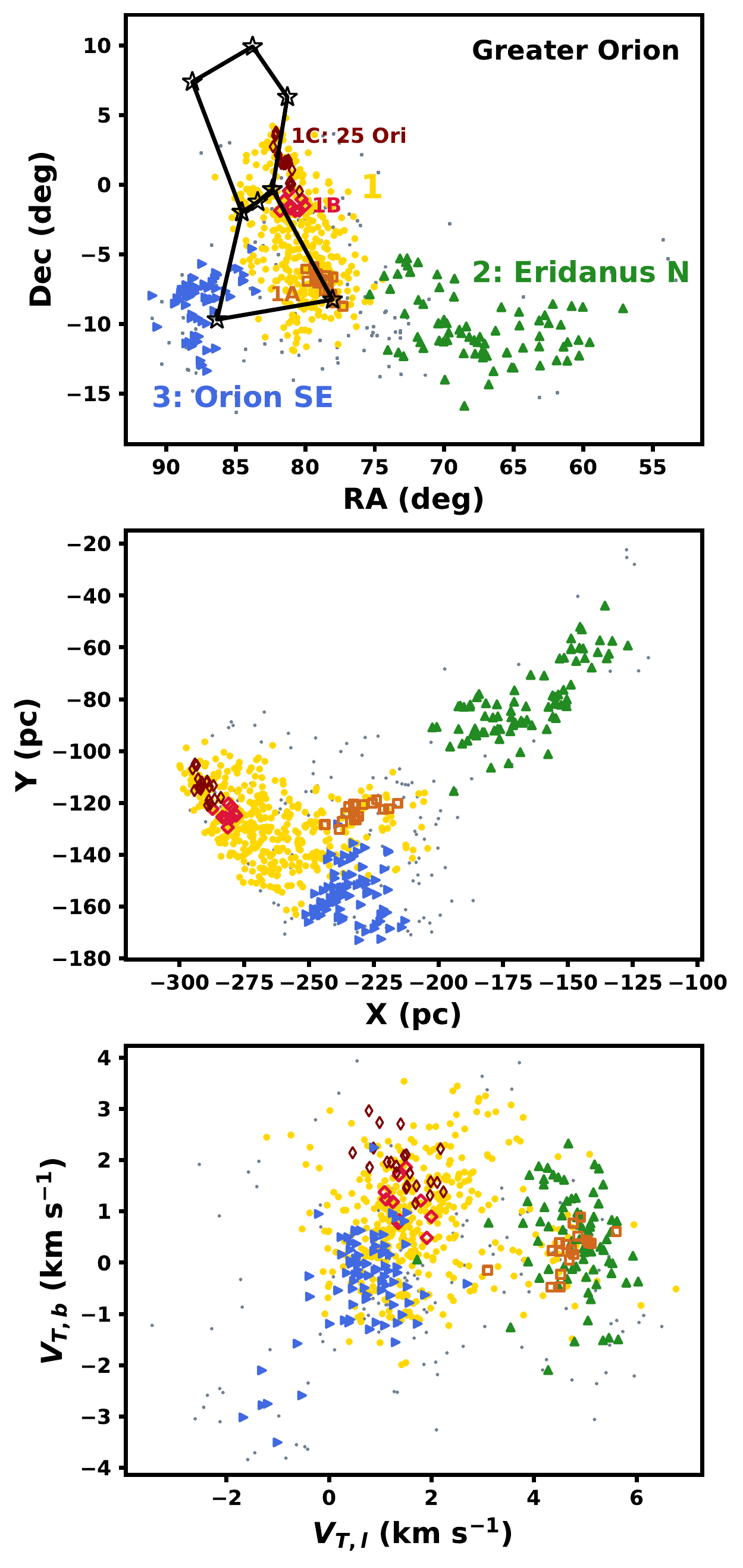}\hfill
\caption{HDBSCAN-identified groups and subgroups in Orion, labelled according to their Orion ID, and shown in RA/Dec, galactic X/Y, and transverse velocity coordinates. The \edit1{empty} diamonds and squares correspond to leaf subclusters of the larger ORI-1 group (yellow circles), while \edit1{filled} icons are top-level EOM clusters. Small grey dots represent unclustered Greater Orion members. The main stars of the Orion constellation are included for reference.}
\label{fig:Orion}
\end{figure}

The Orion OB1 association contains an enormous population of stars in terms of both numbers and extent, while also hosting the Orion Nebula Complex, the nearest site of active high-mass star formation to the Sun \citep{Hillenbrand00,Bally08-ori}. Substructure has long been recognized in the region, with \citet{Blaauw64} dividing the region into four distinct substructures based on location and photometry, labelled Orion OB1a, b, c, and d. All of these Orion OB1 subgroups are centered beyond this project's horizon at 333 pc, however both Orion OB1a and OB1b have components that are nearer \citep{Brown94,Briceno19}. The known populations accessible through this project therefore consist of Orion OB1b, a young ($\sim$5 Myr) population immediately surrounding Orion's Belt, and Orion OB1a, a somewhat older ($\sim$11 Myr) population containing groups to the northwest, such as the 25 Orionis and HD 35762 clusters \citep{Bally08-ori, Briceno19}. All recognized subgroups of Orion OB1a  have significant components within our search horizon, however Orion OB1b has a notably bimodal parallax distribution, resulting in only the nearest component being visible, a subgroup given the name Orion OB1b I by \citet{Briceno19}.

Like in Taurus, we also identify a distributed population beyond the known subgroups in Orion, albeit with much less variety in age. The high-level region we identify as Greater Orion hosts three large HDBSCAN-identified groups, with the largest (ORI-1) subdividing into three small subgroups. Their extents in spatial and velocity coordinates are shown in Figure \ref{fig:Orion}. The mean transverse velocities in these groups never differ from one another by more than 4 km s$^{-1}$, and all groups have tight velocity dispersions of order 1 km s$^{-1}$ or below. Particularly on the far edge of ORI-1, these velocity distributions are found to be nearly identical to those of known young associations near to and beyond our search radius cutoff, such as 25 Orionis and the related subgroups Zari-B$_0$ and Zari-B$_6$ defined by \citet{Zari19}. Some of the more distant members of ORI-1 we identify, including those in subgroups ORI-1B and ORI-1C, overlap with the near edge of Orion OB1a \citep{Briceno19}, with ORI-1C members overlapping directly with known members of the 25 Orionis and HD 35762 subgroups in spatial coordinates. Other sections of ORI-1 overlap with the membership of Orion OB1b I, however most of this group is beyond our search radius limit, leaving the remaining members too sparsely distributed for HDBSCAN to identify a subgroup. Both ORI-1B and ORI-1C, along with the rest of the far edge of ORI-1's main body also appear cut off by our search horizon, as expected given the distance distributions of known Orion subgroups \citep[e.g.,][]{Briceno19}. This indicates that our coverage of known associations in Orion is consistently incomplete. 

The lack of overlap between our extended population in Orion and known Orion subgroups is demonstrated by the DBSCAN-based subclustering analysis of Orion conducted in \citet{Zari19}, which investigated the full range of distances that Orion occupies. The only overlap between our members and those identified with a subgroup by \citet{Zari19} is at the far edge of our ORI-1 group (including ORI-1C), which overlap with the membership for Zari-B in 3-d positions and proper motions. Zari-B$_6$ in particular has a distribution in position/velocity parameter space that overlaps fully with the membership of our subgroup ORI-1C, while the near edge of Zari-B$_{0}$ also overlaps with part of ORI-1's far edge \citep{Zari19}. However, most of Zari-B, as well as the rest of the groups they identify are well beyond our search radius\edit1{, and despite their search window extending to within 200 pc of the sun, all populations identified by \citet{Zari19} are almost entirely located beyond 300 pc}. 

The more nearby populations accessible through this research therefore largely consist of stars well outside of \edit1{well-}known groups, residing in lower-density extensions that methods like the DBSCAN clustering implementation used in \citet{Zari19} are not able to recover due to the inflexibility of DBSCAN in simultaneously handling groups with very different densities and scales. Many of the stars in these extensions are linked directly to the known Orion OB1a and OB1b I populations through membership in ORI-1, forming significant extensions to the region that stretch over 50 pc away from those known sections in the approximate direction of the sun, and towards Rigel in the plane of the sky. \edit1{Some of these nearer populations in ORI-1 as well as ORI-3 appear as part of the Orion D group in \citet{Kounkel18}, however these populations remain obscure in literature, and those in ORI-2 remain completely unknown.} As \edit1{ORI-2 and ORI-3 are both separated from known populations and clearly defined,} we assign them new names: Eridanus North for ORI-2, and Orion Southeast for ORI-3.

Eridanus North forms a major low-density extension to Greater Orion, spanning from the near edge of ORI-1 to within $\sim$200 pc of the Sun, with some members residing more than 150 pc from the known membership in Orion OB1 \citep{Zari19,Briceno19}. While there is a subtle separation between the mean transverse velocities in Eridanus North and the majority of ORI-1 of approximately $\sim$3 km s$^{-1}$, a small subset of ORI-1 stars at the near edge have transverse velocities that overlap with those in Eridanus North. Stars in ORI-1 that have these unique motions form the ORI-1A subgroup, which has spatial positioning consistent with ORI-1, but transverse motions more similar to Eridanus North. As such, ORI-1A appears to represent a bridge between the main Orion Association and the more dispersed extensions into Eridanus North. The stars in Eridanus North appear notably older compared to the known populations in Orion, with an age estimated at $\sim$20 Myr. 

The other new major group we find in the Orion Complex, Orion Southeast is identified as UPK 422 in \citet{Sim19} \edit1{and grouped under Orion D in \citet{Kounkel18}}, but is otherwise undiscussed in the literature. The group has limited separation from from the main body of ORI-1 in transverse velocities but much more visible separation in spatial coordinates. This separation is most evident in sky coordinates, where the group appears to the Southeast of Orion's belt, with a significant central concentration of stars near the border with Monoceros (see Fig. \ref{fig:Orion}). Like in Eridanus North, we find relatively old ages for stars in this group, with an age estimate of $\sim$19 Myr.

Despite the large extensions we find to known populations, ORI-1 is found to have very consistent ages across its three subgroups, all of which are at $\sim$12-14 Myr and within uncertainties of each other. Consequently, ORI-1 appears to be mainly coeval, and no significant hints of age gradients are detectable in the region. One minor exception is a small unclustered group of stars near Orion's belt associated with Orion OB1b I, where the members have individual age estimates consistent with the 5 Myr photometric age estimate for the group from \citet{Briceno19}. Our $\sim$13 Myr age solution for ORI-1C is \edit1{approximately} consistent with \edit1{age estimates given for the HD 35762 and 25 Orionis subgroups, which vary from $\sim$8 Myr in \citet{Briceno19} to 11 and 13 Myr respectively for the similarly-defined ASCC 21 and ASCC 16 subgroups \citep[][]{Kos19}. The 11 Myr age for the broader extended populations in Orion OB1a given by \citet{Briceno19} is also roughly consistent with our bulk age solution for ORI-1, at $\sim$13.3 Myr, as is the 12.3 Myr estimate from \citet{Siess00}.}

\subsection{Perseus} \label{sec:perseus}

\begin{figure}[t]
\centering
\includegraphics[width=8cm]{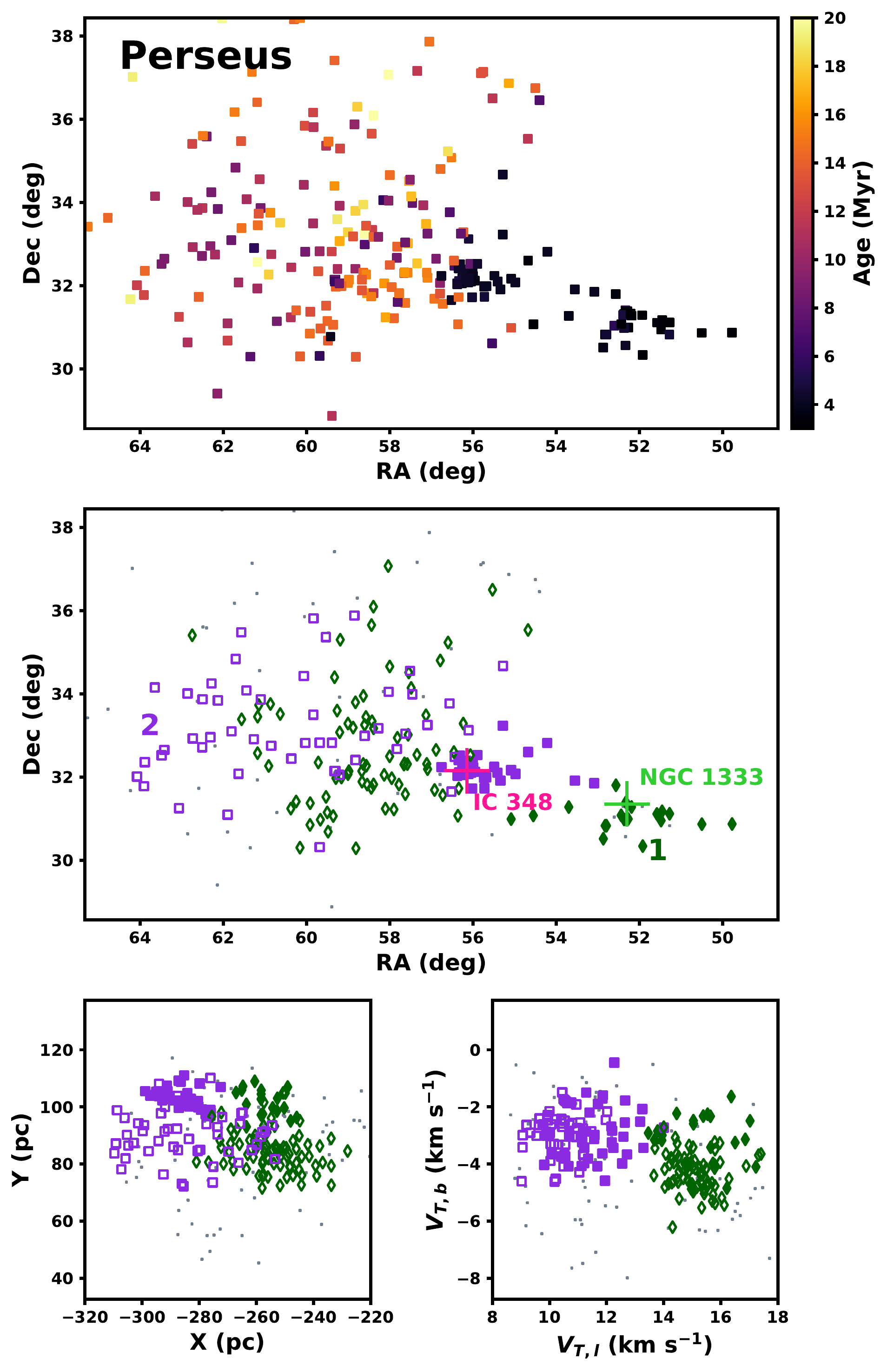}\hfill
\caption{The age distribution and subclustering in Perseus. Ages in RA/Dec space are shown in the top panel, displaying the clear divide between the young populations in the west near the Perseus Molecular Could and older populations to the east. The remaining panels show subclustering in RA/Dec, galactic X/Y, and galactic transverse velocity. The two subclusters are labelled in the middle panel, and objects in the older eastern extensions are marked with \edit1{empty} icons to distinguish them from the young western populations. The centres of IC 348 and NGC 1333 are also marked for reference \citep{OrtizLeon18}. }
\label{fig:perseus}
\end{figure}

Perseus OB2, which we abbreviate to ``Perseus" in this paper, is a young association originally identified as a $\sim$6 degree-wide overabundance of O and B stars near the open cluster IC 348 \citep{Blaauw52}. While IC 348 and the neighboring cluster NGC 1333 have long been suggested as concentrations within the Perseus region \citep[e.g.][]{Blaauw52, Strom74}, the populations in these clusters have since been significantly expanded, revealing an extensive network of recent and continuing star formation centered at a distance of $\sim$300 pc \citep[e.g.,][]{Cernis93, deZeeuw99, Bally08}. The stellar populations in these clusters are contained within the Perseus Molecular Cloud, an extensive region containing numerous sites of active star formation, as evidenced by the presence of dense cores \citep[e.g.][]{Sargent79,Bally08, Rosolowsky08, Kerr19}. As such, most of the young stars we see in the region can be tied to previous formation in that cloud \citep{Bally08}. The presence of a more distributed population reflective of the broader spatial distribution suggested by early works is still accepted in Perseus, having recently been the subject of a substructure analysis from \citet{Pavlidou21} and new WISE observations from \citet{Azimlu15}. However, most populations currently known in Perseus are thought to be related to earlier generations of formation in the current Perseus cloud \citep{Bally08}. The affiliation of these populations to the current star formation activity is reflected in the age estimates, which are consistently at or below 6 Myr \citep{deZeeuw99,Bally08,Pavlidou21}, albeit with a few SED fits from \citet{Azimlu15} suggesting ages closer to 10 Myr. Astrometric measurements have recently established a kinematic division between stars associated with IC 348 and NGC 1333, with the two populations separated by about 5 km s$^{-1}$ in velocity space \citep{OrtizLeon18}

Using HDBSCAN, we identify two spatially overlapping subregions in Perseus, lying along nearly the same line of sight but at different average distances. PER-1 is the nearer of the two at 283 pc, while PER-2 has a mean distance of 314 pc. Both of these subgroups in some capacity relate to the known clusters in Perseus, with PER-1 containing NGC 1333, and PER-2 containing IC 348 \citep{OrtizLeon18}. Both Perseus subgroups, however, contain significant structure beyond the two known young clusters, extending to the east of each cluster's centre. These extensions are fully contiguous with the known elements of each subcluster, and therefore they do not form distinct leaf clusters within PER-1 or PER-2, however the individual age estimates there are distinctly older than those in the known young clusters. The ages in these Perseus subgroups do not appear to form any gradient, instead forming two distinct epochs of star formation in each subgroup, one for the current generation of star formation, and another one in the eastern extensions preceding it. To separately investigate each of these generations, we use simple cuts in the plane of the sky to separate the eastern extensions from the western young clusters. We display the Perseus subgroups in spatial and velocity coordinates in Figure \ref{fig:perseus}, marking the eastern extensions with \edit1{empty} icons. For the sake of identification, we refer to the area around NGC 1333 and IC 348 as PER-1A and PER-2A, respectively, while the corresponding eastern extensions for each are referred to as PER-1B and PER-2B.

PER-1A and PER-2A both have properties consistent with expectations for groups anchored in NGC 1333 and IC 348, respectively. They separate cleanly from one another in velocity space, reflecting the kinematic divisions identified by \citet{OrtizLeon18} between stars associated with each of the young clusters. Despite these kinematic differences, all of the stars we include in these sections are spatially consistent with the known segments of the Perseus Molecular cloud, with PER-2 being limited almost entirely to IC 348 and its immediate surroundings, while PER-1 is considerably less tethered to NGC 1333 alone, as it includes some members more spatially consistent with adjacent clouds such as B1 and L1455, and also excludes some apparent NGC 1333 members from subclustering due to their location at the outskirts of the broader Perseus group. The inclusion of stars from clouds outside NGC 1333 in PER-1 is consistent with the results from \citet{OrtizLeon18}, where stars associated with B1 and L1455 are found to have kinematics consistent with NGC 1333. Both of the young subgroups we find in Perseus have ages that reflect their essentially newborn status at 4-6 Myr, results consistent with current age estimates in the region \citep[e.g.,][]{Bally08}. Due to these very young ages, many of the stars in Perseus are embedded \citep[e.g.,][]{Bally08}, and as such we see a significant downselect of members in these young clumps due to Gaia quality cuts, as was demonstrated in Section \ref{sec:recovery}. A cross-match between Gaia sources and members identified in \citet{Luhman16} by \citet{OrtizLeon18} found 90 and 351 members in those clusters, respectively, significantly more than the 19 and 43 that we find. As such, while both eastern extensions contain populations larger than we identify in their respective young clusters, upon corrections for recovery rates both generations appear to be comparable in size. 
 
The old eastern extensions to Perseus are much more dispersed compared to the more concentrated clusters that we identify near the Perseus Molecular cloud, with both spanning approximately 7$^{\circ}$ in the plane of the sky, or $\sim$35 pc. Both PER-1B and PER-2B have very similar velocities to their younger counterparts, with standard deviations of order 1 km s$^{-1}$, and differences in velocity between old and young components not exceeding 1.5 km$^{-1}$. Due to these kinematic similarities between the young and old populations, PER-1B and PER-2B have clearly separated velocities that reflect the kinematic divisions between their corresponding younger clusters. Despite the kinematic differences seen in each subgroup, PER-1B and PER-2B have remarkably similar age solutions to each other, both at $\sim$17 Myr. The older ages of these eastern extensions may also be reflected in the broad spatial distributions they occupy, as dispersal is expected as time advances \citep[e.g.,][]{Brown97}. The near-identical ages of these two populations, combined with the kinematic differences between them suggests that star formation in those subgroups has progressed essentially in parallel, suggesting that they were assembled by the same initial processes despite their kinematic differences.

The kinematic similarities between the current star-forming regions and their older counterparts suggests that the events that gave rise to the older generations in PER-1 and PER-2 likely originate in the same star-forming processes that led to the current star formation in NGC 1333 and IC 348, respectively. However, due to the significant time lag between the older and younger generations in each subgroup, continuous formation within a static cloud seems unlikely. We present one possible explanation for this lag in Section \ref{sec:dis-sfb}, in which first generation of stars disperses its parent cloud while still being fed by a flow of external material, resulting in another burst of star formation.

\subsection{Sco-Cen Association} \label{sec:sc}

The Sco-Cen Association (also known as Sco-OB2) is an extensive young association covering much of the southern Milky Way, and represents the nearest site of recent high-mass star formation \citep{deZeeuw99}. The association was first identified as an expansive $40\degr \times 70\degr$ overdensity of comoving O and B stars by \citet{Kapteyn14}, with work in the following decades significantly deepening the known membership within this original footprint \citep[e.g.,][]{GutierrezMoreno68}. The first hints of substructure were proposed by \citet{Blaauw64}, which divided the association into three large subregions that are still frequently referenced in literature: Upper Sco (US), Upper Centaurus-Lupus (UCL) and Lower Centaurus-Crux (LCC). The past age estimates vary from between 5 and 11 Myr for Upper Sco \citep[e.g,][]{PriebischZinnecker99, Pecaut12}, to a little under 20 Myr for UCL and LCC \citep[e.g.,][]{Mamajek02}. Recent, more complete surveys of Sco-Cen, especially post-Gaia, have identified thousands of new Sco-Cen candidate members, with current membership estimates of around 10000 \citep[e.g.,][]{Pecaut16}. As the known populations in the region became increasingly complete, \citet{Rizzuto11} noted the presence of a broad population linking all three of the traditional subgroups, suggesting that these traditional divisions are insufficient to capture the interconnected and continuous nature of star formation in the Sco-Cen association. The Ophiuchus and Lupus clouds are now commonly recognized as part of Sco-Cen, and recent work has proposed the inclusion of many other adjacent groups, including Corona Australis (CrA), the Chamaeleon clusters, the TW Hydrae association, and IC 2602 \citep[e.g.,][]{Mamajek99,Mamajek01,Damiani19}

\begin{figure}[h!]
\centering
\includegraphics[width=8.3cm]{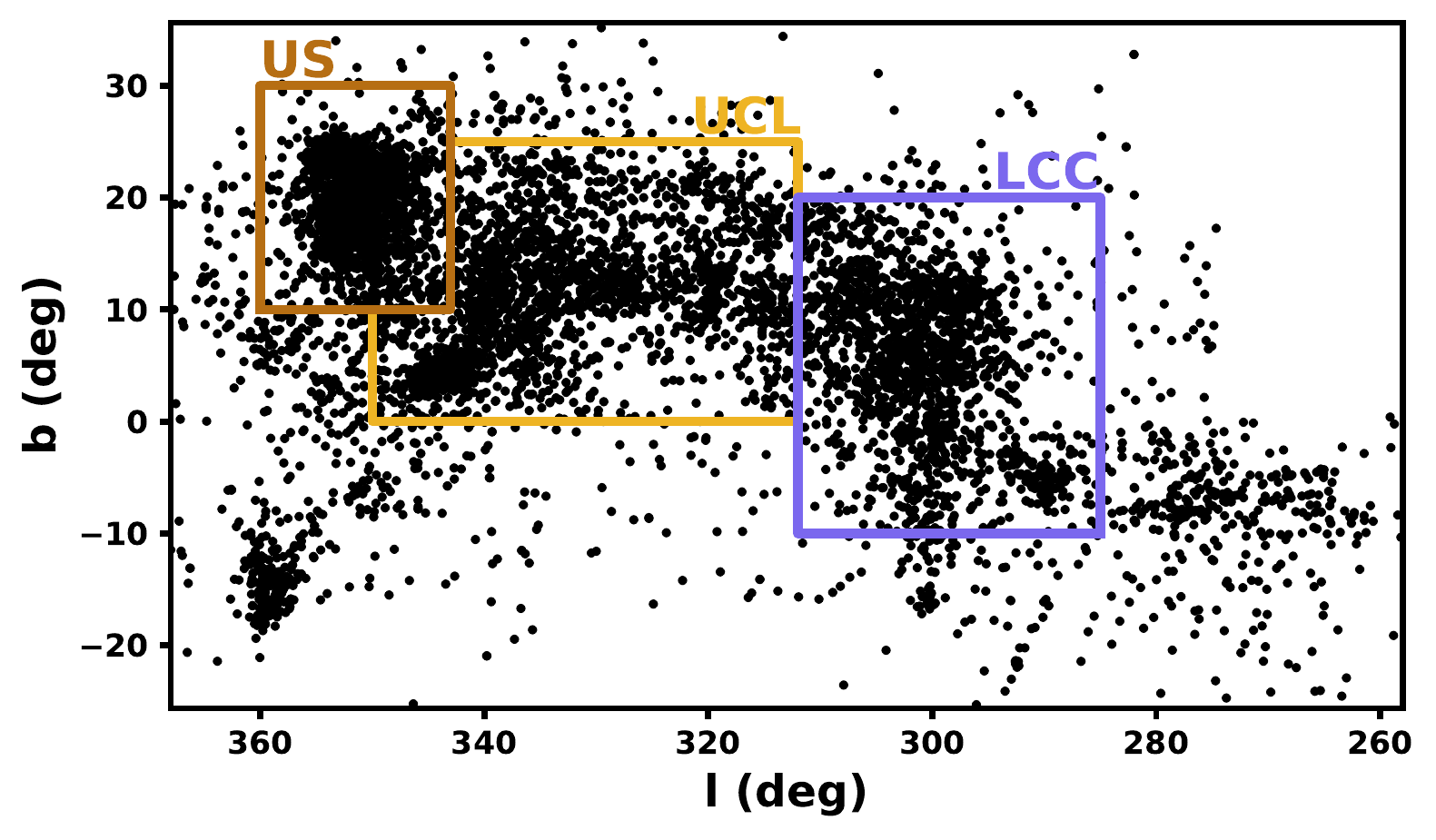}\hfill
\caption{The stars we identify in Sco-Cen, with the \citet{deZeeuw99} subregions plotted over it. }
\label{fig:scocen_all}
\end{figure}

Sco-Cen is by far the largest association within our search radius, containing nearly 7400 candidate young stars. The region thus contains nearly an order of magnitude more stars than the near edge of Greater Orion, which is the next largest group we identify. Figure \ref{fig:scocen_all} plots the populations we identify against the Upper Sco, UCL, and LCC boundaries as defined by \citet{deZeeuw99}. Members of our Sco-Cen population clearly reside in all three of the traditional subregions, with significant extensions beyond the traditional boundaries. Our population's lack of adherence to the three traditional subregions and lack of strong divisions on the boundaries between them support the more continuous view of star formation proposed by \citet{Rizzuto11}, while also meriting a reassessment of the way we divide populations in the region. In this section we present the novel view of Sco-Cen's internal structure that we construct through our subclustering and age distribution analyses, revealing an extensive stellar population with complex substructure and age trends spanning of order 100 pc.

\begin{figure*}[t]
\centering
\includegraphics[width=18cm]{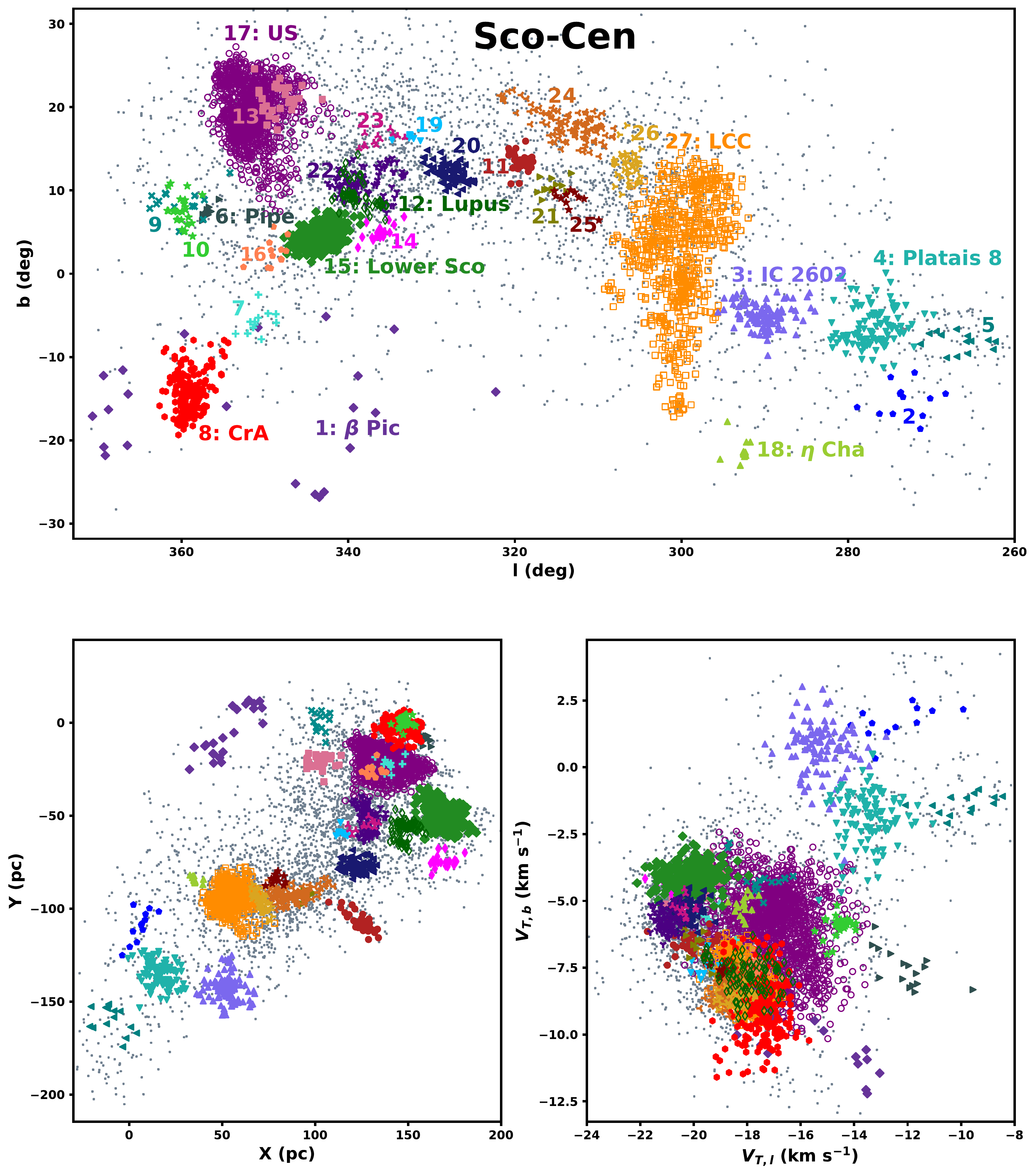}\hfill
\caption{Sco-Cen EOM clustering from HDBSCAN. The top panel displays l/b sky coordinates, bottom left shows galactic X/Y, and bottom right shows l/b transverse velocities. Subgroups are labelled in the top panel, and the corresponding icons and colors are consistent across the three panels. Three regions (US, LCC, and Lupus) have internal subclustering, and those are marked with \edit1{empty} icons. Grey points represent Sco-Cen members not assigned a subcluster.}
\label{fig:scocen_subclustering}
\end{figure*}

\begin{figure*}[t]
\centering
\includegraphics[width=19cm]{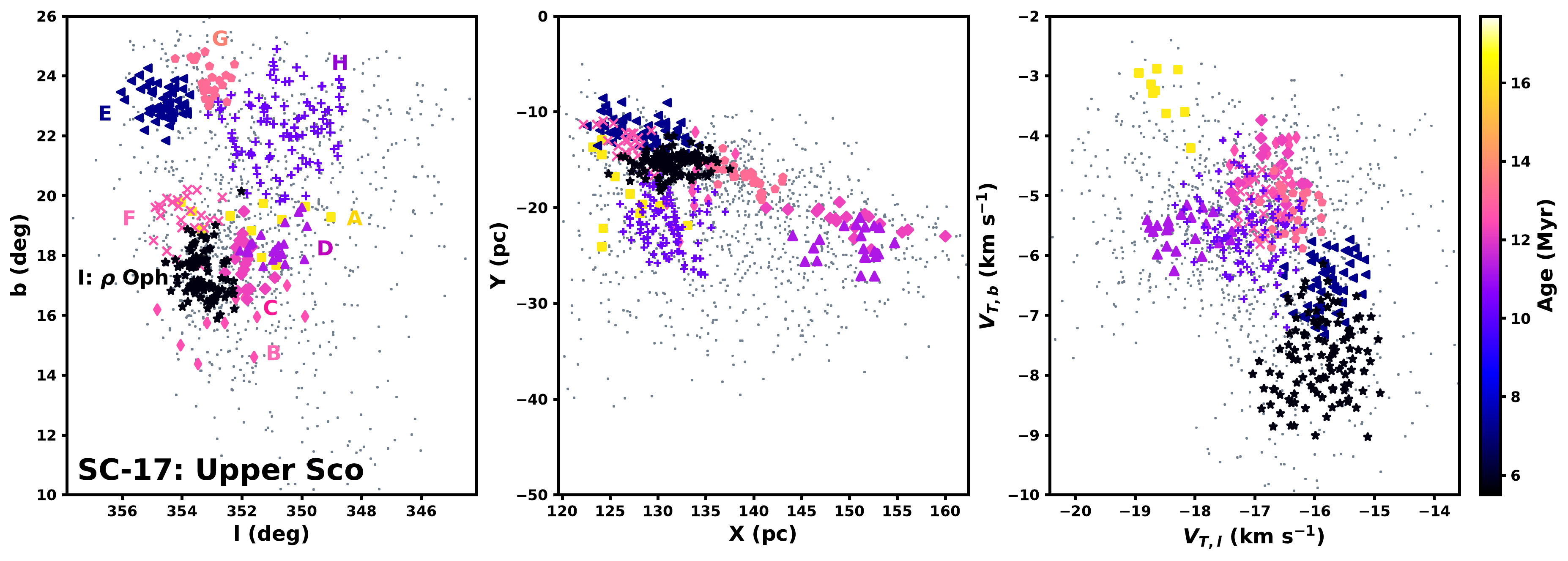}\hfill
\caption{Subclustering in Sco-Cen's Upper Sco subgroup, presented in l/b sky coordinates, X/Y galactic coordinates, and l/b tranverse velocity. US subgroups are marked according to their Sco-Cen Subgroup IDs from Table \ref{tab:sc_sc}, and unclustered US members are shown as small grey dots.}
\label{fig:scsc_us}
\end{figure*}

\begin{figure}[h]
\centering
\includegraphics[width=6.5cm]{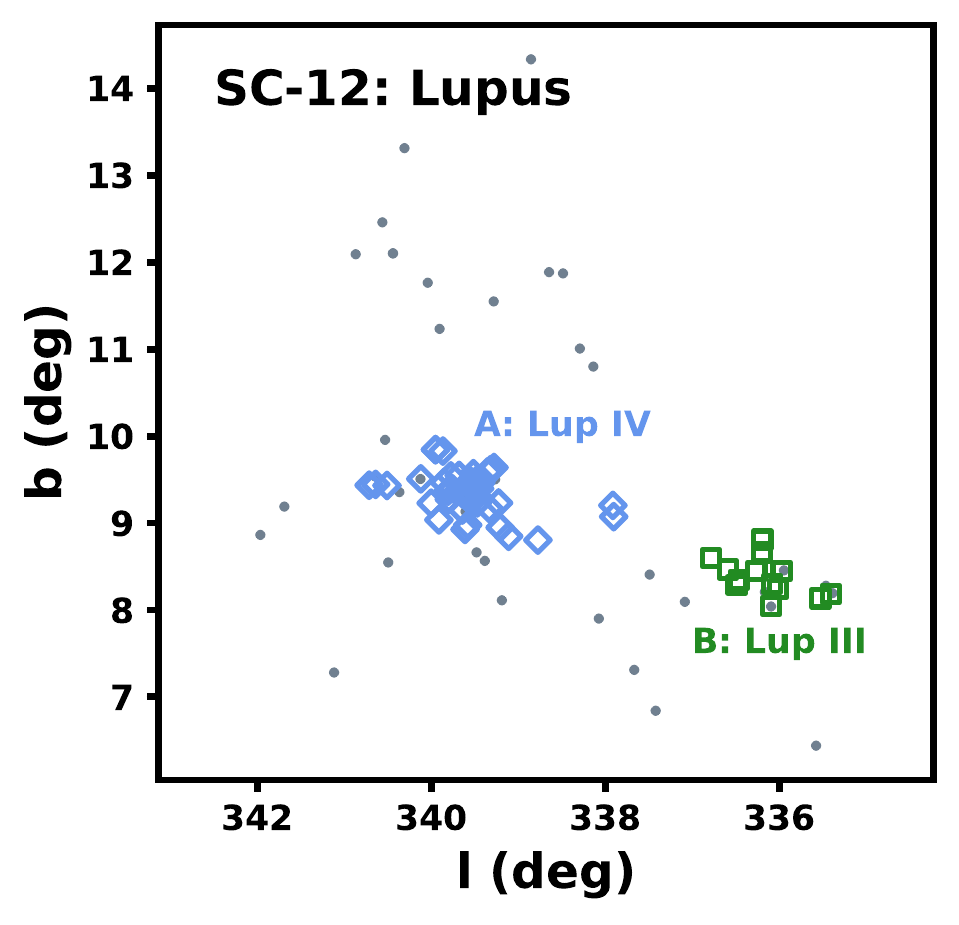}\hfill
\caption{Clustering  within  Sco-Cen's  Lupus  subgroup in l/b sky coordinates. Subclusters are marked, and unclustered Lupus members are  presented as grey dots. Both subgroups are essentially newborn.}
\label{fig:scsc_lupus}
\end{figure}

\begin{figure}[h]
\centering
\includegraphics[width=8.5cm]{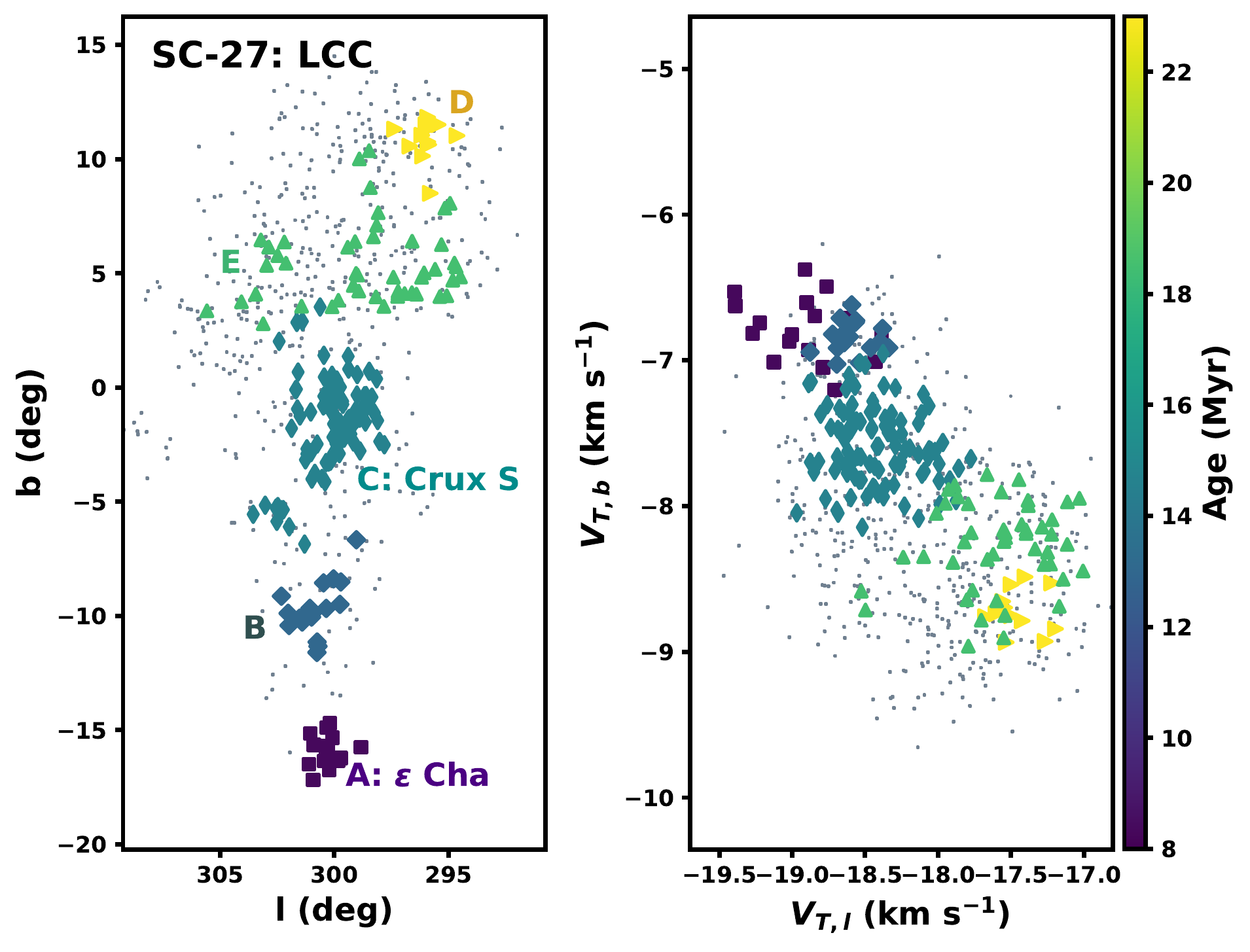}\hfill
\caption{Subclustering in Sco-Cen's Lower Centaurus-Crux subgroup, presented in l/b sky coordinates and l/b transverse velocity. Groups are labelled using the Sco-Cen Subgroup IDs from Table \ref{tab:sc_sc}, and the grey points are unclustered LCC members. }
\label{fig:scsc_lcc}
\end{figure}

\subsubsection{Clustering Overview} \label{sec:sc_sc}

\begin{deluxetable*}{cccccccccccccccc}
\tablecolumns{16}
\tablewidth{0pt}
\tabletypesize{\scriptsize}
\tablecaption{Clustering within the Sco-Cen Association. SC is the Sco-Cen ID, a number from EOM clustering used for identification. Leaf clusters within an EOM cluster are given a subcluster ID, which is given in the LEAF column. NC in the age column marks subclustered groups with non-coeval populations.} \label{tab:sc_sc}
\tablehead{
\colhead{SC} &
\colhead{LEAF} &
\colhead{Name} &
\colhead{N} &
\colhead{RA} &
\colhead{Dec} &
\colhead{l} &
\colhead{b} &
\colhead{$D_{sky}$\tablenotemark{a}} &
\colhead{d} &
\colhead{$\mu_{RA}$} &
\colhead{$\mu_{Dec}$} &
\colhead{$V_{T, l}$} &
\colhead{$V_{T, b}$} &
\colhead{$\sigma_{V_T}$\tablenotemark{b}} &
\colhead{Age} \\
\colhead{} &
\colhead{} &
\colhead{} &
\colhead{} &
\multicolumn{2}{c}{(deg)} &
\multicolumn{2}{c}{(deg)} &
\colhead{(deg)} &
\colhead{(pc)} &
\multicolumn{2}{c}{(mas/yr)} &
\multicolumn{2}{c}{(km/s)} &
\colhead{(km/s)} &
\colhead{(Myr)}
}
\startdata
1  &      & $\beta$ Pictoris &    23 &  279.4 & -43.6 &  351.4 & -16.6 &  29.0$\times$13.2 &     57.4$\pm$10.4 &   9.7 &  -70.3 & -16.0 & -9.8 &   2.1$\times$0.7 &  23.3$\pm$2.1 \\
2  &      &                  &    12 &  120.1 & -59.9 &  273.2 & -15.2 &   5.7$\times$3.9 &     113.6$\pm$7.8 &  -8.3 &   21.9 & -12.4 &  1.7 &   1.3$\times$0.5 &  43.2$\pm$2.6 \\
3  &      & IC 2602          &   101 &  161.8 & -64.6 &  290.1 &  -5.0 &   4.7$\times$2.8 &     152.6$\pm$6.7 & -18.2 &   10.5 & -15.2 &  0.7 &   1.0$\times$0.8 &  40.5$\pm$0.8 \\
4  &      & Platais 8        &   106 &  138.3 & -57.7 &  277.0 &  -6.4 &   5.2$\times$4.5 &     137.5$\pm$6.6 & -16.3 &   13.0 & -13.5 & -1.9 &   1.0$\times$0.8 &  37.0$\pm$0.9 \\
5  &      &                  &    14 &  124.9 & -50.8 &  266.6 &  -8.2 &   5.6$\times$2.2 &     162.3$\pm$6.6 &  -8.7 &    9.8 & -10.0 & -1.4 &   1.0$\times$0.3 &  35.8$\pm$2.0 \\
6  &      & Pipe             &    16 &  257.4 & -27.5 &  356.8 &   7.4 &   1.9$\times$0.3 &     160.2$\pm$1.9 &  -1.2 &  -18.6 & -12.0 & -7.5 &   1.1$\times$0.5 &   3.5$\pm$0.7 \\
7  &      &                  &    12 &  266.7 & -39.7 &  350.9 &  -5.8 &   3.3$\times$2.2 &     141.9$\pm$4.9 &  -7.8 &  -29.4 & -19.7 & -5.6 &   0.6$\times$0.4 &  14.0$\pm$1.2 \\
8  &      & CrA              &   248 &  281.4 & -36.4 &  359.2 & -14.6 &   4.8$\times$2.6 &     150.4$\pm$5.5 &   1.6 &  -27.4 & -17.5 & -8.8 &   1.1$\times$0.7 &  13.3$\pm$0.9 \\
9  &      &                  &    15 &  258.5 & -24.2 &    0.1 &   8.5 &   5.6$\times$3.0 &     104.1$\pm$3.3 & -13.3 &  -33.8 & -17.4 & -4.2 &   0.8$\times$0.4 &  15.6$\pm$1.6 \\
10 &      & Theia 67         &    28 &  259.4 & -24.8 &    0.0 &   7.5 &   3.4$\times$1.8 &     149.3$\pm$3.0 &  -4.8 &  -21.7 & -14.6 & -6.0 &   0.4$\times$0.3 &  20.2$\pm$2.0 \\
11 &      & UPK 606          &    36 &  216.2 & -46.4 &  319.2 &  13.5 &   2.0$\times$1.6 &     169.3$\pm$7.5 & -20.4 &  -16.7 & -20.1 & -6.6 &   0.6$\times$0.4 &  16.9$\pm$1.2 \\
12 &      & Lupus            &   102 &  241.5 & -39.6 &  338.9 &   9.4 &   3.6$\times$2.0 &     160.8$\pm$3.9 & -10.4 &  -23.5 & -18.0 & -7.8 &   0.8$\times$0.6 &   5.9$\pm$0.6\tablenotemark{c}  \\
12 &    A & Lup IV           &    46 &  242.2 & -39.1 &  339.6 &   9.4 &   1.0$\times$0.4 &     159.8$\pm$2.7 & -10.1 &  -23.5 & -17.7 & -7.7 &   0.6$\times$0.5 &   5.4$\pm$0.8 \\
12 &    B & Lup III          &    14 &  239.8 & -42.1 &  336.2 &   8.4 &   0.8$\times$0.4 &     161.2$\pm$2.3 & -11.0 &  -23.4 & -18.1 & -8.0 &   0.4$\times$0.3 &   5.9$\pm$1.3 \\
13 &      &                  &    30 &  240.1 & -24.9 &  348.3 &  20.9 &   3.7$\times$3.3 &     111.9$\pm$5.7 & -21.2 &  -32.3 & -19.9 & -4.4 &   0.6$\times$0.3 &  25.9$\pm$1.5 \\
14 &      &                  &    20 &  242.9 & -44.6 &  336.2 &   5.0 &   2.9$\times$1.2 &     185.3$\pm$4.2 & -13.8 &  -19.9 & -20.8 & -4.4 &   0.4$\times$0.2 &  14.1$\pm$1.2 \\
15 &      & Lower Sco        &   370 &  250.2 & -39.7 &  343.5 &   4.5 &   3.3$\times$1.8 &     176.3$\pm$6.3 & -12.1 &  -21.3 & -20.0 & -4.1 &   0.6$\times$0.4 &  18.8$\pm$0.6 \\
16 &      &                  &    12 &  256.5 & -36.7 &  348.9 &   2.5 &   3.6$\times$2.2 &     134.0$\pm$3.7 &  -7.2 &  -29.1 & -17.5 & -7.5 &   0.4$\times$0.2 &  18.5$\pm$1.6 \\
17 &      & Upper Sco        &  1478 &  243.7 & -23.6 &  351.8 &  19.4 &   6.7$\times$4.7 &     146.4$\pm$9.3 & -10.5 &  -23.6 & -16.9 & -5.8 &   1.5$\times$1.1 &  11.3$\pm$0.3\tablenotemark{c} \\
17 &    A &                  &    11 &  243.9 & -23.9 &  351.7 &  19.1 &   3.2$\times$1.4 &     135.4$\pm$3.2 & -15.9 &  -24.7 & -18.6 & -3.3 &   0.4$\times$0.2 &  16.2$\pm$2.0 \\
17 &    B &                  &    11 &  247.0 & -25.7 &  352.3 &  15.8 &   3.0$\times$1.5 &     140.4$\pm$1.5 & -11.5 &  -23.2 & -16.7 & -4.5 &   0.4$\times$0.4 &  12.5$\pm$1.0 \\
17 &    C &                  &    18 &  245.1 & -24.7 &  351.9 &  17.7 &   1.6$\times$0.7 &     159.3$\pm$4.0 & -10.4 &  -20.5 & -16.8 & -4.6 &   0.4$\times$0.4 &  12.2$\pm$0.6 \\
17 &    D &                  &    22 &  244.0 & -24.9 &  351.0 &  18.3 &   1.5$\times$0.9 &     160.7$\pm$2.8 & -11.0 &  -22.4 & -18.2 & -5.6 &   0.4$\times$0.3 &  11.2$\pm$0.9 \\
17 &    E &                  &    54 &  242.7 & -19.1 &  354.7 &  23.1 &   1.2$\times$0.9 &     140.0$\pm$3.3 &  -8.8 &  -24.2 & -15.8 & -6.5 &   0.4$\times$0.3 &   7.2$\pm$0.7 \\
17 &    F &                  &    25 &  245.3 & -22.2 &  354.0 &  19.2 &   1.5$\times$1.3 &     135.3$\pm$2.3 & -11.6 &  -25.0 & -16.9 & -5.1 &   0.4$\times$0.2 &  12.6$\pm$1.1 \\
17 &    G &                  &    29 &  241.1 & -19.6 &  353.1 &  23.9 &   1.1$\times$0.7 &     152.9$\pm$2.3 & -10.2 &  -21.4 & -16.4 & -5.3 &   0.3$\times$0.3 &  13.2$\pm$2.8 \\
17 &    H &                  &   102 &  240.6 & -22.4 &  350.6 &  22.3 &   2.4$\times$2.2 &     143.3$\pm$2.6 & -11.9 &  -23.7 & -17.1 & -5.6 &   0.6$\times$0.5 &  10.2$\pm$0.7 \\
17 &    I & $\rho$ Oph       &   110 &  246.4 & -23.9 &  353.3 &  17.4 &   1.5$\times$1.0 &     138.7$\pm$2.6 &  -7.2 &  -25.9 & -15.9 & -7.7 &   0.7$\times$0.5 &   5.7$\pm$0.4 \\
18 &      & $\eta$ Cha       &    17 &  132.3 & -79.0 &  292.7 & -21.3 &   2.3$\times$1.7 &      98.3$\pm$1.7 & -30.7 &   26.1 & -18.1 & -5.2 &   0.3$\times$0.2 &   8.2$\pm$1.7 \\
19 &      &                  &    10 &  229.9 & -37.8 &  332.6 &  16.4 &   1.7$\times$0.5 &     134.6$\pm$2.4 & -19.0 &  -26.9 & -19.6 & -7.5 &   0.4$\times$0.3 &  15.3$\pm$2.8 \\
20 &      &                  &    77 &  227.7 & -44.0 &  327.8 &  12.1 &   3.2$\times$1.7 &     147.4$\pm$4.9 & -20.8 &  -21.7 & -20.3 & -5.4 &   0.4$\times$0.3 &  22.8$\pm$1.0 \\
21 &      &                  &    12 &  212.5 & -50.5 &  315.5 &  10.5 &   2.7$\times$1.6 &     134.7$\pm$2.5 & -25.9 &  -19.6 & -19.6 & -6.9 &   0.9$\times$0.3 &  23.9$\pm$2.9 \\
22 &      &                  &   102 &  239.5 & -38.9 &  338.1 &  10.9 &   4.9$\times$2.9 &     140.6$\pm$4.6 & -18.0 &  -27.1 & -20.9 & -5.7 &   0.4$\times$0.3 &  22.7$\pm$1.1 \\
23 &      &                  &    12 &  233.6 & -35.9 &  336.3 &  16.1 &   3.7$\times$1.1 &     143.6$\pm$4.3 & -19.1 &  -24.2 & -20.3 & -5.3 &   0.3$\times$0.2 &  21.5$\pm$2.1 \\
24 &      &                  &   108 &  206.8 & -43.8 &  313.5 &  17.9 &   6.9$\times$2.5 &     134.3$\pm$6.6 & -25.7 &  -19.3 & -18.7 & -8.3 &   0.5$\times$0.3 &  20.8$\pm$1.1 \\
25 &      &                  &    11 &  209.8 & -52.5 &  313.3 &   9.0 &   3.5$\times$1.5 &     117.2$\pm$2.5 & -29.3 &  -22.2 & -18.9 & -7.6 &   0.2$\times$0.1 &  20.7$\pm$2.0 \\
26 &      &                  &    40 &  198.1 & -49.7 &  306.4 &  13.0 &   3.5$\times$2.0 &     123.0$\pm$5.3 & -30.3 &  -17.2 & -18.4 & -8.5 &   0.5$\times$0.4 &  21.7$\pm$1.1 \\
27 &      & LCC              &   697 &  187.3 & -60.1 &  300.1 &   2.5 &  13.9$\times$6.0 &     109.5$\pm$6.4 & -36.1 &  -12.2 & -18.1 & -8.0 &   0.8$\times$0.4 &  NC \\
27 &    A &    $\epsilon$ Cha  &    17 &  180.1 & -78.5 &  300.3 & -15.9 &   1.3$\times$1.1 &     101.9$\pm$1.2 & -41.2 &   -5.8 & -18.9 & -6.8 &   0.3$\times$0.2 &   8.3$\pm$1.0 \\
27 &    B &                  &    17 &  186.0 & -72.4 &  300.8 &  -9.7 &   2.7$\times$1.4 &     102.0$\pm$2.1 & -39.7 &   -9.9 & -18.6 & -6.8 &   0.1$\times$0.1 &  13.0$\pm$1.4 \\
27 &    C & Crux S\tablenotemark{$\dagger$}          &   100 &  186.5 & -64.4 &  300.2 &  -1.6 &   4.1$\times$2.2 &     106.8$\pm$2.3 & -37.7 &  -11.3 & -18.4 & -7.6 &   0.3$\times$0.2 &  14.6$\pm$0.8 \\
27 &    D &                  &    12 &  181.9 & -51.4 &  296.0 &  10.9 &   1.8$\times$1.3 &     114.8$\pm$2.2 & -34.4 &  -10.2 & -17.5 & -8.7 &   0.2$\times$0.1 &  23.0$\pm$2.3 \\
27 &    E &                  &    48 &  185.0 & -57.1 &  298.7 &   5.4 &   6.0$\times$3.4 &     109.3$\pm$2.6 & -35.5 &  -11.7 & -17.6 & -8.3 &   0.4$\times$0.3 &  18.5$\pm$0.7 \\
-\tablenotemark{d} & - & TW Hydrae &  19 & 179.7 & -36.2 & 290.9 &  25.1 & 14.7$\times$6.2&     62.9$\pm$10.9 & -67.8 &  -23.0 & -17.6 & -11.0 &  1.3$\times$0.5 &  10.8$\pm$0.9 \\
\enddata
\tablenotetext{a}{On-sky spatial extent in galactic l/b, in the form of the RMS in major axis $\times$ minor axis, when fit with a bivariate Gaussian.}
\tablenotetext{b}{The radial extent of the velocity distribution, in the form of the RMS in semi-major axis $\times$ semi-minor axis, when fit with a bivariate Gaussian.}
\tablenotetext{c}{Has substructure, but most of the region is close enough to coeval that a global age estimation can be meaningful.}
\tablenotetext{d}{Defined through crossmatching the Sco-Cen candidate young stars with known TW Hydrae members, not HDBSCAN clustering.}
\tablenotetext{\dagger}{name is newly assigned by this paper.}
\vspace*{0.1in}
\end{deluxetable*}

We find a remarkable diversity of substructures in Sco-Cen with 27 EOM clusters, three of which subdivide into a total of 16 additional leaf clusters (see Table \ref{tab:sc_sc}). Our top-level clustering results in Sco-Cen are presented in Figure \ref{fig:scocen_subclustering}, while clustering within the three EOM clusters with substructure occupy Figures \ref{fig:scsc_us}, \ref{fig:scsc_lupus}, and \ref{fig:scsc_lcc}. Compared to the previously discussed Greater Taurus Association, most of Sco-Cen has a very compact velocity distribution, especially given its size. Excluding the groups to the far west, all subregions we identify have mean velocities within 7 km s$^{-1}$ of each other, a velocity distribution comparable to the spread in transverse velocities expected from projecting the UVW velocity vector for Upper Sco onto the plane of the sky \citep{Luhman20}. As such, projection effects may dominate the observed velocity spread. The remaining groups can extend up to $\sim$ 10 km s$^{-1}$ away from the centre of the main distribution, although these extended populations remain spatially and kinematically contiguous with the main body of Sco-Cen. 

Most of the denser regions of Sco-Cen fall within the \citet{deZeeuw99} boundaries, however 13 of the 27 subgroups we identify in Sco-Cen exist at least partially outside of the \citet{deZeeuw99} boundaries displayed in Figure \ref{fig:scocen_all}. The groups that overflow the traditional Sco-Cen boundaries include the southern tip of SC-27, which is the largest group falling mostly within the traditional boundaries of LCC. \edit1{SC-27 also matches closely with the population referred to as the Crux Moving Group (CMG) in \citet{Goldman18}, a proposed LCC subgroup. However, given the lack of larger-scale substructure identified in Sco-Cen we find that SC-27 and by extension the CMG are better described as revised extents for LCC, rather than distinct populations within}. Upper Sco also has its own EOM cluster with substructure in the form of SC-17; however the remaining traditional subregion, UCL, does not, instead being loosely comprised of seven clusters with no parent EOM cluster. This lack of identified higher-level structure disputes UCL's existence as a physical structure, although the label remains useful to refer to the area more generally.  One of these subgroups contained within the \citet{deZeeuw99}-defined extent of UCL is SC-12, or Lupus, which is the third and final group with substructure we identify. The known subregions of Lupus III and Lupus IV correspond to the defined lower-level subgroups of SC-12B and SC-12A, respectively, and some members of Lupus II also appear in SC-12, although these are too sparse for HDBSCAN to identify a subcluster \citep{Galli13}. The lack of stars detected in Lupus II and near-absence of Lupus I is likely due to heavy extinction in much of that group \citep{Comeron08}, which leads to dramatically higher rates of rejection based on Gaia quality cuts (see Section \ref{sec:recovery}). 

LCC (SC-27) contains five subgroups, the largest of which is SC-27C, a dense group containing nearly one-seventh of the total population in LCC. This substantial subgroup does not appear to have been previously identified independently from LCC, so we name it the Crux South Group, after its parent constellation. Some of the Chamaeleon clusters overlap with LCC and Sco-Cen more generally near LCC's south edge, with SC-27A matching with catalogues for the known $\epsilon$ Chamaeleontis cluster \citep[e.g.,][]{Murphy13}. The $\eta$ Chamaeleontis cluster is located near $\epsilon$ Chamaeleontis, however there is little connecting it to the rest of LCC, hence it being identified as separate under the ID SC-18. Chamaeleon I and II, the other two previously-known Chamaeleon-associated clusters also share kinematics with Sco-Cen, although they are much farther away and apparently non-contiguous with the rest of Sco-Cen, so HDBSCAN does not include them as Sco-Cen subgroups. The relation between these more distant Chamaeleon clusters and Sco-Cen is discussed in Section \ref{sec:prph_cha}. Aside from the aforementioned notable features, the rest of LCC contains relatively sparse and small subgroups. 

Upper Sco (SC-17) is the largest subgroup identified in Sco-Cen, and it also contains the most substructure, with nine subgroups. The largest of these, SC-17I, is centered on $\rho$ Ophiuchi, which is located adjacent to sites of active star formation in and around the L1688 cloud \citep[e.g.][]{Kerr19}. The rest of Upper Sco, however, appears to be composed of largely new substructure. SC-17 subgroups C and D represent populations distributed nearly 20 pc behind the active star-forming central region around $\rho$ Oph, while subgroups E, G, and to a lesser extent F form an arc of dense and populous subgroups located to the north of $\rho$ Oph. The remainder of the subregions are very close to SC-17I in spatial coordinates, but reside up to $\sim$5 km s$^{-1}$ away from SC-17I in transverse velocity space, hence their division by HDBSCAN. 

In the remainder of Sco-Cen within the \citet{deZeeuw99} boundaries, we identify a wealth of diverse subgroups stretching from LCC through to the traditionally-defined extent of UCL. The largest of the groups in this region by far is Lower Sco (SC-15), a known clustered group to the area's southeast (\citealp{Mamajek13}, \citealp[and also presented as the V1062 Sco Moving Group in][]{Roser18}). However, Lower Sco is a relative outlier in spatial coordinates, representing a significant protrusion behind most other groups in the area. SC-14 also protrudes behind the rest of Sco-Cen in a similar way to Lower Sco and holds an overlapping velocity distribution, possibly indicating a link between the two. SC-11, identified as UPK 606 in the \citet{Sim19} catalog, represents another clear protrusion behind Sco-Cen, although this one is much further west. The remainder of the groups in this region follow a fairly contiguous arc of young substructures, with a slight lower-density gap between the groups more closely connected to LCC (SC-24, 21, 25, and 26) and those that fall within the traditional UCL region (SC-12, 22, 23, 19, 20). SC-19 is a uniquely dense subgroup, with most stars within 15 arcmin, making it worth revisiting as small bound cluster. 

The subgroups we identify outside of the \citet{deZeeuw99} boundaries for Sco-Cen almost entirely fall into two regions: The southwest, which contains the far south of SC-27 as well some more western groups including IC 2602, and the southeast, which contains structure leading to and including Corona Australis. We refer to the subset of the southwestern extension that extends to IC 2602 and beyond as the IC 2602 branch, which includes IC 2602 (SC-3), along with SC-2, SC-4, and SC-5. This is a region that has somewhat weaker connection to the rest of Sco-Cen, with velocities that are visibly distinct from but still contiguous with the rest of Sco-Cen (see Fig \ref{fig:scocen_subclustering}). The southeastern extensions are mostly much closer to the known regions of Sco-Cen in velocity space, and are anchored by the large Corona Australis Association. This is a well-known group, however it has not typically been linked to Sco-Cen due to its apparent lack of any structure connecting it to the rest of the association. We identify this previously unknown linking structure in the form of five subgroups in the space between Upper Sco, Lower Sco, and Corona Australis (SC-7, SC-16, SC-6, and SC-10), along with further, low-density populations in between subgroups. SC-9 is in the same area in the plane of the sky, but is actually projected in front of these other groups. The inclusion of CrA as a new sub-region of Sco-Cen is therefore well supported by both the kinematics from Gaia and the structural layout of the region. 

The last of these outlying groups, SC-1, appears to correspond to the $\beta$ Pictoris Moving Group. Known $\beta$ Pic member $\eta$ Telescopii is the brightest member we identify \citep{Zuckerman01}, while ten of the remaining 22 SC-1 members are labelled as $\beta$ Pic members by \citet{Shkolnik18}, found distributed fairly evenly throughout our population. Our recognition of $\beta$ Pic members is however limited to stars more distant that about $\sim$40 pc, as geometric effects badly skew the transverse velocity vector for clustering nearer populations. While a link between $\beta$ Pic and Sco-Cen has been proposed in the past \citep{Preibisch08}, this claim has been made on the basis of common motion, rather than the presence of stellar populations linking them. Through our clustering analysis, we show that low-density connections between $\beta$ Pictoris and the rest of Sco-Cen do exist, representing the first direct evidence that $\beta$ Pic is fully contiguous with Sco-Cen. 

\subsubsection{TW Hydrae}

\begin{figure}[t!]
\centering
\includegraphics[width=8.2cm]{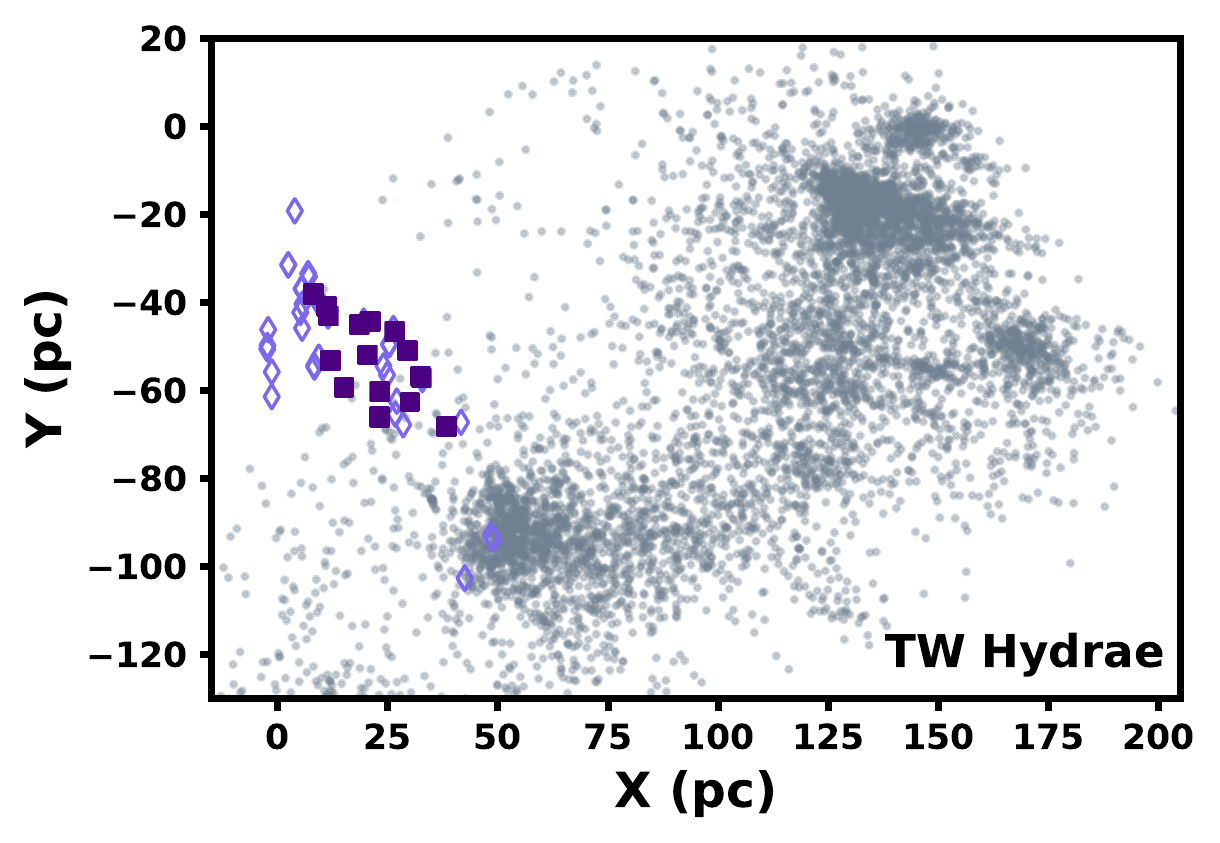}\hfill
\caption{Our members for Sco-Cen (light grey), compared to TW Hydrae Association (TWA) Members \citep{Song02,Reid03,Gagne17}. dark squares are in both our Sco-Cen set and TWA, while the light \edit1{empty} diamonds are other TWA members.}
\label{fig:twa}
\end{figure}

The TW Hydrae Association (TWA) is a very nearby young stellar population, with an average distance of $\sim$60 pc \citep[][]{Reid03}. Due to its proximity, TWA is very well-studied, and detailed investigations of its membership and kinematics have strongly suggested a connection to the Sco-Cen association, particularly through LCC, the nearest traditional subregion \citep[e.g.,][]{Mamajek99}. While we do not directly identify a subgroup consistent with TWA due to the region's low relative density, many TWA members appear in our Sco-Cen population. In Fig \ref{fig:twa}, we cross-match currently known TWA members \citep{Song03,Reid03,Gagne17} with our complete Sco-Cen population, revealing that nearly half of that known sample is contained within our population of photometrically young Sco-Cen members. All appear along an extended spur of stars that reaches from the near side of LCC in the direction of the Sun, following a distribution fully contiguous with the rest of Sco-Cen. The identification rates for the known TWA members within our Sco-Cen sample appear to taper off for objects on the region's western edge with a distance of less than 40-50 pc from the Sun; however, this is likely due to the transverse velocities, which are increasingly sensitive to geometric effects from their position in space as they get closer to the Sun, therefore making clustering more difficult, especially for objects farther from the line of sight with the rest of the Sco-Cen members. Regardless, the overlap in these populations that we uncover represents the most direct evidence to date of a physical link between the TW Hydrae Association and the Lower Centaurus-Crux subregion of Sco-Cen, a link that has been suggested by multiple recent papers \citep[e.g.,][]{Mamajek01, Murphy15}.

\subsubsection{Age Structure}

\begin{figure*}[t]
\centering
\includegraphics[width=19cm]{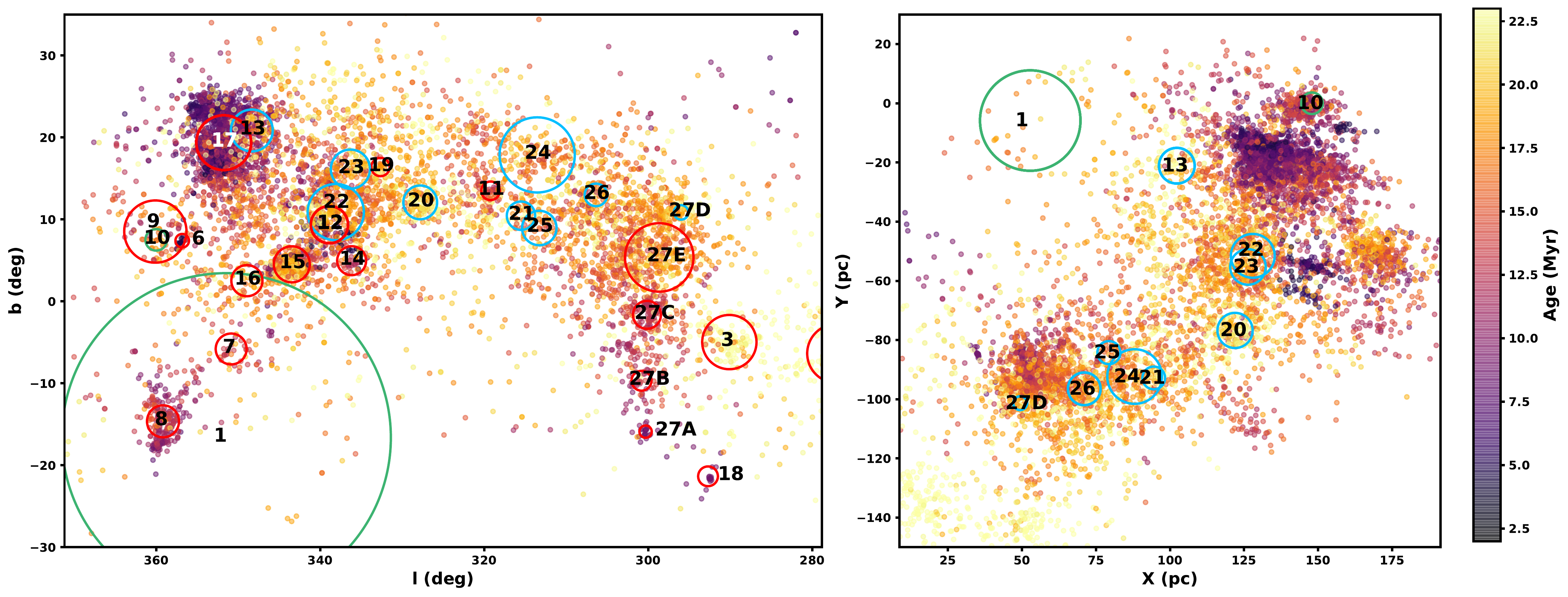}\hfill
\caption{Sco-Cen members plotted in l/b spatial coordinates and X/Y galactic coordinates, color-coded by the median age fit of the 10 nearest neighbors. \edit1{The groups presented in Fig. \ref{fig:scocen_subclustering} are labelled and marked with circles that represent the locations and effective radii of each group, with the subclusters in LCC shown as opposed to the top-level group}. Most of those outlines are in red, however groups not on the IC 2602 branch that have Age $>$ 20 Myr are marked in blue or green, with the blue outlines marking groups we associate with the Libra-Centaurus Arc (LCA).}
\label{fig:scocen_ages}
\end{figure*}

\begin{figure}[t]
\centering
\includegraphics[width=4.1cm]{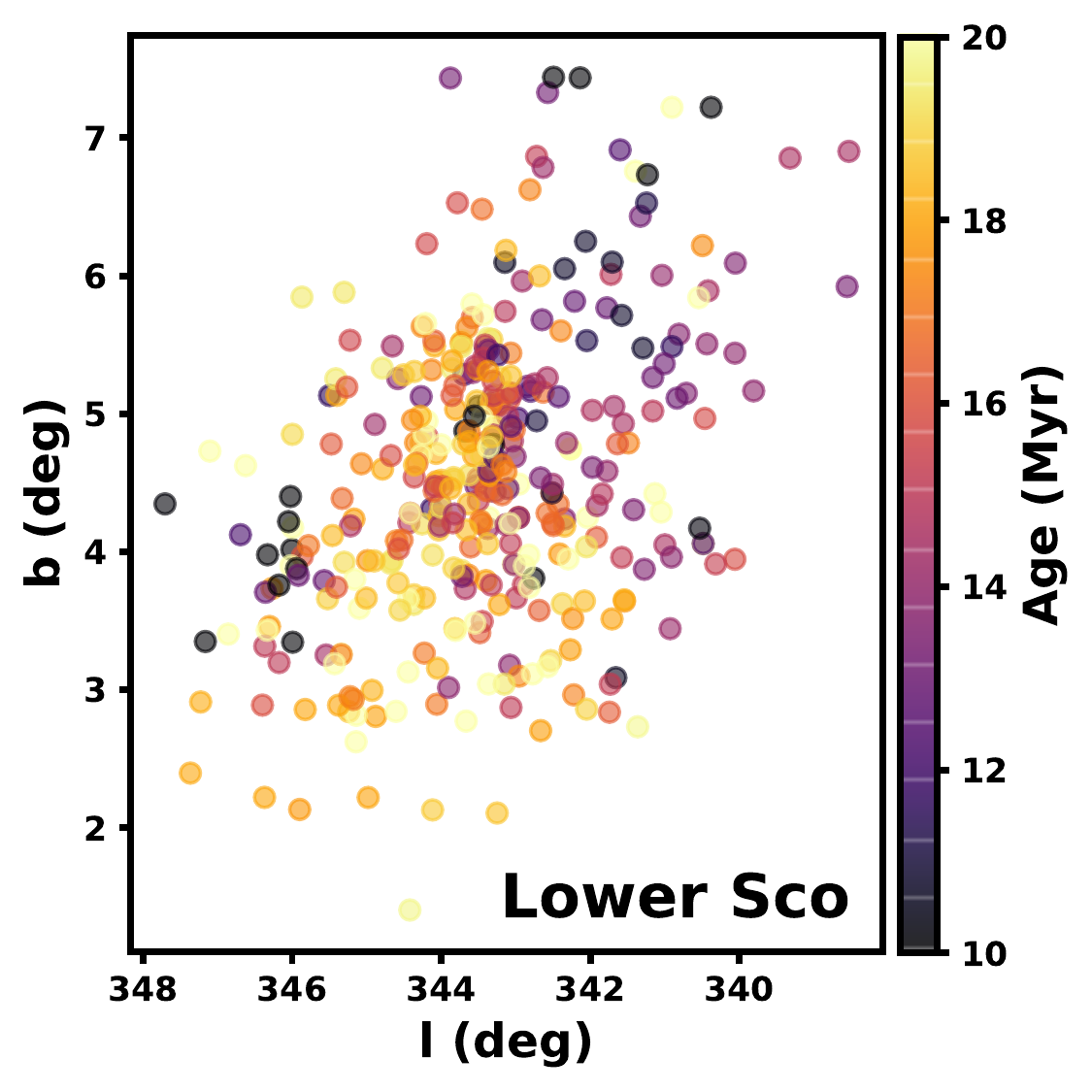}\hfill
\includegraphics[width=4.3cm]{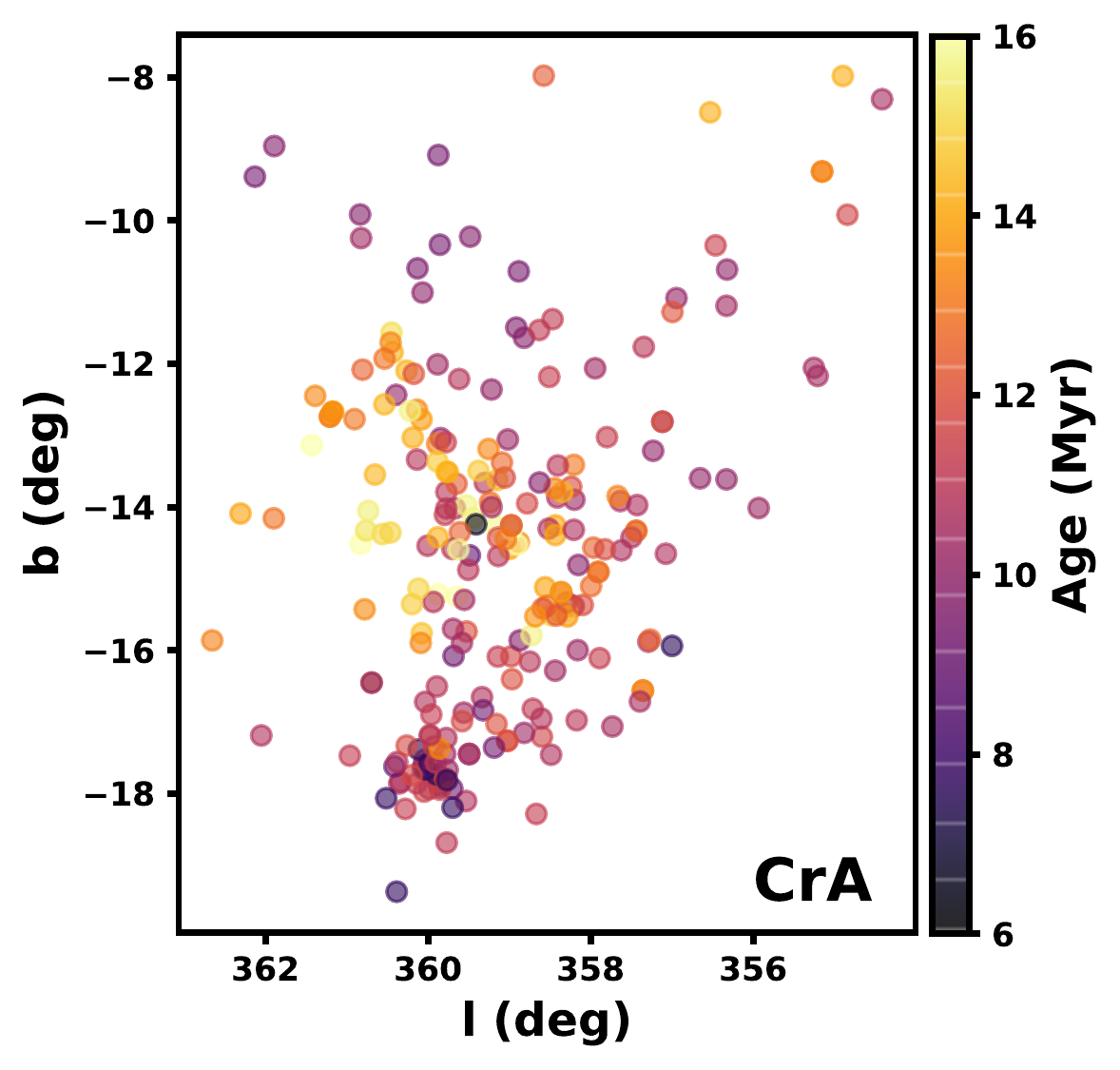}\hfill
\caption{The age distributions in Corona Australis and Lower Sco according to the median age of the 10 nearest neighbors. Both show strong visible gradients, with Lower Sco younger in the direction of Lupus, and CrA younger towards a tight clump at the far south, which is associated with recent formation in the Corona Australis Molecular Cloud.}
\label{fig:age_closeups}
\end{figure}

As shown in Figure \ref{fig:scocen_ages} and Table \ref{tab:sc_sc}, ages in the Sco-Cen association range from essentially newborn, such as those in Lupus or the $\rho$ Ophiuchi complex, to over 20 Myr old. Aside from the older populations around IC 2602, all groups with ages older than 20 Myr fall along the northern edge of Sco-Cen, following an arc of old star formation stretching from the north edge of LCC, across the northern and western edge of UCL (including SC-20 SC-21, SC-22, SC-23, SC-24, SC-25, and SC-26), and to SC-13 in the foreground of Upper Sco at the border with Libra. This arc of old star formation will hereafter be referred to at the Libra-Centaurus Arc (LCA), in reference to the constellations at the approximate endpoints of this structure. This arc of old star formation appears to represent a possible starting point for star formation in the Sco-Cen association, as the age distributions in much of the rest of the association can be explained by sequential star formation originating in the LCA. 

In most cases, the farther a star is away from the LCA, the younger it is found to be. This trend is most plainly evident in LCC, which has a smooth age gradient from galactic north to south, clearly visible in Figures \ref{fig:scocen_ages} and \ref{fig:scsc_lcc}. SC-27D, which we consider part of the LCA, has an age of 23 Myr, while more southerly groups range from SC-27E at $\sim$18.5 Myr in the north, to $\epsilon$ and $\eta$ Chamaeleontis at $\sim$8 Myr in the south. Other features that protrude from the LCA are similarly young, such as SC-11, which has an age of $\sim$17 Myr, about 6 Myr younger than the adjacent LCA groups. The TW Hydrae association, which extends towards the sun from northern LCC, is also young compared to adjacent LCA groups, with an age estimated at 8-10 Myr in existing literature \citep{Murphy15,Weinberger13,Ducourant14}, and $\sim$11 Myr in our isochronal age fit. Corona Australis and surrounding groups may also have an age gradient towards the LCA. The southernmost tip of CrA is associated with the CrA dark clouds and is therefore essentially newborn \citep[e.g., see][]{Galli20}, while the age solution for the broader population centered closer to the LCA is $\sim$13 Myr, and SC-7 and SC-16, two separate groups even nearer to the LCA have ages of $\sim$14 Myr and $\sim$18.5 Myr, respectively. However, the stellar populations connecting the LCA to CrA are much more sparse and the distances between subgroups are much greater, making this connection less clear. We investigate Sco-Cen's age trends in the context of propagation rates for sequential star formation in Section \ref{sec:dis-seqprop}.

\edit1{These smooth age gradients towards the LCA are not universal, however, particularly for Upper Sco and groups in the direction of Lupus and Lower Sco}. These different age patterns come in the form of both internal age structure, typically with younger stars in the denser regions, such as in Upper Sco and CrA, and larger gradients that oppose those off the LCA, such as the area between Lupus and Lower Sco. The age gradients in Corona Australis and Lower Sco are given close-ups in Figure \ref{fig:age_closeups}. The sections of UCL closest to the LCA are older as expected, showing a subtle age gradient like in LCC, but some of the more distant groups such as Lupus, which contains essentially newborn stars, break that smooth progression. Lupus represents an anomalously young group for its location in Sco-Cen, and the youth of the region does not continue past Lupus, as Lower Sco and SC-14 have ages of $\sim$ 19 and 14 Myr, respectively. There is, however, an age gradient in Lower Sco that is younger in the direction of Lupus, suggesting that the two events may be directly connected. 

The other very young region in Sco-Cen, Upper Sco, is considerably younger than its surroundings, and the transition from the LCA to the adjacent younger populations is quite abrupt. This region is generally younger towards its centre, with the large central $\rho$ Ophiuchi group being the youngest. However, most of our Upper Sco ages outside $\rho$ Oph, including our Upper Sco at-large solution that assumes common ages across the entire region, converge between 10 and 13 Myr, a result consistent with the Upper Sco age solutions from \citet{Pecaut12} and \citet{Pecaut16}, and more recently supported by \citet{Sullivan21}. The only subgroup exceeding this age range at $\sim$16 Myr, SC-17A, is also one of the nearest subgroups to the LCA, so its older age is consistent with our narrative of star formation propagating away from the LCA. The first bursts of star formation in Upper Sco may have therefore originated relatively early in the history of Sco-Cen, with activity peaking $\sim$10-13 Myr ago.

The IC 2602 branch appears to contain some of the oldest stars in our sample, however we treat these groups with caution when considering them in the context of star formation in Sco-Cen more generally. The relatively different velocities in this area may suggest a degree of removal from the star formation processes in the rest of Sco-Cen, and the lack of smooth age gradients leading off the IC 2602 branch supports this. This region may therefore represent an older outbreak of star formation that pre-dates the formation of much of the rest of Sco-Cen, making its influence on the more recent star formation events in Sco-Cen doubtful.

\edit1{Our overview of the ages in Sco-Cen suggests that smooth, sequential star formation originating along the Libra-Centaurus Arc explains most of the age patterns in Sco-Cen, presenting a much more continuous picture of star formation compared to the set of discrete triggered star-forming events proposed in \citet{Preibisch08}. However, a few notable regions do not fully adhere to these age trends, requiring further investigation to make sense of those outliers. \citet{Krause18} proposes a star formation scenario that explains the presence of two of these age trend outliers (Lupus and $\rho$ Oph), suggesting that Sco-Cen formed within an extended gas overdensity such that when the earliest generation formed it preferentially pushed gas out perpendicular to the gas overdensity's long axis, creating superbubbles that later engulfed sparser material along the long axis and guided its collapse. The coeval populations we find along nearly the entire length of Sco-Cen in the form of the LCA are largely inconsistent with this star formation sequence, as our results demand a much broader star formation genesis than the individually coeval star-forming events discussed in \citet{Krause18}. However, this surround-and-squash explanation may contribute to the methods of formation elsewhere in Sco-Cen, particularly in regions like Upper Sco and Lupus that are somewhat removed from both the LCA and the associated age gradients. Lower Sco may serve an important role in applying the surround-and-collapse concept to our updated view of Sco-Cen, as the group is both relatively old and lies opposite to the LCA relative to Lupus and to a lesser extent Upper Sco. As such, a surround-and-collapse pattern guided by superbubbles from the LCA and Lower Sco may provide an explanation for the existence of both of these sites, while the rest of Sco-Cen appears well explained by a sequential star formation pattern, as discussed in Section \ref{sec:dis-seqprop}.}

\subsubsection{Chamaeleon} \label{sec:prph_cha}

\begin{figure}[t]
\centering
\includegraphics[width=8.5cm]{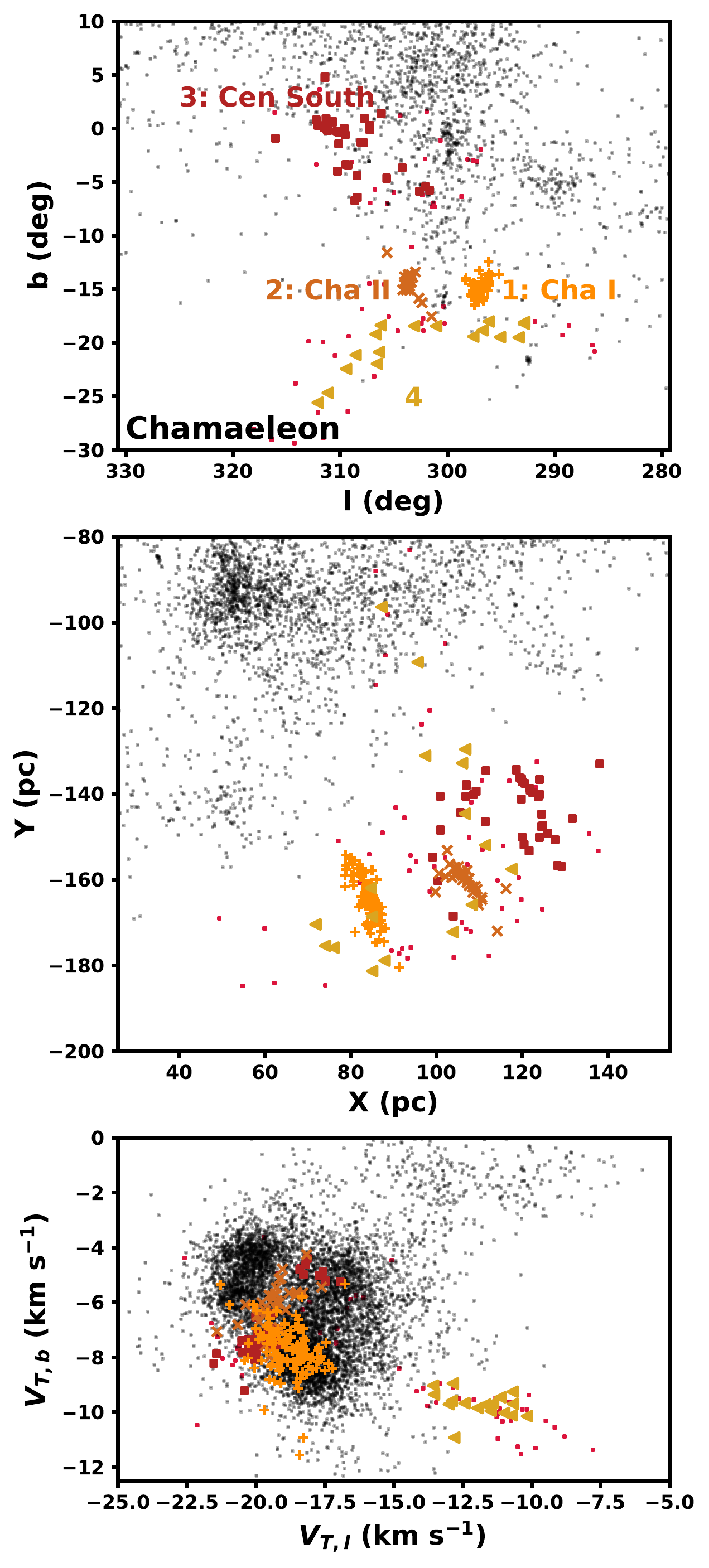}\hfill
\caption{Clusters in the Chamaeleon Complex, which hold similar kinematics to Sco-Cen. The panels, from top to bottom, show l/b sky coordinates, XY galactic coordinates, and the galactic transverse velocity. Subgroups are marked in the top panel, while small red points represent non-clustered Chamaeleon members, and small black points indicate Sco-Cen members.}
\label{fig:periph_cha}
\end{figure}

While two of the nearby young clusters in Chamaeleon were identified as Sco-Cen-affiliated by HDBSCAN, the somewhat more distant clusters were not, including the clusters associated with the actively star-forming Chamaeleon I and II clouds \citep{Luhman08}. We identify these known clusters as CHA-1 and CHA-2, respectively, and they are nestled about 80 pc behind LCC and the near Chamaeleon clusters in the plane of the sky \citep{LopezMarti13}. CHA-3 represents a third relatively compact and populous group in this region, and due to its distinctness as a clustered extension to Chamaeleon, we assign it the name Centaurus South. The last Chamaeleon subgroup, CHA-4, has limited overlap with Theia 94 from \citet{Kounkel19}, and has a much sparser spatial distribution compared to the other, much denser Chamaeleon clusters. Its identification as an independent group can be attributed to its notably different velocity distribution, which like the other Chamaeleon groups has relatively tight $\sim$1 km s$^{-1}$ velocity dispersion, despite its considerably larger extent. The age solutions for the Cha I and II clusters are essentially newborn as expected, with Centaurus South being older at $\sim$11 Myr. CHA-4 is by far the oldest of the groups we identify in Chamaeleon, at $\sim$37 Myr.

As shown in Fig. \ref{fig:periph_cha}, despite their physical separation, Cha I, Cha II, and Centaurus South all have similar transverse velocities to both Sco-Cen and each other. The radial velocities for the Centaurus South group cannot be checked due to a lack of past coverage, however radial velocities for Cha I and II have been reported as 12.9$\pm$1.6 km s$^{-1}$ and 14.6$\pm$1.2 km s$^{-1}$ respectively \citep{LopezMarti13}. This is very similar to the values in LCC, the nearest Sco-Cen subgroup, which has a radial velocity of $\sim$12 km s$^{-1}$ \citep{deZeeuw99}. The similarities of the kinematics of these peripheral clusters to the stars in Sco-Cen has previously been used to suggest that the material forming stars in Chamaeleon may be related to that in Sco-Cen \citep{Preibisch08}, however we find no evidence for the existence of lower-density stellar populations between the two associations that might suggest a direct structural link.

\subsection{Minor Groups} \label{sec:minorgroups}

Aside from Taurus, Perseus, Orion, Sco-Cen, and Chamaeleon, 22 other small clusters and groups are also identified in this paper. We assign all of these top level groups a name in accordance with either past literature or their location in the sky, depending on whether the group has been firmly established in the literature yet. Associations that are known but have not yet been assigned an at-large name in literature are assigned one that is descriptive of their location in the sky. While many of these groups have catalog identifications in \citet{Kounkel19} or \citet{Sim19}, their youth and proximity makes them worth highlighting, hence the inclusion of distinctive non-catalog names. The names used to refer to each of these groups are listed in Table \ref{tab:allsky_clusters}, along with basic information on each cluster.

Some of these groups are firmly connected to known objects, including open clusters IC 2391, NGC 2451A, ASCC 123, and the Pleiades, as well as somewhat larger and better known associations, including Perseus OB3, the Cepheus Flare region, and both Vela subgroups. Other groups, such as Cepheus Far North and Monoceros Southwest are present in literature but not yet well-established (see the discussion in Sections \ref{sec:cfn} and \ref{sec:mon}), while the remaining 12 are either absent from the literature completely or only noted as part of very large clustering surveys, such as \citet{Kounkel19} and \citet{Sim19}. The \citet{Sim19} catalog includes Taurus-Orion III as UPK 385, as well as Carina-Musca subgroup CM-2, which is identified as UPK 569. The Theia clusters from \citet{Kounkel19} include another five of the ten remaining clusters: Ophiuchus Southeast (Theia 70), Canis Major North (Theia 69), Aquila East (Theia 53), Taurus-Orion I (Theia 230), and Taurus-Orion II (Theia 116). Taurus-Orion IV also is partially represented in \citet{Kounkel19} as Theia 71, although the Theia group is limited to only the southern half of TO-IV in galactic coordinates. Lyra, Cepheus-Cygnus, and Cerberus (named after the obsolete constellation in eastern Hercules where it resides) are included as part of larger ``string" groups as outlined in \citet{Kounkel19}, but are not identified as distinct groups as we do in this work. Finally, the Fornax-Horologium group, which is located at a high-southern galactic latitude, appears to be completely absent from current literature.

Due to the large number of groups we identify, we do not perform a detailed overview of all of them as we have for Sco-Cen, Orion, Perseus, and Taurus. Instead, we only perform a deep investigation on those that show clear signs of substructure, including Perseus OB3, Monoceros Southwest, Vela-CG4, Cepheus Far North, and Carina-Musca, which are provided in the subsections below. The remaining subgroups should be adequately described by the statistics in Table \ref{tab:allsky_clusters}, and may receive more detailed followup in future publications.

\subsubsection{Perseus OB3} \label{sec:perob3}

Perseus OB3, also known as the $\alpha$ Persei cluster or Melotte 20, is a long-recognized nearby association identifiable by its wealth of O and B-type members, which form an overdensity clearly visible with binoculars or even the naked eye under dark skies \citep[e.g.,][]{Boss1910,Eddington1910}. This region has been well-studied over the past century, with the original $<$20 members identified in earlier works expanded to $\sim$300 by the first CCD photometry of the cluster \citep[e.g.,][]{Prosser92}, and later covered by wide-field astrometric studies \citep[e.g.,][]{Deacon04}. Hipparcos measurements of Per OB3 members provided distance estimates of $\sim$180 pc \citep{Mermilliod97,deZeeuw99}, a result that has changed relatively little in recent years \citep[e.g.,][]{Lodieu19}. While most members lie within a $\sim$3$^{\circ}$-wide core in the plane of the sky, a considerably larger halo has been uncovered in previous works \citep[e.g.,][]{Rasmuson1921,deZeeuw99,Lodieu19}, which was tied to the much larger \citep[and somewhat controversial, see][]{Crawford63} Cas-Tau group by \citet{Blaauw56}. The age estimates for the region have varied considerably, although most appear to fall between $\sim$50-80 Myr \citep[e.g.,][]{Meynet93,Basri99}

\begin{figure*}[t]
\centering
\includegraphics[width=16cm]{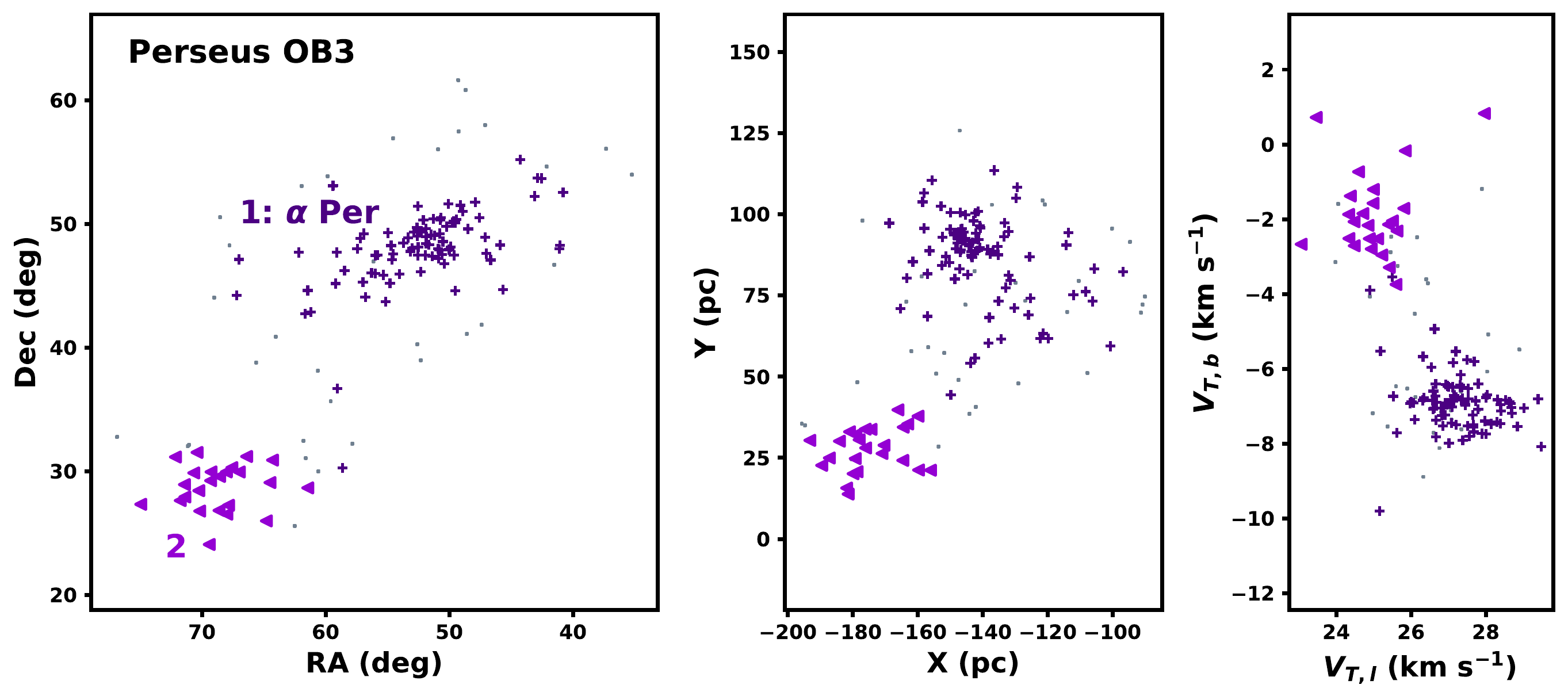}\hfill
\caption{Subgroups in the Perseus OB3 Association, plotted in RA/Dec, galactic X/Y, and galactic transverse velocity, and labelled according to their POB3 ID in Table \ref{tab:minor_sc}. Small grey dots mark unclustered Perseus OB3 members. POB3-1 corresponds to the known $\alpha$ Persei cluster, while the other subgroup, POB3-2, is a newly discovered, younger extension.}
\label{fig:aper}
\end{figure*}

As expected, we do identify a significant halo of stars belonging to Perseus OB3, although our extended population contains much more structure than previously identified, with an additional subcluster $\sim$20$^{\circ}$ to the southeast of $\alpha$ Persei \citep{deZeeuw99}. The extensions that we find beyond the core of Perseus OB3 are shown in Figure \ref{fig:aper}, where both the $\alpha$ Persei cluster (POB3-1) and the new southeastern extension (POB3-2) are shown in spatial and proper motion coordinates. POB3-2 is contiguous with the rest of Perseus OB3 (POB3-1/$\alpha$ Per) through low-density branches of stars connecting the two, however it is physically closer to sections of Greater Taurus, located within 30 pc of the centre of GT-9A in spatial coordinates compared to $\sim$70 pc for $\alpha$ Persei. The connection between POB3-2 and $\alpha$ Persei drawn by HDBSCAN is therefore most strongly rooted in kinematics, as POB3-2 is within only $\sim$5 km s$^{-1}$ of $\alpha$ Persei in transverse velocity, compared to $\sim$10 km s$^{-1}$ for the nearest Taurus groups. While the low density bridge of young stars connecting the two components of this region supports their connection, the significant distance between the two components in spatial coordinates ($\sim$75 pc) casts doubt on whether this connection is real. POB3-2 has near-nonexistent coverage in radial velocity, so we currently lack the 3-dimensional velocities necessary to confirm or reject the status of POB3-2 as a genuine subgroup of Perseus OB3. 

The age we derive for POB3-2 is considerably younger than $\alpha$ Persei, at $\sim$30 Myr compared to $\sim$50 Myr for the main cluster. Our $\sim$50 Myr age solution for $\alpha$ Persei is at the lower edge of the generally accepted age range of $\sim$50-80 Myr \citep[e.g.,][]{Meynet93,Basri99}, however large discrepancies would not be surprising in this case, as $\alpha$ Persei has a notably super-solar metallicity ([Fe/H]=+0.18) which may reduce the accuracy of our solar metallicity model \citep{Pohnl10}. Regardless, the fact that the age of POB3-2 is younger is robustly supported by our age solutions. 

\begin{deluxetable*}{cccccccccccccccc}
\tablecolumns{16}
\tablewidth{0pt}
\tabletypesize{\scriptsize}
\tablecaption{Subclustering inside smaller regions. The parent top-level cluster is reported in the TLC column, and the ID column gives the EOM subcluster, with Perseus divided further into its older and younger A and B subregions.} \label{tab:minor_sc}
\tablehead{
\colhead{TLC} &
\colhead{ID} &
\colhead{Name} &
\colhead{N} &
\colhead{RA} &
\colhead{Dec} &
\colhead{l} &
\colhead{b} &
\colhead{$D_{sky}$\tablenotemark{a}} &
\colhead{d} &
\colhead{$\mu_{RA}$} &
\colhead{$\mu_{Dec}$} &
\colhead{$V_{T, l}$} &
\colhead{$V_{T, b}$} &
\colhead{$\sigma_{V_T}$\tablenotemark{b}} &
\colhead{Age} \\
\colhead{} &
\colhead{} &
\colhead{} &
\colhead{} &
\multicolumn{2}{c}{(deg)} &
\multicolumn{2}{c}{(deg)} &
\colhead{(deg)} &
\colhead{(pc)} &
\multicolumn{2}{c}{(mas/yr)} &
\multicolumn{2}{c}{(km/s)} &
\colhead{(km/s)} &
\colhead{(Myr)}
}
\startdata
PER & 1A & NGC 1333       &  19 &  52.3 &  31.0 &  158.5 & -20.7 &   2.2$\times$0.6 &     296.9$\pm$8.3  &   7.4 &   -8.2 &  15.3 &  -2.9 &   1.2$\times$0.7 &   6.0$\pm$3.2 \\
PER & 1B &                &  77 &  58.5 &  32.8 &  161.7 & -16.0 &   3.1$\times$2.4 &     279.3$\pm$12.1 &   6.4 &  -10.0 &  15.1 &  -4.3 &   0.8$\times$0.6 &  17.5$\pm$0.9 \\
PER & 2A &  IC 348        &  42 &  55.8 &  32.2 &  160.2 & -18.0 &   1.3$\times$0.5 &     320.6$\pm$7.2\tablenotemark{c}  &   4.7 &   -6.3 &  11.6 &  -2.9 &   1.0$\times$0.9 &   4.7$\pm$0.5 \\
PER & 2B &                &  50 &  60.3 &  33.2 &  162.5 & -14.7 &   4.1$\times$2.4 &     308.5$\pm$16.3\tablenotemark{c} &   3.8 &   -6.3 &  10.4 &  -3.0 &   0.7$\times$0.7 &  17.1$\pm$1.1 \\
POB3 & 1 & $\alpha$ Per   &  89 &  52.5 &  48.1 &  148.4 &  -6.6 &   8.5$\times$4.8 &     166.3$\pm$15.5 &  23.5 &  -27.0 &  27.2 &  -6.9 &   1.0$\times$0.7 &  49.7$\pm$0.5 \\
POB3 & 2 &                &  24 &  68.7 &  28.7 &  171.0 & -12.6 &   5.3$\times$3.8 &     181.7$\pm$9.1  &  17.4 &  -23.4 &  25.0 &  -1.9 &   1.2$\times$0.9 &  30.0$\pm$2.5 \\
MSW &    1 &              &  12 &  93.5 &   3.1 &  205.9 &  -6.9 &   1.9$\times$1.5 &     322.4$\pm$4.9  &  -4.2 &   -3.9 &  2.3  &  -8.4 &   0.9$\times$0.7 &  25.3$\pm$2.6 \\
MSW &    2 & NGC 2232     &  23 &  97.0 &  -4.8 &  214.5 &  -7.4 &   1.6$\times$0.9 &     323.2$\pm$7.9\tablenotemark{c}  &  -4.7 &   -1.9 & -0.6  &  -7.7 &   0.8$\times$0.3 &  27.8$\pm$1.2 \\
MSW &    3 &              &  16 &  98.8 &  -4.2 &  214.9 &  -5.6 &   3.9$\times$2.4 &     223.4$\pm$8.0  &  -7.3 &   -3.6 & -0.1  &  -8.6 &   0.7$\times$0.4 &  24.8$\pm$2.2 \\
MSW &    4 & LP 2439      & 123 & 103.0 &  -5.9 &  218.3 &  -2.5 &   3.1$\times$2.1 &     284.2$\pm$15.0 &  -7.3 &   -2.6 & -1.4  & -10.3 &   0.8$\times$0.5 &  27.0$\pm$0.7 \\
VCG4 &   1 & UPK 535      &  21 & 127.0 & -52.4 &  268.7 &  -8.1 &   3.2$\times$2.2 &     310.9$\pm$8.9  & -13.0 &    3.6 & -15.4 & -12.6 &   1.2$\times$0.4 &  32.2$\pm$1.5 \\
VCG4 &   2 &              &  29 & 105.3 & -53.8 &  264.0 & -20.2 &   9.6$\times$2.5 &     198.7$\pm$8.3  & -12.7 &   11.8 & -14.4 &  -7.8 &   0.9$\times$0.7 &  37.8$\pm$1.3 \\
VCG4 &   3 &              &  10 & 112.3 & -46.2 &  258.4 & -13.2 &   4.2$\times$2.1 &     258.7$\pm$2.8  & -12.4 &    7.6 & -14.9 &  -9.7 &   0.7$\times$0.4 &  39.2$\pm$2.0 \\
VCG4 &   4 & Cr 135/UBC 7 & 145 & 107.5 & -37.2 &  248.5 & -12.6 &   3.0$\times$2.6 &     285.0$\pm$21.9 & -10.1 &    6.8 & -14.0 &  -8.5 &   1.1$\times$0.6 &  34.8$\pm$0.5 \\
VCG4 &   5 &              &  48 & 119.0 & -44.5 &  259.1 &  -8.2 &   4.3$\times$1.2 &     321.6$\pm$6.9\tablenotemark{c}  & -11.6 &    4.5 & -14.9 & -11.8 &   0.9$\times$0.6 &  32.1$\pm$1.0 \\
CFN &    1 &              & 174 & 334.4 &  76.1 &  113.7 &  17.9 &  15.3$\times$4.5 &     171.3$\pm$15.9 &  16.2 &    5.6 &  14.5 &  -2.8 &   2.3$\times$0.9 &  23.8$\pm$1.3 \\
CFN &    2 &              &  14 & 319.6 &  70.2 &  106.7 &  14.5 &   3.9$\times$2.4 &     235.1$\pm$7.0  &  10.1 &    6.4 &  12.9 &  -3.5 &   0.8$\times$0.3 &  25.8$\pm$3.7 \\
CHA &    1 & Cha I        & 112 & 167.2 & -76.8 &  297.1 & -15.1 &   1.5$\times$0.8 &     191.2$\pm$5.2  & -22.5 &    0.5 & -18.9 &  -7.7 &   1.0$\times$0.7 &   4.6$\pm$0.4 \\
CHA &    2 & Cha II       &  33 & 195.7 & -77.4 &  303.6 & -14.5 &   2.2$\times$0.8 &     199.1$\pm$4.4  & -20.2 &   -7.3 & -19.3 &  -6.0 &   0.9$\times$0.5 &   3.7$\pm$0.8 \\
CHA &    3 & Cen South\tablenotemark{$\dagger$}    &  39 & 206.6 & -63.4 &  309.2 &  -1.3 &   7.0$\times$3.7 &     186.6$\pm$8.2  & -20.0 &  -12.3 & -19.6 &  -7.0 &   1.5$\times$0.4 &  11.4$\pm$1.9 \\
CHA &    4 &              &  17 & 195.6 & -80.7 &  302.1 & -20.2 &  14.0$\times$3.0 &     192.7$\pm$18.4 &  -8.9 &   -8.3 & -11.9 &  -9.7 &   1.1$\times$0.4 &  36.5$\pm$1.7 \\
CM  &    1 &              &  23 & 133.8 & -63.8 &  280.1 & -12.1 &   6.1$\times$1.2 &     208.6$\pm$5.9  & -27.4 &   14.7 & -28.3 & -12.0 &   0.7$\times$0.5 &  33.5$\pm$2.5 \\
CM  &    2 &              &  75 & 163.9 & -70.7 &  293.5 & -10.3 &   7.2$\times$2.0 &     254.5$\pm$21.3 & -25.0 &    0.9 & -27.7 & -12.4 &   1.0$\times$0.6 &  33.9$\pm$0.7 \\
\enddata
\tablenotetext{a}{On-sky spatial extent in galactic l/b, in the form of the RMS in major axis $\times$ minor axis, when fit with a bivariate Gaussian.}
\tablenotetext{b}{The radial extent of the velocity distribution, in the form of the RMS in semi-major axis $\times$ semi-minor axis, when fit with a bivariate Gaussian.}
\tablenotetext{c}{has a member within 1 pc of the search horizon, and therefore members are likely present beyond that limit.}
\tablenotetext{\dagger}{name is newly assigned by this paper.}
\vspace*{0.1in}
\end{deluxetable*}

\subsubsection{Monoceros Southwest} \label{sec:mon}

Most of the region that we refer to as Monoceros Southwest (MSW) is a relatively recent discovery, with NGC 2232, a small young open cluster being its only well-established component. Until recently, relatively little interest has been dedicated to the area around NGC 2232, and as such the coverage of its parent environment has been quite limited. \citet{Claria72} performed the first major photometric survey and age estimate for the cluster, while \citet{Lyra06} significantly expanded on and refined this work, introducing the first CCD survey of the cluster, and computing a photometric age of $\sim$25-32 Myr. Suggestions of a broader population in the area first emerged through \citet{Zari18}, where a small 5-7$^{\circ}$-wide clump of stars with a tight proper motion distribution is identified out of the sample of photometrically young Gaia DR2 stars they investigate. This cluster was later characterized by \citet{Liu19}, where it is assigned the name LP 2439. \citet{Pang20} revisits the Monoceros region, identifying objects related to the two known clusters by applying the machine learning algorithm StarGO \citep{Zhen18} to a population of candidates within a 100 pc radius of NGC 2232 and a 2.8 mas yr$^{-1}$ proper motion radius centered between LP 2439 of NGC 2232. The result is the identification of four subgroups - NGC 2232, LP 2439, and two new, smaller groups, which are referred to by their colors of the \citet{Pang20} figures - green and purple. \citet{Pang20} finds all four of these groups to be approximately coeval, with an age of approximately 25 Myr, matching the earlier solutions from \citet{Lyra06}

\begin{figure}[t]
\centering
\includegraphics[width=8cm]{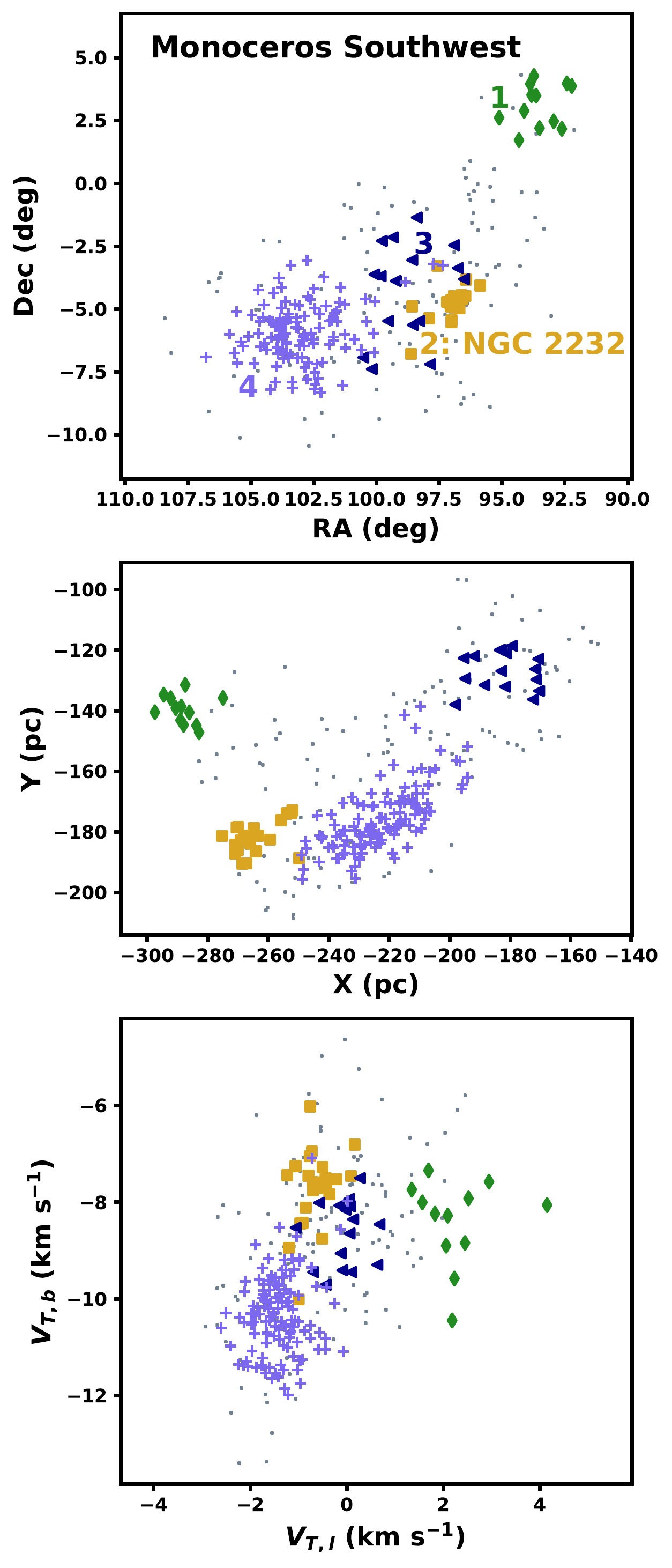}\hfill
\caption{Subgroups in Monoceros Southwest (MSW), plotted in RA/Dec, galactic X/Y, and galactic transverse velocity. Each subgroup is labelled according to their MSW ID in Table \ref{tab:minor_sc}, and small grey dots mark unclustered MSW members. }
\label{fig:monoceros}
\end{figure}

In our work we identify a large, substructured group in Monoceros Southwest, which contains just under 300 stars, divided into four subgroups. The results of subclustering in MSW are shown in spatial and velocity coordinates in Figure \ref{fig:monoceros}. The most compact of the subgroups we identify, MSW-2, corresponds to NGC 2232, while the larger and more dispersed subgroup MSW-4 corresponds to LP 2439 \citep{Liu19}. The group we refer to as MSW-1 matches the green group in \citet{Pang20}, while the purple group, which is sightly more dispersed, shares some membership with our top-level MSW population between NGC 2322 and MSW-1, but is not assigned a subgroup. The methods employed by \citet{Pang20} for the inclusion of potential members are much more permissive than those we use, only culling objects with inconsistent photometry, rather than our approach, which demands that photometry clearly indicate youth. This approach appears to work well for \citet{Pang20} in Monoceros Southwest, as the region appears to be relatively uncontaminated and easy to separate from the nearby background, enabling somewhat more detailed subclustering capable of revealing the more tenuous purple group. However, the remaining subgroup that we identify, MSW-3, is not identified in \citep{Pang20}, likely due to it falling partially outside of the 100 pc search radius around NGC 2232 used in that paper. This result highlights the complementary strengths of our survey relative to more targeted studies. While tailoring investigations to an individual group can be very effective at expanding on known populations, assumptions made to isolate that group can result in broader, more dispersed populations being missed, which our more photometrically selective but spatially and kinematically indiscriminate survey excels at finding.

The overall velocity distribution in Monoceros Southwest is similar to that of Orion. All subclusters are within 4 km s$^{-1}$ of each other in transverse velocity, and the velocity dispersions within groups are consistently below 1 km s$^{-1}$. Our age fits for each subgroup are all between 24 and 28 Myr, which, combined with our bulk age estimate of 25.5 Myr, is in agreement with the 25 Myr coeval age solution suggested in the literature \citep[e.g.,][]{Pang20, Lyra06}.

\subsubsection{Vela} \label{sec:vela}

\begin{figure}[t]
\centering
\includegraphics[width=8cm]{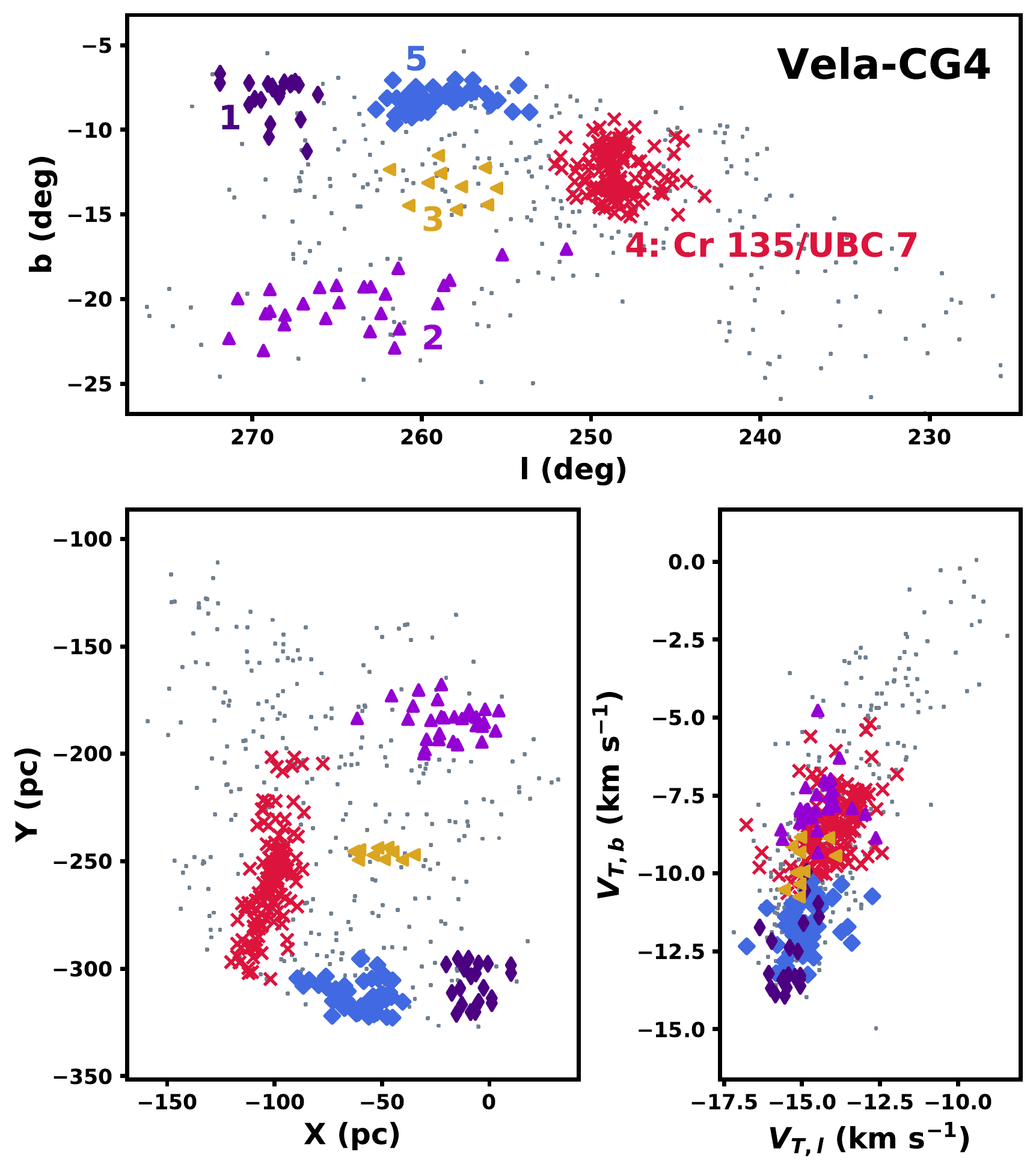}\hfill
\caption{Subgroups in Vela-CG4 \citep{CantatGaudin19}, shown in l/b galactic sky coordinates, galactic X/Y, and galactic transverse velocity. Each Vela-CG4 subcluster is labelled according to the VCG4 IDs from Table \ref{tab:minor_sc}, and small grey dots mark unclustered Vela-CG4 members.}
\label{fig:VCB}
\end{figure}

The association we now refer to as Vela OB2 was first identified by \citet{Kapteyn14} and received its first major kinematic analysis from \citet{deZeeuw99}, with the latter paper confirming both the common motions of the members and the inclusion of the open clusters NGC 2547 and $\gamma$ Velorum. Later work revealed that the region's kinematic substructure is divided into two distinct populations: one connected to the $\gamma$ Velorum cluster, and another connected to NGC 2547 \citep{Jeffries14,Damiani17}. \edit2{Two of the} most recent major substructure analyses of Vela were performed in \citet{CantatGaudin19b} and \citet{CantatGaudin19}, which identify significantly more substructure. By using the UPMASK unsupervised classification method, \citet{CantatGaudin19} identified seven kinematically distinct substructures in the Vela complex. Most of the larger Vela OB2 association they identify extends far beyond the 333 pc search radius used by our survey, limiting the subgroups we can cover, although sections of \citet{CantatGaudin19} groups IV and VII (hereafter Vela-CG4 and Vela-CG7) are close enough to be included in this work. Vela-CG4 is the older of the two groups at $\sim$34 Myr and contains NGC 2547, while Vela-CG7 has an age closer to 14 Myr and contains the $\gamma$ Velorum cluster.

Vela-CG7 is the smaller of the two Vela subgroups we identify, and our candidate members overlap with the \citet{CantatGaudin19} region in proper motions and spatial coordinates. The group's limited size is not inherent, however, as its membership is thought to fall almost entirely outside of our 333 pc search radius. \citet{CantatGaudin19} displays a much more extensive population, which is centred on the $\gamma$ Velorum cluster at a distance of $\sim$350 pc. Vela-CG7 does have known substructure outside of our search radius, however the section we identify consists of only Vela-CG7 component A, as originally defined in \citet{CantatGaudin19b}. As a result we do not identify substructure, with the entire group having a tight velocity distribution of $\sim$1 km s$^{-1}$. Our $\sim$14 Myr age fit for Vela-CG7 also agrees with the age solution plotted in \citet{CantatGaudin19} Figure 7 for the combined Vela-CG7 components A+B. Due to the lack of subclustering within our search radius, we do not include a separate plot for Vela-CG7, although it is visible through the top-level clustering analysis in Figure \ref{fig:allsky_clustering}. 

Unlike Vela-CG7, Vela-CG4 has considerable substructure within our search radius. The full extent of Vela-CG4 as defined in \citet{CantatGaudin19} contains numerous known subclusters both inside and outside of our search radius, including NGC 2451B, NGC 2547, Collinder 140, Collinder 135, and UBC 7. Of these known subgroups, only the adjacent open clusters of Collinder 135 and UBC 7 \citep{Kovaleva20,CastroGinard18} lie entirely within our search radius, together comprising what we call VCG4-4. In total, we identify five subgroups in Vela-CG4, including Collinder 135/UBC 7, which are shown in spatial and velocity coordinates in Figure \ref{fig:VCB}. Like most subgroups identified in other regions, the velocity dispersions of the Vela-CG4 subclusters are all near 1 km s$^{-1}$, although stars in the region follow a linearly-stretched transverse velocity distribution, spanning up to 15 km s$^{-1}$ along the v$_{T,b}$ axis. The mean velocities for all identified subgroups, however, all remain within 5 km s$^{-1}$ of each other, and most of this stretched velocity distribution can likely be attributed to geometric effects from the $\sim$800 square degree extent of the region. 

As the top-level HDBSCAN implementation defines it, Vela-CG4 contains a very broad distribution of stars, spanning over 150 pc across in both the X and Y galactic coordinates, excluding membership beyond our search radius. \edit2{Structures that extend far beyond the core Vela-CG4 regions have previously been proposed by \citet{Beccari20}, however those that we identify are considerably closer and further to the galactic south. }The two most distant groups (VCG4-1 and VCG4-5) are located closest to the dense, clustered regions in Vela-CG4 beyond our search radius such as NGC 2547 \citep{CantatGaudin19}, while VCG4-1 also matches spatially and kinematically with \citet{Sim19} group UPK 535. These distant groups connect through lower-density stellar populations to Collinder 135 and UBC 7 (VCG4-4), and through similar low-density populations to VCG4-2 and VCG4-3, overdensities located at $\sim$200 and $\sim$260 pc, from the sun, respectively. The ages we derive in Vela-CG4 are consistent with the groups being mostly coeval, with an age spread of just over 7 Myr \edit2{and a bulk age of $\sim$ 34 Myr, similar to the 35 Myr estimate for the populations discussed in \citep{Beccari20}}. \citet{CantatGaudin19} provides age solutions for UBC 7 and Collinder 135 separately at $\sim$35 and 40 Myr, respectively, in agreement with our $\sim$35 Myr age solution.

The largest and closest of the new nearby extensions to Vela-CG4 that we find is VCG4-2, which extends in the direction of the Sco-Cen association and some of the other clusters in Carina, including IC 2391 and NGC 2451A. The individual groups and clusters located in this region host mutually discontiguous transverse velocities, typically separated by $>$10 km s$^{-1}$, however they form a visually identifiable structure stretching between the IC 2602 branch of Sco-Cen (see Section \ref{sec:sc_sc}) and Collinder 135/UBC 7 in Vela. The individual components in this structure also host relatively similar ages, spanning between $\sim$30-50 Myr. It is therefore possible that interactions between these individual star formation events may have occurred and contributed to their mutual evolution, without the gravitational influence between them being large enough to for the resulting clusters to merge. 

\subsubsection{Cepheus Far North} \label{sec:cfn}

A few different papers have noted a population of nearby young stars in Northern Cepheus previously \citep[e.g., see][]{Tachihara05,Klutsch08,Oh17,Frasca18}, although only \citet{Klutsch20} has defined a substantial population with which preliminary analyses have been performed. The extent of these populations as explored by \citet{Klutsch20} is relatively limited, consisting primarily of a clump centred on (X,Y) = (-75, 125) pc, with most of the 32 confirmed members within 20 pc of that location. They estimated an approximate age of 10-20 Myr for the region \citep{Klutsch20}

\begin{figure}[t]
\centering
\includegraphics[width=8cm]{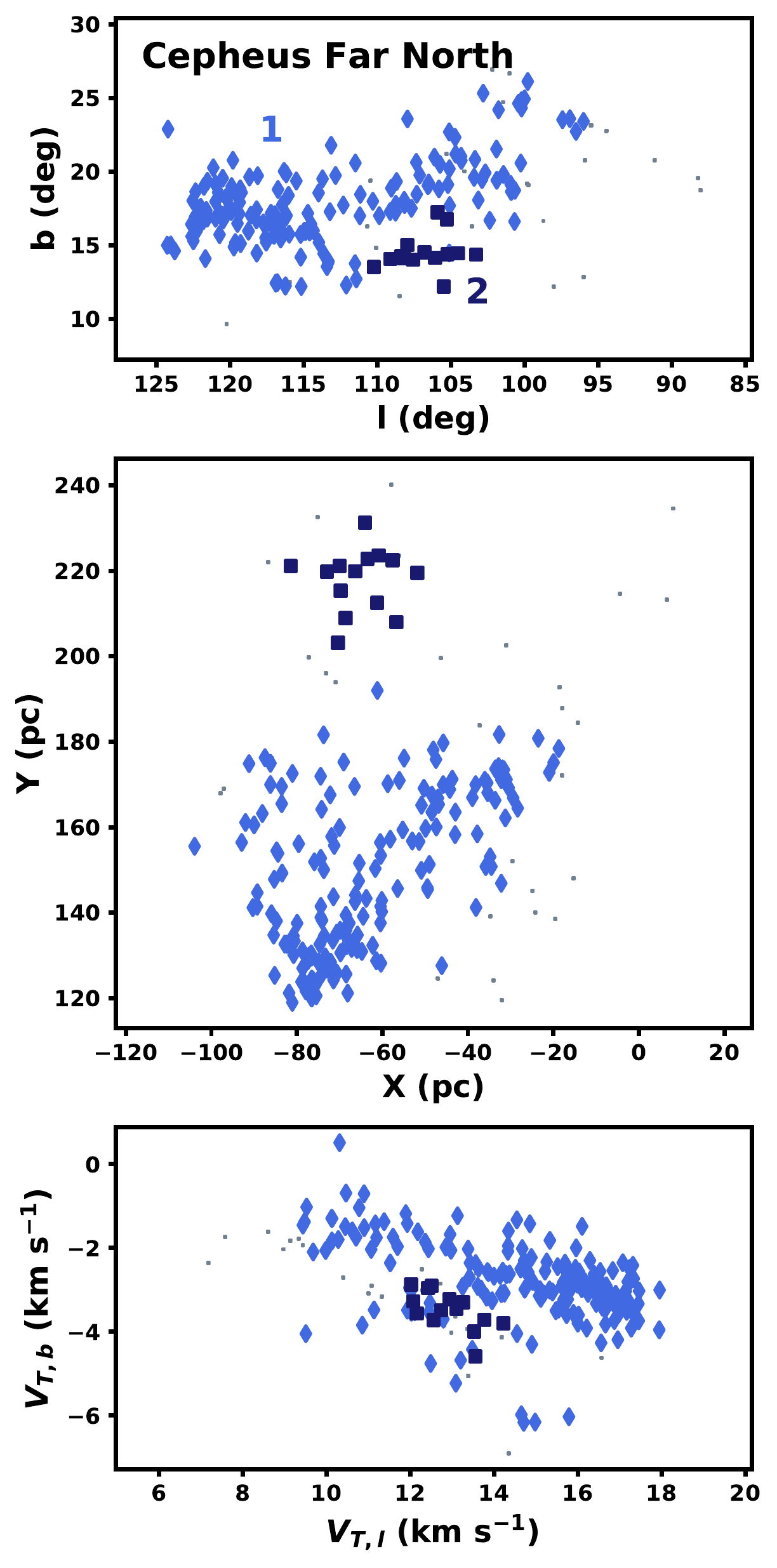}\hfill
\caption{Cepheus Far North plotted in l/b galactic sky coordinates, galactic X/Y, and galactic transverse velocity. The two subgroups are labelled according to their CFN IDs from Table \ref{tab:minor_sc}, and unclustered CFN members are marked with small grey dots. We identify 219 stars in the association, a significant expansion over the sample of 32 centered around (X,Y)=(-75,125) pc noted in \citet{Klutsch20}.}
\label{fig:CFN}
\end{figure}

The population we identify in the region is significantly larger compared to what is presented in \citet{Klutsch20}, with the membership expanded to 219 members spanning more than 100 pc at its longest axis. The clump containing most of the \citet{Klutsch20} members is included in this population, resting at its near edge, while we identify large numbers of stars extending far beyond this clump, with a mean distance of $\sim$180 pc. The group is primarily located within the bounds of Northern Cepheus, but also overflows significantly into Draco, reaching near the North Ecliptic Pole and overlapping with the TESS continuous viewing zone at its western edge \citep{TESS15}.  To distinguish it from more distant and better-known features associated with Cepheus \citep[e.g., see][]{Kun08}, we refer to this region as Cepheus Far North (CFN). The region's extent in spatial and velocity coordinates is displayed in Figure \ref{fig:CFN}, along with a subclustering analysis. HDBSCAN divides this group into two subgroups: one for the main body of the region (CFN-1) and one that is more distant and much less populous (CFN-2). We derive age solutions for CFN-1 and CFN-2 of $\sim$24 and 26 Myr respectively, suggesting that the star formation event that created CFN-2 may slightly predate star formation in the rest of the region, but the measurements are consistent with the populations being coeval. This age solution for CFN-1 is similar to the estimate of 10-20 Myr given by \citet{Klutsch20}, which is based on a subset of stars in the denser central region of CFN-1.

\subsubsection{Carina-Musca} \label{sec:CaM}

\begin{figure}[t]
\centering
\includegraphics[width=8cm]{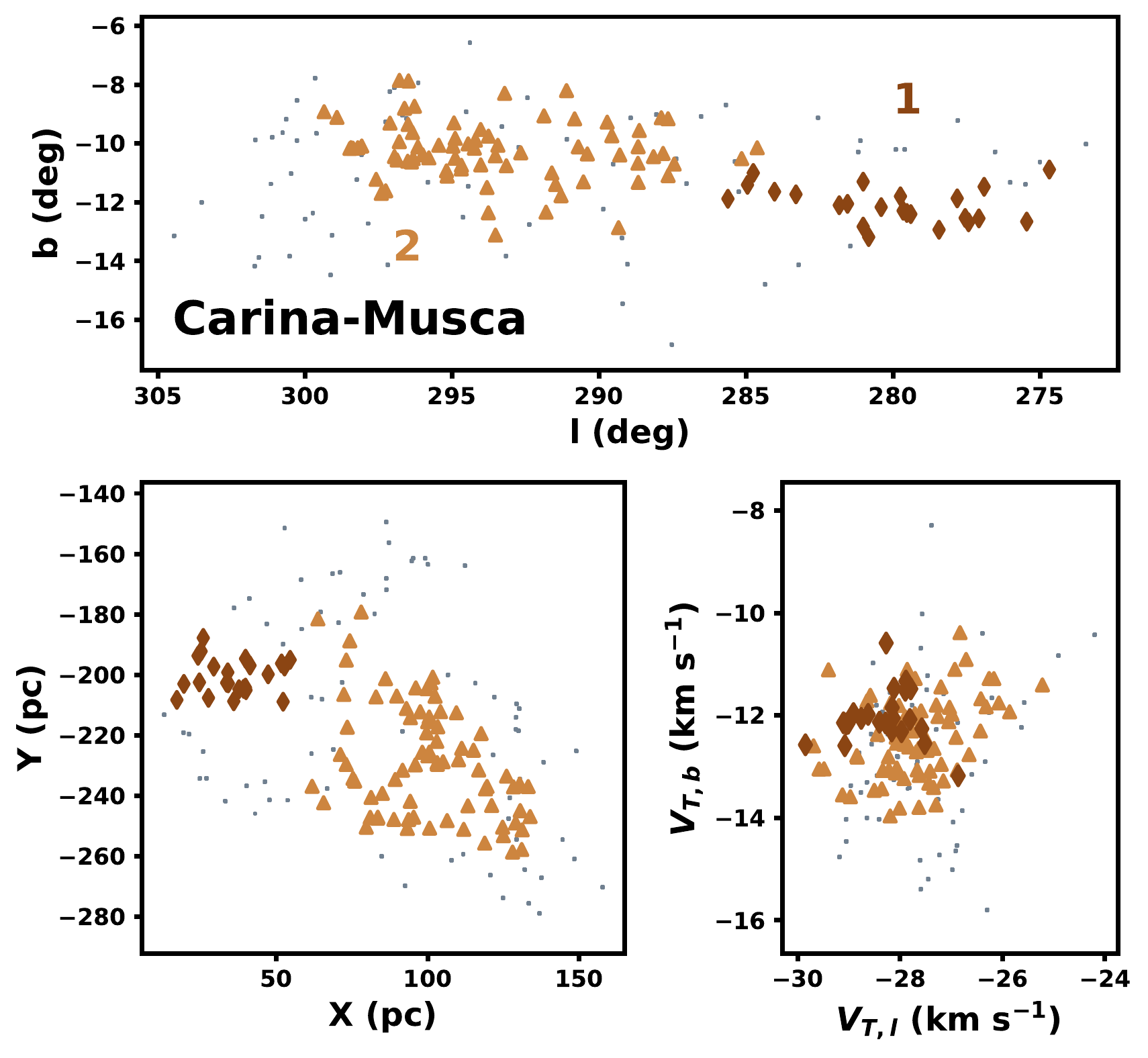}\hfill
\caption{Subgroups in the Carina-Musca association, shown in l/b galactic sky coordinates, galactic X/Y, and galactic transverse velocity. Each Carina-Musca subcluster is labelled according to the CM IDs from Table \ref{tab:minor_sc}, and unclustered Carina-Musca members are marked by small grey dots. CM-1 is not identified in literature, nor is the top-level structure enclosing the two subgroups.}
\label{fig:CaM}
\end{figure}

The final minor association with visible substructure that we identify is the Carina-Musca Association. At 168 members, the population of identified members in this group is quite large, bigger than any of the individual open clusters as well as Perseus OB3. It contains two subgroups: CM-1, which is relatively close to the clusters that bridge the gap between Vela and Sco-Cen, and CM-2, which is nearly 100 pc more distant on average. The association as a whole has very limited recognition in the literature, with the only exception being the densest region of CM-2, which appears as UPK 569 in \citet{Sim19}. The kinematics of the region are very consistent, with a velocity dispersion of only $\sim$1 km s$^{-1}$ for the entirety of Carina-Musca despite its large size and substructure. The ages, all at $\sim$33 Myr, enforce the internal consistency of properties within Carina-Musca, with an incredibly tight pre-main sequence, and less than 1 Myr separating the age fits for the two subgroups.

Carina-Musca also hosts what appear to be filamentary stellar structures crisscrossing the association, especially in CM-2, where the X/Y panel of Figure \ref{fig:CaM} shows two completely star-free voids centered around (X,Y) = (84, -221) and (104, -238) pc, completely surrounded by thin linear stellar overdensities. This apparent preservation of filamentary structure further enforces the very small internal velocity dispersions, as structures with significant gravitational interactions would not be expected to survive for the $\sim$33 Myr age of the region. The Carina-Musca association may therefore serve as a useful case-study for star formation in a large-scale turbulence dominated region, a possibility explored in more depth in Section \ref{sec:dis-lsssf}.

\section{Local Patterns in Star Formation} \label{sec:discussion}

\subsection{Large-Scale Simultaneous Star Formation} \label{sec:dis-lsssf}

Multiple sites of recent star formation we uncover, including Vela-CG4, Monoceros Southwest, Carina-Musca, and to a lesser extent Cepheus Far North show remarkably consistent and effectively identical ages for their subgroups despite their immense size, all of which exceed 100 pc across along at least one axis. These groups are also notable for their very low densities, especially for Carina-Musca, which has 168 members over an area more than 100 pc wide, compared to the approximately coeval newborn groups in the Taurus Molecular Cloud, which have an even larger membership despite covering only $\sim$40 pc from end to end. While some regions we discuss host denser subgroups, such as Collinder 135/UBC 7 in Vela-CG4, all such regions also host subgroups with densities at least comparable to that of the TMC populations. The dearth of gravitational influence inferred by these low densities is supported by the morphology of the regions we observe, all of which show evidence of filamentary structures in the form of extended stellar overdensities, including in Monoceros Southwest, where filamentary structure was independently identified in \citet{Pang20}. Small virialized clusters have been shown to disperse almost entirely over the $>$20 Myr lifetimes of these regions \citep{Moeckel12}, so the fact that these tenuous stellar structures remain intact suggests a lack of internal gravitational interactions, effectively allowing the stars to sit in a near-stationary state relative to each other for tens of millions of years. In Carina-Musca this near-stationary state is reflected in a very tight $\sim$1 km s$^{-1}$ tangential velocity dispersion for the entire association, although the relative motions of stars are difficult to ascertain without radial velocities. 

Due to the apparent weakness of the gravitational forces in these regions, global gravitational collapse is likely a poor explanation for these structures \citep[e.g.,][]{VasquezSemadeni19,Krause20}. Instead, turbulence-driven collapse seems more likely. In this scenario, rather than collapse being driven directly from the self-gravity of the cloud, the cloud's internal turbulence and associated shocks are able to produce local density fluctuations capable of creating dense cores \citep[][]{Klessen00,Krause20,MacLow04}. The cores generated by these turbulent processes tend to form along filaments and sheets, not unlike the stellar structures left behind in Carina-Musca and elsewhere \citep[e.g.,][]{MacLow04}. A turbulent environment is capable of not only driving the generation of overdensities and star-forming cores, but also promoting these cores' collapse \citep[e.g.,][]{Padoan02,Hennebelle08}. Sustained turbulence in the surrounding environment applies a turbulent pressure to dense cores, driving collapse in locations with weak gravitational binding \citep{McKee02,Field11}. Recent work on dense cores has shown that in the sparser environments present in Taurus, external cloud turbulence likely has a dominant influence on the collapse of dense cores, relative to more massive environments like Orion A, where self-gravitation and cloud weight pressure are more important \citep{Kirk17,Kerr19}. The sparse groups we identify here may therefore serve as a larger-scale analog to the star formation events visible in Taurus, which could make for useful comparisons with new modelling. It remains unclear whether simultaneous turbulent collapse on 100 pc scales is possible in isolation or whether a significant injection of turbulence, such as from a high-speed supernova shock, is required to produce the approximately coeval formation times we observe \citep[e.g.,][]{Padoan16}. More detailed coverage, including radial velocities and improved ages, will be necessary to better constrain the dynamical histories of groups like Carina-Musca and subsequently assess their origins. 

\subsection{Evidence for Sequential Star Formation}\label{sec:dis-seqprop}

\begin{figure}[t]
\centering
\includegraphics[width=7.1cm]{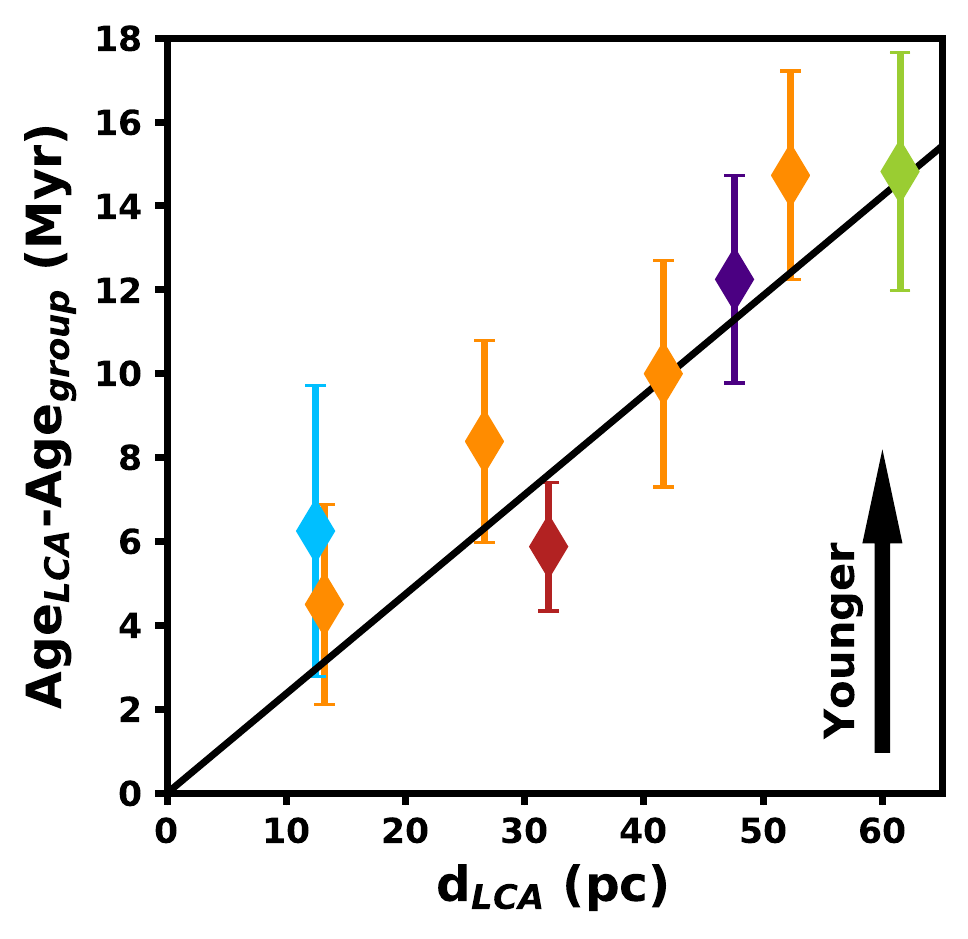}\hfill
\caption{The difference in age relative to the LCA, the oldest arc of subclusters in Sco-Cen (excluding the IC 2602 branch), as a function of distance to the nearest LCA component group. Color schemes match those in figures \ref{fig:scocen_subclustering} and \ref{fig:twa}, with LCC sub-groups in orange, TW Hydrae in indigo, SC-19 in blue, $\eta$ Cha in green, and SC-11 in red. Younger clusters are consistently located further from the LCA, suggesting a gradual progression of star formation outwards from the LCA.}
\label{fig:seq_prop}
\end{figure}

Age gradients are left behind by multiple different star formation mechanisms, and the patterns can provide important hints as to the star formation history of a region \citep[e.g.,][]{Elmegreen77,Krause20}. While most of our regions have either tenuous or nonexistent age gradients, Sco-Cen is an exception, most notably hosting gradients from older ages along the Libra-Centaurus Arc (LCA) to progressively younger ages farther away. This gradient appears to apply across all subgroups in and around LCC, as well as all LCA-adjacent groups west of Lupus. We present the age differences between these young populations and the nearest LCA subgroup as a function of distance in Figure \ref{fig:seq_prop}, and we find a clear linear trend between age and distance. We apply a least squares fit to these results, only allowing the slope to vary, as the intercept must be at zero to allow the LCA to have ages and distances of zero relative to itself. The result upon inverting that slope is a star formation propagation speed of 4.12 $\pm$ 0.19 km s$^{-1}$ (4.12 $\pm$ 0.19 km s$^{-1}$ = 4.21 $\pm$ 0.19 pc Myr$^{-1}$), with a reduced $\chi^2$ of 0.55. To our knowledge, this result is the first large-scale empirical measurement of a star formation propagation speed, providing a useful comparison for theoretical models. 

Shocks from supernovae or feedback from energetic young clusters are commonly suggested as triggers of star formation, creating self-gravitating overdensities as they pass through a molecular cloud, forming stars \citep[e.g.,][]{Elmegreen02,Dale07}. In many known regions, the most active sites of star formation appear on the periphery of a shell carved out by radiation from the most recent generation of star formation \citep[e.g.,][]{Phelps97,Lefloch00,Lee07,Zavagno10, Elmegreen00}. While our stellar populations do not show clear radial trends originating in any one location, we do find a coeval semicircular population in the form of the LCA, possibly consistent with triggered formation from an supernova shell compressing a large gas cloud. However, the populations adjacent to the LCA in Sco-Cen increase in age as a function of distance from the nearest component of the LCA, a result inconsistent with spherical shock propagation, or any other form of shock such as those generated through the collision of molecular clouds \citep{Elmegreen11}. This close trend between the age and position relative to the LCA suggests that rather than a single external shock creating a propagating overdensity, the triggers for these later generations may originate directly in those earlier generations of stars. Substantial populations of B stars are known to exist in the LCA and LCC, and O stars likely existed before ending their shorter main sequence lifetimes as supernovae, so they serve as a likely driver for these observed patterns \citep[e.g.,][]{deZeeuw99,Rizzuto11}. 

The propagation of star formation prompted by the influence of earlier generations is referred to as sequential star formation, first suggested in \citet{Blaauw64} and described in detail by \citet{Elmegreen77}. In this process, radiation pressure from O and B stars powers an ionization shock front, which can generate overdensities capable of supporting star formation. In a large star-forming region, sequential star formation can continue for millions of years, forming new generations of stars until the cloud is exhausted \citep{Elmegreen77}. The precise speed at which sequential star formation is expected to advance has yet to be predicted in the literature, especially on the large scales we observe, due to the lack of simulations covering the range in space and time covered by Sco-Cen. Observational studies of these speeds are also largely absent on the scales we considered, with the existing work only considering gradients on scales of a few pc or Myr \citep[e.g.,][]{Getman18,Lim18}. While we are currently unable to make comparisons, the star formation progression speed may provide a useful probe of gas properties in the parent cloud, as properties such as magnetic fields and internal turbulence are expected to increase the cloud's resistance to collapse and potentially slow the progression of any sequential star formation event \citep[e.g.,][]{BallesterosParedes20}. Through these inferences of cloud properties, the importance of additional processes that regulate them can also be assessed, such as turbulence excited by stellar feedback, which can inject energy into the cloud ahead of the current generation of star formation \citep{Offner18}.

Regardless of the inferences that can be drawn about gas properties and dynamics from the rate of star formation progression, the configuration of star formation near the LCA in Sco-Cen is consistent with sequential star formation, and represents the strongest indication to date that this process occurs on timescales of $\sim$20 Myr. Other features in Sco-Cen also show gradients potentially consistent with these mechanisms, such as those in Lower Sco and towards Corona Australis. However, more complete and precise age mapping is necessary to compare them to our result in the LCA. Expanding the sample of sites with evidence of sequential star formation will be important to ascertain whether the mechanisms that regulate propagation speed are consistent across clouds or whether fundamental changes to the cloud properties can slow or accelerate sequential propagation. 

\subsection{Repeated Star Formation Bursts}\label{sec:dis-sfb}

Multiple groups we identify show star-forming events with clear age separation, but essentially identical kinematics and similar spatial extents. The clearest example of this is in Perseus, where both subgroups do not have HDBSCAN-defined substructure but do have very clear separation in age, with essentially newborn ages around the Perseus Molecular Cloud in the west, and more dispersed stars with ages around 17 Myr in the east. GT-9 has subregions separated enough to be identified as independent leaf structures, but there we identify the same age structure: one essentially newborn subgroup in GT-9A, and a second, $\sim$12 Myr old subgroup in GT-9B. Both have no visible gradient in age, ruling out a continuous sequential star formation event. A few other sites show similar patterns, such as Perseus OB3 and a few other subgroups in Taurus, but the larger mutual separations between subgroups in spatial and velocity coordinates makes the connections between these different-age subgroups weaker relative to those in GT-9 and Perseus.

After star formation takes place in a dense cloud, it is generally expected that the remaining gas present around the young stars will be dispersed by stellar feedback from newly formed stars \citep[e.g.,][]{Walch12,Dobbs14}. If the initial collapse only affects part of the cloud, a sequential star-forming event may begin to propagate across the cloud as newly formed stars collect gas at higher densities in a shock that moves across the region, as described in Section \ref{sec:dis-seqprop} \citep{Elmegreen77}. Otherwise, these feedback mechanisms are expected to disrupt future star formation \citep{Walch12,Grudic20b}. In Perseus (and to a lesser extent GT-9), however, we observe time-separated bursts of star formation with no age gradient between them, as well as the near-identical kinematics and a close spatial connection. The internal consistency of these groups makes it unlikely that these star formation bursts originated in fully distinct environments, suggesting that processes from within the same environment were able to force a period of dormancy between the star formation events. 

Gas dispersal by stellar feedback is a likely cause for the end of star formation in the initial star-forming events \citep[e.g.,][]{Walch12}. However, each of the groups we discuss here appear to have been able to re-accumulate enough gas afterwards to support a new burst of star formation. One possible explanation for these secondary star formation bursts is that gas collection by flows that fed the initial star-forming event continued after the dispersal of the gas at the original collection point. In this scenario, star formation at the first star-forming site ended once stellar feedback dispersed the gas there, but new gas continued to stream in along the original flows. These streams created new collection points capable of supporting a second burst of star formation, like those in Perseus and GT-9. This process may be comparable to the generation of molecular clouds through the collision of convergent flows discussed in \citet{Dobbs14}, with gas from stellar feedback recycling into the next episode of star formation. \edit2{A similar mechanism was proposed in \citet{Kroupa18} to explain discrete star-forming events in Orion, however the dynamical ejection of high-mass stars that they suggest to permit the continuation of star formation after disruption is unlikely to have significant influence in the dynamically cooler Perseus region. }The spatial and temporal scales spanned by the stars in our study have \edit2{however} not been well-sampled in simulations due to limitations in computing power and dynamic range. Our possible explanation of co-spatial episodic bursts \edit2{therefore} requires new simulations to explore feasibility and develop a rigorous theoretical model

\subsection{Defining Top-Level Regions}\label{sec:dis-subgroups}

In our HDBSCAN implementation, the subgroups that we identify are included as part of the same larger region when they are connected to other subgroups in spatial and velocity coordinates through stellar overdensities. While the presence of apparent physical and kinematic connections is a commonly used and logical standard in clustering \citep[e.g.,][]{Zari19,Kounkel19}, it does not exclude the possibility of false connections emerging. For example, close interactions between two unrelated groups after formation may leave a trail of stars linking them, which our HDBSCAN implementation could group together at the top level. Alternatively, unrelated groups could be made contiguous through the ejection of members from a cluster's core, extending its envelope. 

Falsely connected associations would be physically contiguous, but have centers that are widely separated in spatial and velocity coordinates. Since we do not include age or traceback in our clustering, we are unable to investigate whether subclusters with wider spacing in spatial and velocity coordinates and therefore apparently weaker mutual connections were likely closer at the time of formation. A few regions we identify have subgroups with these seemingly weak connections to one another. These include Perseus OB3, which has wide separations between subgroups in space, velocity, and age, Sco-Cen, which has a significant velocity and age division between its main body and the IC 2602 branch, and Taurus, which has many subgroups with significant separations from one another by multiple metrics. All of these sets of subgroups are contiguous with one another through lower-density stellar populations, however their mutual separations in spatial and velocity coordinates, as well as significant age gaps make their physical association with one another more uncertain. 
 
Ruling out false connections between subgroups will require the addition of radial velocity measurements to the transverse velocity vector we use for clustering. Transverse velocities are plane-of-sky measurements, and they are therefore vulnerable to projection effects that can hide genuine velocity difference or induce spurious ones. The current implementation suffers greatly from these projection effects, contributing to the velocity scatter in Taurus and elsewhere, and preventing our identification of any known groups within $\sim$50 pc of the sun. The addition of radial velocity measurements would eliminate all projection effects while also completing the velocity vector, allowing 6-dimensional space-velocity clustering and providing a robust assessment of kinematic similarity between groups, regardless of their distance from one another in the plane of the sky.
 
Having this complete 3d velocity vector also enables long-term traceback of the stellar motions \citep[e.g.,][]{Kraus17}. Recent developments have been made at improving traceback in young stellar populations \citep[e.g.,][]{Crundall19}, so after the completion of follow-up radial velocity observations, the complete history of these populations can be established, and possibilities for a common origin can be better assessed. Age determinations (such as from the luminosity of the Lithium Depletion Boundary) will also be necessary to exclude the possibility of reddening anomalies causing subgroups to be falsely marked as young,  while also helping to establish their location at the time of formation in relation to other subgroups. This traceback may also help to establish groups that have differing velocities, but may have had interactions with other groups near the time of formation, perhaps as part of opposing colliding gas flows. Therefore, while the top-level regions we identify in this work should not be treated as entirely homogeneous in formation origin, they do connect subgroups where a common origin is likely and mutual association can be robustly evaluated as more accurate ages and complete 3-d velocity information becomes available. 

\section{Conclusion} \label{sec:conclusion}

We have photometrically identified $>3 \times 10^4$ probable young stars within a distance of 333 pc, and assessed the distinct groups and clusters among these stars using the HDBSCAN density-based hierarchical clustering algorithm. Through this method we have significantly expanded the census of known young stars in the Solar neighborhood, identifying new structures, and revealing new star formation patterns that trace the progression of star formation, particularly in some of the large and sparse groups that are found distributed throughout the Solar neighborhood. The key conclusions and discoveries reached through this work are:

\begin{enumerate}
    \item We identify 27 robust young groups and associations within 333 pc of the Sun at the top level. Of these, 15 have significant presence in the literature, eight have only been recognized as part of the large catalogs from \citet{Kounkel19} and \citet{Sim19} (Carina-Musca, Ophiuchus Southeast, Canis Major North, Aquila East, and Taurus-Orion I, II, III, and IV), and the remaining four (Cerberus, Lyra, Cepheus-Cygnus, and Fornax-Horologium) are identified as distinct groups for the first time.
    \item Ten of these groups show significant substructure, which is manifested in the form of HDBSCAN-defined excess of mass (EOM) and leaf-level subgroups within each of these larger structures. We provide complete overviews of the structure in each of these cases: Sco-Cen, Greater Taurus, Greater Orion, Perseus, Chamaeleon, Vela-CG4, Perseus OB3, Monoceros Southwest, Cepheus Far North, and Carina-Musca. 
    \item Nearly all large known groups contain substructures that we newly identify, including Eridanus North in Greater Orion, Sco-Cen subgroups SC-7, 10, and 16, which connect the region's known extent to Corona Australis, and Centaurus South, a somewhat older clustered extension to the Chamaeleon complex.
    \item We identify a complex network of subgroups in Taurus, linking the current sites of active star formation directly to stars considerably older than 10 Myr.
    \item The two Perseus subgroups, as well as GT-9 in Taurus, all contain an older and a younger stellar population, with near-identical kinematics and close spatial connections between these separate generations. Stellar feedback from the earlier star formation burst may have prompted a period of dormancy in the region during which gas continued to collect, eventually permitting a second burst of star formation.
    \item We have identified the Libra-Centaurus Arc (LCA), an old semicircular structure along the northern edge of Sco-Cen that appears to represent the first star formation in the region and a possible trigger for later star formation nearby. We also observe a smooth age gradient towards nearby subgroups, which appears to indicate the presence of sequential star formation with a propagation speed of 4.12 $\pm$ 0.19 km s$^{-1}$ (4.21 $\pm$ 0.19 pc Myr$^{-1}$).
    \item The presence of large, sparse, and approximately coeval populations in Monoceros Southwest, Vela-CG4, Cepheus Far North, and Carina-Musca may indicate mechanisms for large-scale ($\gtrsim$ 100 pc) simultaneous star formation in turbulence-dominated environments.
\end{enumerate}

\acknowledgments
ACR was supported as a 51 Pegasi b Fellow though the Heising-Simons Foundation. SSRO acknowledges support from the National Science Foundation (NSF) Career Grant 1748571 and NSF AAG 1812747. RMPK was supported in part by NASA grant NSSC18K0405 through the ADAP program. RMPK acknowledges the use computational resources at the Texas Advanced Computing Center (TACC) at the University of Texas at Austin, which was used for the more computationally intensive operations in this project. 

\vspace{5mm}
\facilities{Gaia}

\software{Astropy \citep{2013A&A...558A..33A},  HDBSCAN \citep{McInnes2017}, Matplotlib \citep{Hunter07}, SciPy \citep{Scipy20}}

\appendix

\section{Group Isochrone Fits} \label{appendix}

\edit1{Here we present the color-magnitude diagrams and corresponding age fits for each group and subgroup we identify in this work, as described in Section \ref{sec:groupage}). One example is given in Figure \ref{fig:cmdexemplar}, and the remaining fits are available in the online version of this paper.}

\setcounter{figure}{0}
\renewcommand{\thefigure}{A\arabic{figure}}

\figsetstart
\figsetnum{A1}
\figsettitle{Isochrone fits for each group}

\figsetgrpstart
\figsetgrpnum{A1.1}
\figsetgrptitle{Cepheus Flare}
\figsetplot{age_fit_1.pdf}
\figsetgrpnote{Isochrone fits for the group, with the best fit isochrone represented by the blue curve. The stars used in the fit are marked with black diamonds, and those that are removed due to RUWE and weight cuts are marked as red dots. The group name and age are both labelled on the figure.}
\figsetgrpend

\figsetgrpstart
\figsetgrpnum{A1.2}
\figsetgrptitle{Taurus-Orion I}
\figsetplot{age_fit_3.pdf}
\figsetgrpnote{Isochrone fits for the group, with the best fit isochrone represented by the blue curve. The stars used in the fit are marked with black diamonds, and those that are removed due to RUWE and weight cuts are marked as red dots. The group name and age are both labelled on the figure.}
\figsetgrpend

\figsetgrpstart
\figsetgrpnum{A1.3}
\figsetgrptitle{Ophiuchus Southeast}
\figsetplot{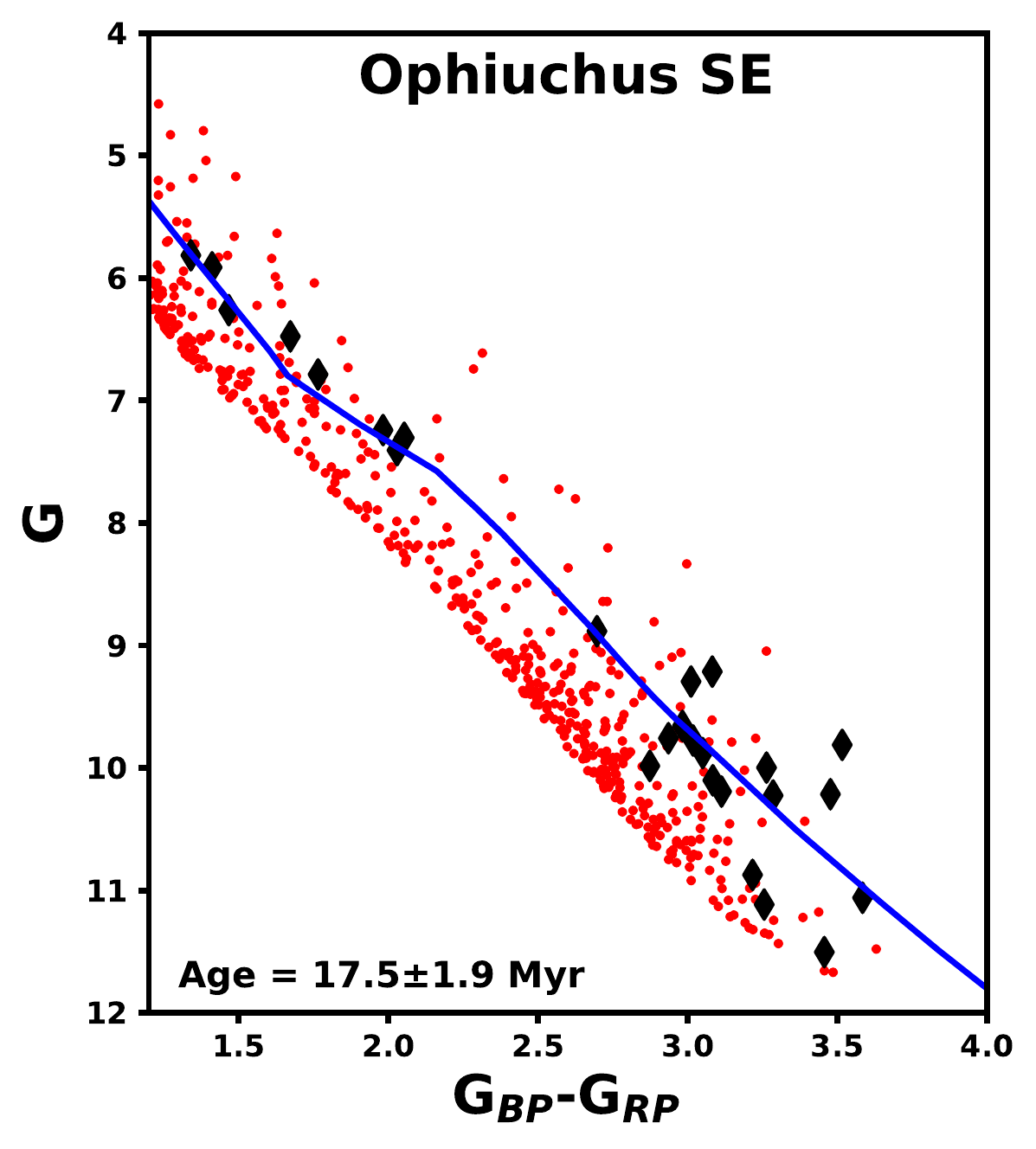}
\figsetgrpnote{Isochrone fits for the group, with the best fit isochrone represented by the blue curve. The stars used in the fit are marked with black diamonds, and those that are removed due to RUWE and weight cuts are marked as red dots. The group name and age are both labelled on the figure.}
\figsetgrpend

\figsetgrpstart
\figsetgrpnum{A1.20}
\figsetgrptitle{Monoceros Southwest}
\figsetplot{age_fit_26.pdf}
\figsetgrpnote{Isochrone fits for the group, with the best fit isochrone represented by the blue curve. The stars used in the fit are marked with black diamonds, and those that are removed due to RUWE and weight cuts are marked as red dots. The group name and age are both labelled on the figure.}
\figsetgrpend

\figsetend

\begin{figure}[h!] 
\centering
\includegraphics[width = 2.4in]{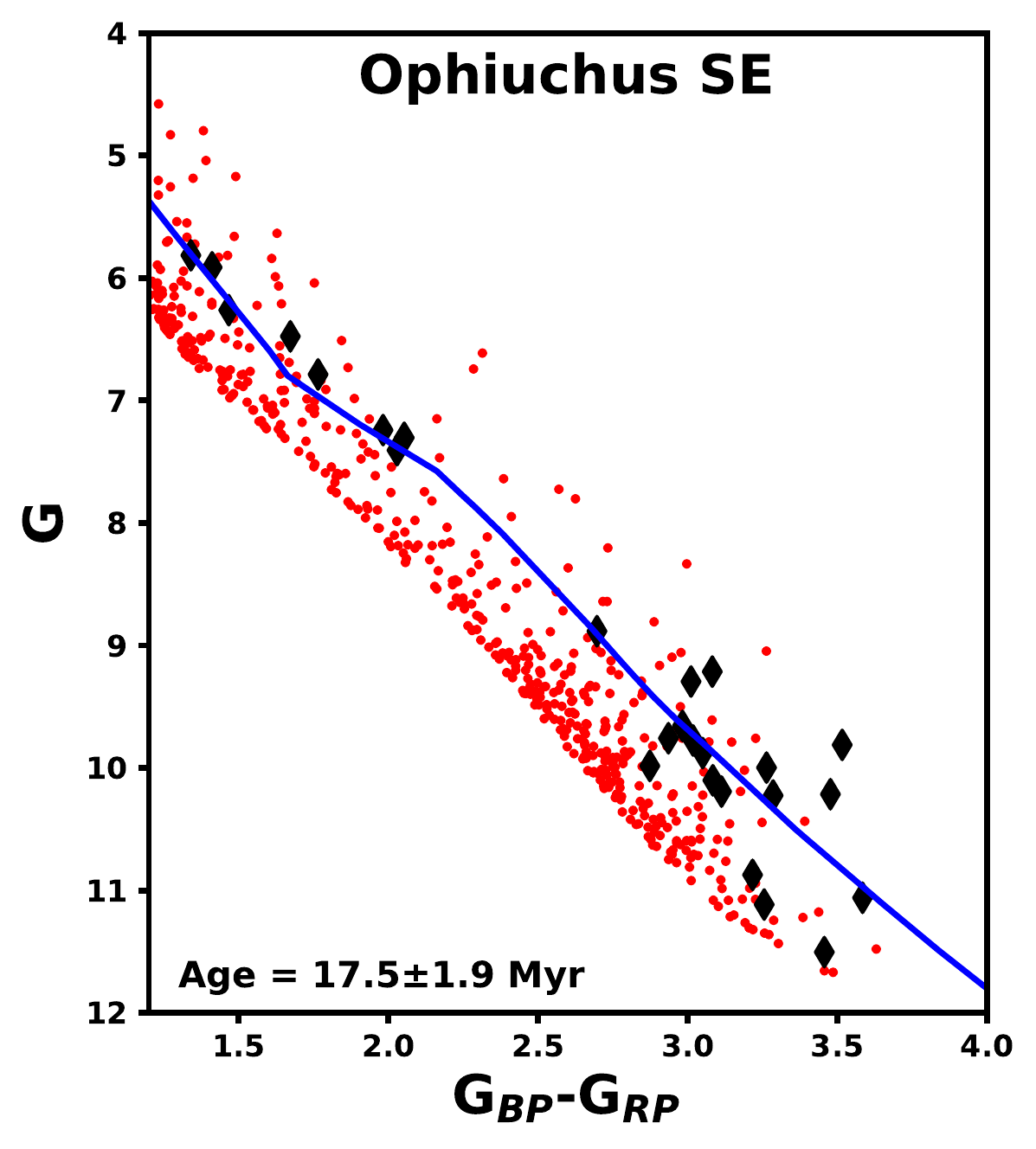}
\caption{\edit1{Isochrone fits for each group, with the best fit isochrone represented by the blue curve. The stars used in the fit are marked with black diamonds, and those that are removed due to RUWE and weight cuts are marked as red dots. The group name and age are both labelled on the figure. The remaining fits for each group are available in the online version of this paper.}}\label{fig:cmdexemplar}
\end{figure}

\bibliography{sample63}{}
\bibliographystyle{aasjournal}

\end{document}